\documentclass[%
 reprint,
superscriptaddress,
 amsmath,amssymb,
 aps,
prx
]{revtex4-2}

\usepackage{graphicx}% Include figure files
\usepackage{dcolumn}% Align table columns on decimal point
\usepackage{bm}% bold math
\usepackage{color,soul}
\usepackage{braket}
\usepackage{hyperref}% add hypertext capabilities
\newcommand{\blue}[1]{\textcolor{black}{#1}}

\newcommand{\rev}[1]{\textcolor{black}{#1}}

\begin{document}

%\preprint{APS/123-QED}

\title{
%Twisted bilayer Bi$_2$(Te,Se)$_3$
%Quantum spin Hall effect in twisted bilayer Bi$_2$(Te,Se)$_3$
%Quantum spin Hall effect in twisted bilayer Bi$_2$(Te$_{1-x}$Se$_x$)$_3$
Quantum spin Hall effect from multi-scale band inversion in twisted bilayer Bi$_2$(Te$_{1-x}$Se$_x$)$_3$
%Moir\'e-scale quantum spin Hall phase in twisted bilayer Bi$_2$(Te,Se)$_3$
}% Force line breaks with \\
%\thanks{A footnote to the article title}%

\author{Ikuma Tateishi}  \email{ikuma.tateishi@riken.jp}
 \affiliation{RIKEN Center for Emergent Matter Science, Wako, Saitama 351-0198, Japan}
\author{Motoaki Hirayama} \email{hirayama@ap.t.u-tokyo.ac.jp}
  \affiliation{RIKEN Center for Emergent Matter Science, Wako, Saitama 351-0198, Japan}
 \affiliation{Department of Applied Physics, University of Tokyo, Tokyo 113-8656, Japan}
  \affiliation{JST, PRESTO, Hongo, Bunkyo-ku, Tokyo 113-8656, Japan}
%\collaboration{MUSO Collaboration}%\noaffiliation

% \homepage{http://www.Second.institution.edu/~Charlie.Author}
%\affiliation{
% Second institution and/or address\\
% This line break forced% with \\
%}%
%\affiliation{
% Third institution, the second for Charlie Author
%}%
%\author{Delta Author}
%\affiliation{%
% Authors' institution and/or address\\
% This line break forced with \textbackslash\textbackslash
%}%

%\collaboration{CLEO Collaboration}%\noaffiliation

\date{\today}% It is always \today, today,
             %  but any date may be explicitly specified

\begin{abstract}
Moir\'e materials have become one of the most active fields in material science in recent years due to their high tunability, and their unique properties emerge from the Moir\'e-scale structure modulation. Here, we propose twisted bilayer Bi$_2$(Te$_{1-x}$Se$_x$)$_3$ as a new Moir\'e material where the Moir\'e-scale modulation induces a topological phase transition. We show, in twisted bilayer Bi$_2$(Te$_{1-x}$Se$_x$)$_3$, a topological insulator domain and a normal insulator domain coexist in the Moir\'e lattice structure, and edge states on the domain boundary make nearly flat bands that dominate the material properties. The edge states further contribute to a Moir\'e-scale band inversion, resulting in Moir\'e-scale topological states. There are corresponding Moir\'e-scale edge states and they are so to speak ``edge state from edge state", which is a unique feature of twisted bilayer Bi$_2$(Te$_{1-x}$Se$_x$)$_3$. Our result not only proposes novel quantum phases in twisted bilayer Bi$_2$Te$_3$-family, but also suggests the twisting of stacking sensitive topological materials paves an avenue in the search for novel quantum materials and devices.

\end{abstract}

%\pacs{Valid PACS appear here}% PACS, the Physics and Astronomy
                             % Classification Scheme.
%\keywords{Suggested keywords}%Use showkeys class option if keyword
                              %display desired
\maketitle

%\tableofcontents

\section{Introduction \label{sec:main_intro}}

The twisted van der Waals heterostructure materials, or Moir\'e materials, have been studied very intensively in recent years as a platform for exploring novel quantum phases \cite{cao2018correlated,cao2018unconventional,PhysRevX.8.031089,yankowitz2019tuning,sharpe2019emergent,jiang2019charge,polshyn2019large,lu2019superconductors,kerelsky2019maximized,choi2019electronic,xie2019spectroscopic,PhysRevLett.124.076801,serlin2020intrinsic,stepanov2020untying,saito2020independent,hunt2013massive,dean2013hofstadter,PhysRevX.8.031087,PhysRevB.85.195458,PhysRevLett.109.196802}. In those materials, Moir\'e superlattices are formed by the lattice misalignment with a small twist angle, and the Moir\'e superlattices produce flat electric bands and various strongly correlated phases. In particular, the experimental reports on the magic-angle twisted bilayer graphene have stimulated this field \cite{cao2018correlated,cao2018unconventional}. They reported that the bilayer graphene stacked with a twist angle $1.08^\circ$ (magic angle), which has been known to have flat bands near the Fermi level, shows correlated insulating phases and superconducting phases when the filling factor is tuned. Because the behavior resembles the phase diagram of the high-temperature cuprate superconductors, twisted bilayer graphene has attracted great attention. The unique feature of the Moir\'e material is its high tunability. The twist angle is a tunable parameter specific to Moir\'e materials. Furthermore, because the system is two-dimensional (2D) and has a large Moir\'e unit cell, the filling factor can be tuned easily and significantly. This high tunability allows us to find various quantum phases in a single Moir\'e material. Inspired by the twisted bilayer graphene, Moir\'e systems of some other layered materials have been studied. In the twisted bilayer transition metal dichalcogenides (TMD), nearly flat bands have been theoretically predicted on the valence band edge, and an experimental signature of a correlated insulator phase has been reported \cite{tang2020simulation,regan2020mott,wang2020correlated,PhysRevLett.122.086402,PhysRevLett.125.186803,xu2020correlated,huang2021correlated,PhysRevB.103.125146,PhysRevB.103.155142,PhysRevLett.127.037402,PhysRevLett.121.026402,jin2021stripe}. Also for other materials, such as hexagonal boron nitride (hBN), the existence of nearly flat bands has been suggested \cite{PhysRevLett.124.086401,lian2020flat}.
Although the tunability of Moir\'e materials is remarkable, research of Moir\'e materials has concentrated on the layered materials above. Hence, the physics that describes their low-energy electronic states can be qualitatively categorized into two groups, semimetallic one (graphene) and insulating ones (TMD, hBN).
%The layered materials that are used in the previous studies of the Moir\'e materials can be roughly categorized into two groups based on the physics that describe their low-energy electronic states, semimetallic one (graphene) and insulating ones (TMD, hBN). 

In this paper, we theoretically propose twisted bilayer Bi$_2$(Te$_{1-x}$Se$_x$)$_3$ (Fig.\ref{fig:main1_1}) as a new Moir\'e material that hosts novel low-energy electronic states described by a topological phase transition and corresponding topological edge states. Three-dimensional bulk Bi$_2$(Te$_{1-x}$Se$_x$)$_3$ is one of the van der Waals heterostructure materials, and is well known as a typical strong topological insulator \cite{zhang2009topological,xia2009observation,chen2009experimental,PhysRevLett.105.266806}. For the thin film Bi$_2$(Te$_{1-x}$Se$_x$)$_3$ case, it has been suggested that the topological invariant strongly depends on the number of stacked layers \cite{PhysRevB.81.041307}. Therefore, topological phase transitions are expected to occur when the stacking order or interlayer distance is changed. Generally in a Moir\'e material, the local stacking order and interlayer distance are modulated by the lattice misalignment \cite{PhysRevLett.109.196802}. Combining the stacking modulation and the stacking sensitive topological insulator, we propose a Moir\'e material with mixed topological insulator domain and normal insulator domain.

%The rest of this paper is organized as follows: In Sec.II, 

\section{Effective model for Moir\'e materials \label{sec:main_model}}
\begin{figure*}
    \centering
    \includegraphics[width=\textwidth]{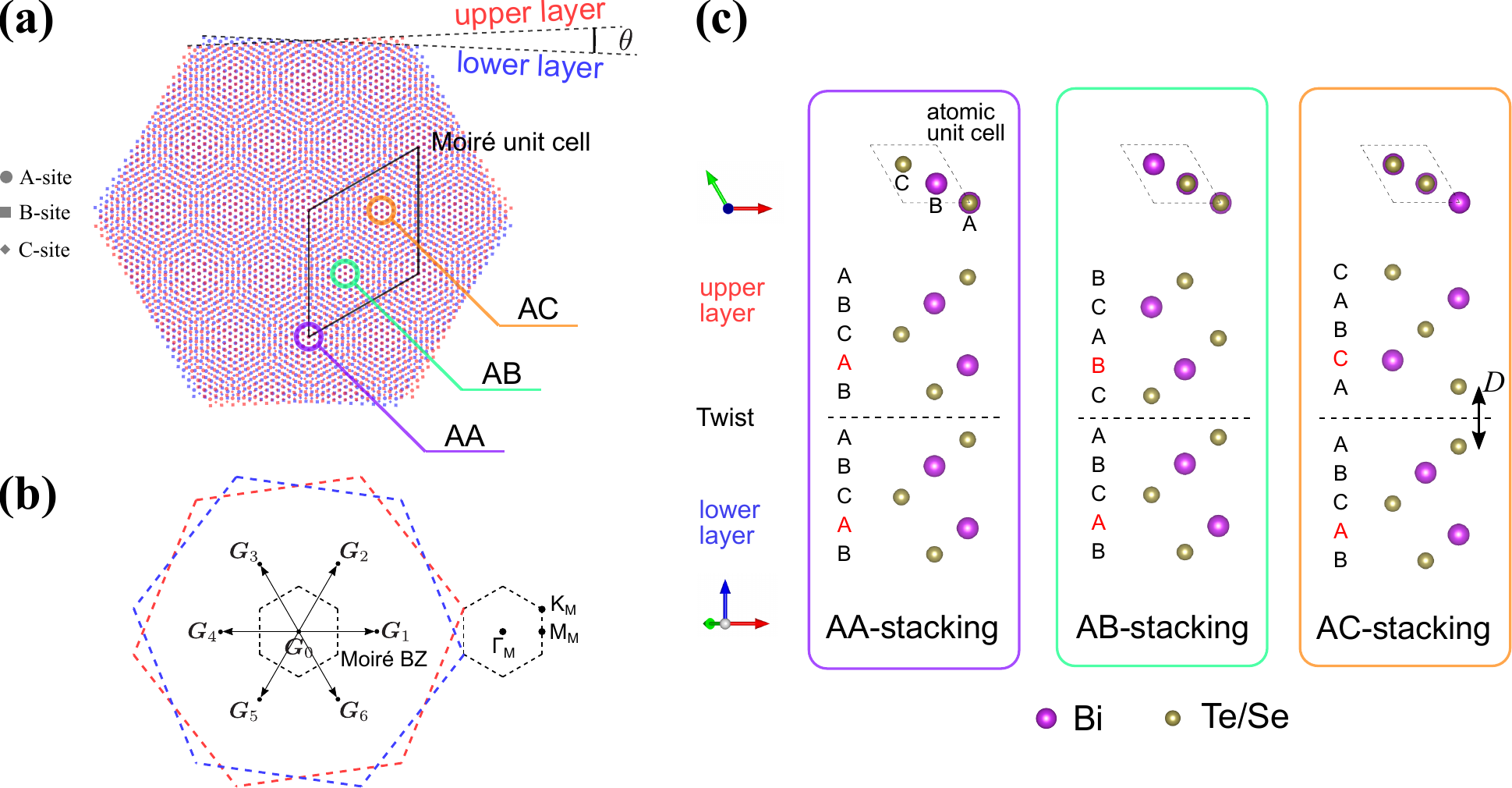}
    \caption{Lattice structure of twisted bilayer Bi$_2$(Te$_{1-x}$Se$_x$)$_3$. (a) Schematic picture of the Moir\'e pattern in twisted bilayer Bi$_2$(Te$_{1-x}$Se$_x$)$_3$. There are three in-plane atomic sites, A-, B-, and C-site, in the untwisted Bi$_2$(Te$_{1-x}$Se$_x$)$_3$. The atomic lattices in the upper and lower layers are drawn with red and blue markers, respectively. The solid rhombus represents the Moir\'e unit cell. The three sampling points used in the effective model are denoted with circles. (b) Schematic picture of the Moir\'e BZ. The red and blue dashed hexagons are twisted atomic BZ of the upper and lower layers, respectively. $\bm{G}_0 \sim \bm{G}_6$ are the Moir\'e reciprocal lattice vectors that we used in the effective model. (c) Top and horizontal views of the atomic lattice structures of untwisted bilayer Bi$_2$(Te$_{1-x}$Se$_x$)$_3$ for AA- (left), AB- (middle), and AC-stacking (right), drawn by VESTA \cite{vesta}. Each of the upper and lower layers has five atoms in the atomic unit cell.}
    \label{fig:main1_1}
\end{figure*}

%\noindent{\bf Effective model for Moir\'e materials}\\
In this paper, to avoid confusion, an atomic-scale lattice structure in an untwisted system are explicitly referred to as ``{\it atomic lattice}" and a Moir\'e superlattice structure in a twisted system is referred as ``{\it Moir\'e lattice}" (Fig.\ref{fig:main1_1}). The words ``atomic" and ``Moir\'e" are used in the same way for other terms, such as atomic(Moir\'e) Brillouin zone (BZ).

The Moir\'e electronic states are calculated with an effective model within a small-angle approximation \cite{bistritzer2011moire,jung2014ab}.
The Hamiltonian of the effective model is given as
\begin{equation}
    \begin{split}
        H &= \int d\bm{k} \sum_{\alpha\sigma,\beta'\sigma'} \sum_{\bm{G}_l} t_{\bm{k},\bm{k}-\bm{G}_l,\bm{G}_l}^{\alpha\sigma,\beta'\sigma'}~ c_{\alpha\sigma,\bm{k}}^\dagger c_{\beta'\sigma', \bm{k}-\bm{G}_l}^{\phantom{\dagger}} ,
        \label{eq.main.H}
    \end{split}
\end{equation}
where $\alpha\sigma$ and $\beta'\sigma'$ are orbital-spin indices, and $c^\dagger$ ($c$) is a creation (annihilation) operator.  The sum $\sum_{\bm{G}_l}$ is taken over seven Moir\'e reciprocal lattice vectors $\bm{G}_0 \sim \bm{G}_6$ (Fig. \ref{fig:main1_1}(b)). The $t_{\bm{k},\bm{k}-\bm{G}_l,\bm{G}_l}^{\alpha\sigma,\beta'\sigma'}$ are determined to satisfy
\begin{equation}
    \begin{split}
        \sum_{\bm{G}_l} e^{i \bm{G}_{l} \cdot \bm{r}_j} ~t^{\alpha\sigma,\beta'\sigma'}_{\bm{k},\bm{k}-\bm{G}_{m},\bm{G}_{l}} = &~ h_{\bm{r}_j}^{\alpha\sigma,\beta'\sigma'} ( \bm{k}-\bm{G}_m/2 ) ,
    \label{eq.main.t}    
    \end{split}
\end{equation}
where $h_{\bm{r}_j}^{\alpha\sigma,\beta'\sigma'} ( \bm{k} )$ is the matrix elements calculated from hopping parameters around position $\bm{r}_j$ (See Appendix \ref{sec:SM_moire}, \ref{sec:SM_model}, and \ref{sec:SM_numerical} for more detail of the model derivation). The local atomic lattice structure around $\bm{r}_j$ is approximated by an untwisted lattice with a particular stacking order, and thus $h_{\bm{r}_j}^{\alpha\sigma,\beta'\sigma'} ( \bm{k} )$ and the electronic states on it are also estimated by calculation in the untwisted lattice. We take three three sampling points of $\bm{r}_j$, and interpolate the intermediate region by the discrete Fourier transform. We call the three stacking orders at the three sampling points AA-, AB-, and AC-stacking (Fig. \ref{fig:main1_1}(a),(c)). The AB-stacking is the most stable one and thus it is realized in the 3D bulk Bi$_2$(Te$_{1-x}$Se$_x$)$_3$. With the three sampling points approximation, the explicit definition of $t_{\bm{k},\bm{k}-\bm{G}_l,\bm{G}_l}^{\alpha\sigma,\beta'\sigma'}$ are given as
\begin{equation}
    \begin{split}
        t^{\alpha\sigma,\beta'\sigma'}_{\bm{k},\bm{k},\bm{G}_{0}} =& \frac{1}{3} \left[ h_{\bm{r}_{AA}}^{\alpha\sigma,\beta'\sigma'}(\bm{k}) + h_{\bm{r}_{AB}}^{\alpha\sigma,\beta'\sigma'}(\bm{k}) + h_{\bm{r}_{AC}}^{\alpha\sigma,\beta'\sigma'}(\bm{k}) \right] , \\
        t^{\alpha\sigma,\beta'\sigma'}_{\bm{k},\bm{k}-\bm{G}_{l},\bm{G}_{l}} =& \frac{1}{9} \left[  h_{\bm{r}_{AA}}^{\alpha\sigma,\beta'\sigma'}(\bm{k}-\bm{G}_{l}/2) \right. \\
        &~~~~~ + e^{-i \frac{2\pi}{3} } h_{\bm{r}_{AB}}^{\alpha\sigma,\beta'\sigma'}(\bm{k}-\bm{G}_{l}/2) \\
        &~~~~~ \left. + e^{i \frac{2\pi}{3} } h_{\bm{r}_{AC}}^{\alpha\sigma,\beta'\sigma'}(\bm{k}-\bm{G}_{l}/2) \right] ~~ (l=1,3,5), \\
        t^{\alpha\sigma,\beta'\sigma'}_{\bm{k},\bm{k}-\bm{G}_{l},\bm{G}_{l}} =& \frac{1}{9} \left[  h_{\bm{r}_{AA}}^{\alpha\sigma,\beta'\sigma'}(\bm{k}-\bm{G}_{l}/2) \right. \\
        &~~~~~ + e^{i \frac{2\pi}{3} } h_{\bm{r}_{AB}}^{\alpha\sigma,\beta'\sigma'}(\bm{k}-\bm{G}_{l}/2) \\
        &~~~~~ \left. + e^{-i \frac{2\pi}{3} } h_{\bm{r}_{AC}}^{\alpha\sigma,\beta'\sigma'}(\bm{k}-\bm{G}_{l}/2) \right] ~~ (l=2,4,6) .
    \end{split}
\end{equation}
$h_{\bm{r}_{AA}}^{\alpha\sigma,\beta'\sigma'}(\bm{k})$, $h_{\bm{r}_{AB}}^{\alpha\sigma,\beta'\sigma'}(\bm{k})$, and $h_{\bm{r}_{AC}}^{\alpha\sigma,\beta'\sigma'}(\bm{k})$ are matrix elements of Hamiltonians calculated in the AA-, AB-, and AC-stacking untwisted bilayer Bi$_2$(Te$_{1-x}$Se$_x$)$_3$, respectively.

%In numerical calculation, we use finite basis within a momentum cutoff $k_c$ as
%\begin{equation}
%    \left\{ c_{\alpha\sigma,\bm{k}+\bm{v}_{mn}} \left|\bm{k} \in \mathrm{Moir\acute{e}~BZ},~ \bm{v}_{mn}=m \bm{G}_1 + n \bm{G}_2 , ~ m,n \in \mathbb{Z},~|\bm{v}_{mn}|<k_c \right. \right\} ,
%\end{equation} 
%and a finite dimension matrix
%\begin{equation}
%    \begin{split}
%        H(\bm{k}) = \left(
%        \begin{array}{cccccc}
%            t_{\bm{k},\bm{k},\bm{0}} & & & & \mathrm{if}~\bm{v}_{mn}-\bm{v}_{m'n'}=\bm{G}_l & \\
%             & \ddots & & & \downarrow & \\
%             & & t_{\bm{k}+\bm{v}_{mn},\bm{k}+\bm{v}_{mn},\bm{0}} & \cdots & t_{\bm{k}+\bm{v}_{mn},\bm{k}+\bm{v}_{m'n'},\bm{G}_l} & \\
%             & & & \ddots & \vdots & \\
%             & & & & t_{\bm{k}+\bm{v}_{m'n'},\bm{k}+\bm{v}_{m'n'},\bm{0}} & \\
%             & & & & & \ddots \\
%        \end{array}
%        \right) .
%    \end{split}
%\end{equation}

%\noindent{\bf First-principles calculation}\\
\begin{figure*}
    \centering
    \includegraphics[width=\textwidth]{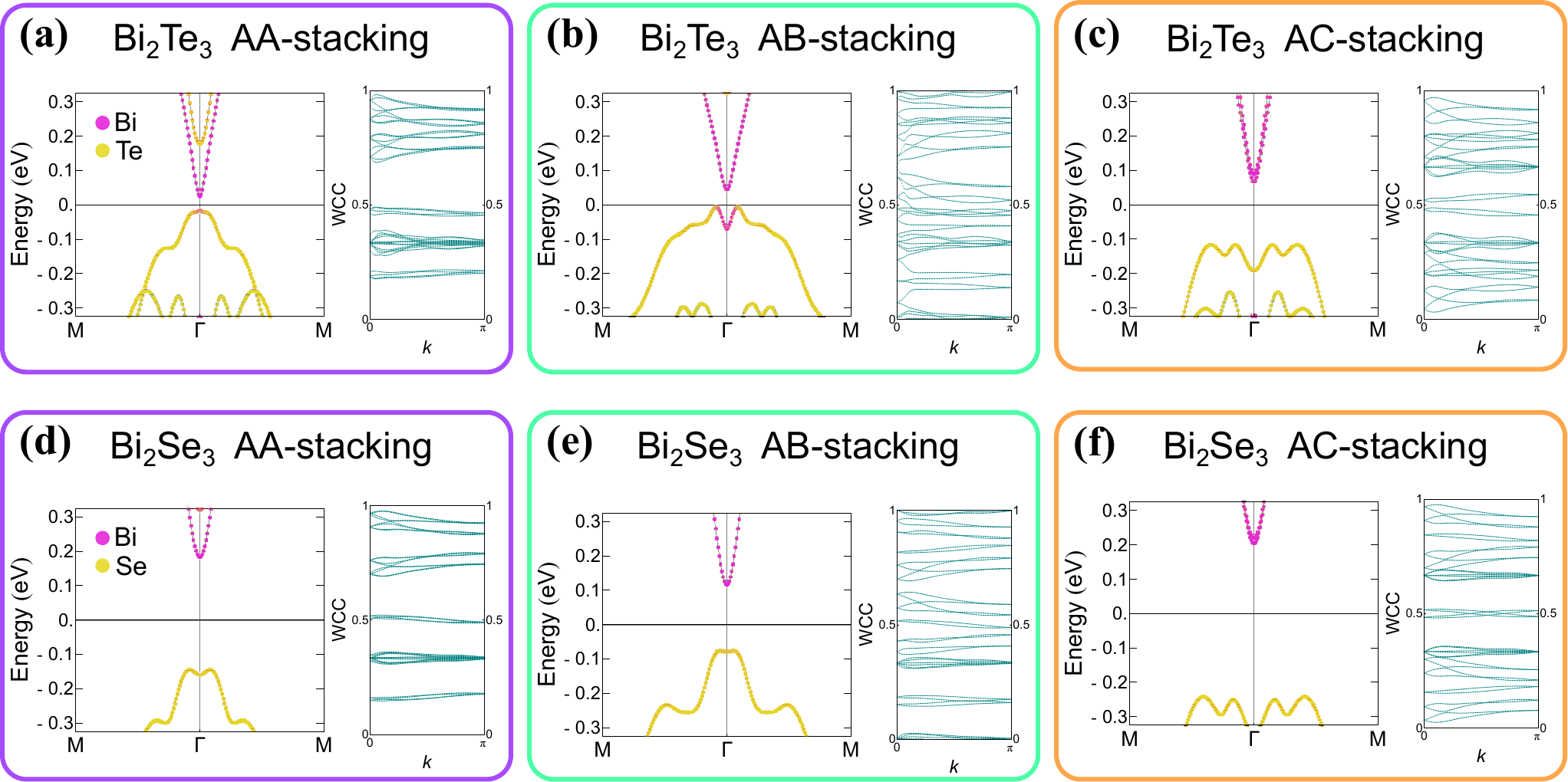}
    \caption{Electronic band structure and Wilson loop spectra of the untwisted bilayer Bi$_2$Te$_3$ and Bi$_2$Se$_3$ for each stacking orders. (a), (b), and (c) are for AA-, AB-, and AC-stacking bilayer Bi$_2$Te$_3$, and (d), (e), and (f) are for those of bilayer Bi$_2$Se$_3$, respectively. The magenta and yellow dots in the band structure figures represent the projected weight on the $p_z$ orbitals of Bi and Te/Se atoms, respectively. Only AB-stacking bilayer Bi$_2$Te$_3$ (b) is a topological insulator, and the others are normal insulators.}
    \label{fig:main1_2}
\end{figure*}

\begin{table}[]
    \centering
    \begin{tabular}{cccc}
        \hline
        Bi$_2$Te$_3$ & ~~~AA~~~ & ~~~AB~~~ & ~~~AC~~~ \\
        \hline
        in-plane lat. const. $a$ (\AA) & 4.40 & 4.41 & 4.39 \\
        interlayer distance $D$ (\AA) & 3.18 & 2.73 & 3.98 \\
        \hline
        \\
        \hline
        Bi$_2$Se$_3$ & AA & AB & AC \\
        \hline
        in-plane lat. const. $a$ (\AA) & 4.15 & 4.16 & 4.15 \\
        interlayer distance $D$ (\AA) & 3.05 & 2.75 & 3.78 \\
        \hline
    \end{tabular}
    \caption{Optimized in-plane lattice constant $a$ and interlayer distance $D$ for bilayer Bi$_2$Te$_3$ and Bi$_2$Se$_3$ with three stacking orders. The AA-, AB-, and AC-stacking are defined in Fig.\ref{fig:main1_1}.}
    \label{tab:opt_lat}
\end{table}

$h_{\bm{r}_{AA}}^{\alpha\sigma,\beta'\sigma'}(\bm{k})$, $h_{\bm{r}_{AB}}^{\alpha\sigma,\beta'\sigma'}(\bm{k})$, and $h_{\bm{r}_{AC}}^{\alpha\sigma,\beta'\sigma'}(\bm{k})$ are obtained by the first-principles calculation. All first-principles calculations are implemented in the Vienna ab initio simulation package (VASP)~\cite{vasp}. We use the projector augmented wave (PAW) potential sets recommended by VASP and set the kinetic-energy cutoff to 500 eV. For each of the three stacking orders, we optimize the lattice structure and obtain electronic band structures. The lattice optimizations are performed in the strongly constrained and appropriately normed (SCAN) meta-generalized gradient approximation for short- and intermediate-range interactions with the long-range vdW interaction (rVV10) with the spin-orbit interaction~\cite{PhysRevX.6.041005}.
%By placing a sufficiently large vaccume in the c-axis direction ($c\sim40$\AA), the layered materials are calculated.
%The lattice optimizations are performed for both of the in-plane and normal directions.
%\blue{$a_{\mathrm{Sb_2Te_3,AA}}=xxx$, $a_{\mathrm{Sb_2Te_3,AB}}=xxx$, and $a_{\mathrm{Sb_2Te_3,AC}}=xxx$.}
%Because the stacking dependence of the in-plane lattice constant is small, we neglect it and use an averaged value in the effective model calculations.
We calculate the electronic band structure in the B3LYP with the VWN3-correlation~\cite{doi:10.1063/1.464913}. 
%Based on the electronic band structures calculated with the optimized lattices, we
We construct Wannier functions \blue{for the Bi and Te/Se $p$ orbitals} with WANNIER90 package \cite{wannier90}.
We use $8 \times 8 \times 1$ $k$-mesh for the lattice optimization and $9 \times 9 \times 1$ $k$-mesh for the electronic calculation and the Wannier function.
\blue{The obtained band structures reproduce well the results of angle-resolved photoemission spectroscopy (ARPES)~\cite{PhysRevB.85.195442,Zhang2010} and the GW approximation~\cite{PhysRevB.93.205442}.}
The Fermi level is set in the averaged Hamiltonian of three stacking orders $t^{\alpha\sigma,\beta'\sigma'}_{\bm{k},\bm{k},\bm{G}_{0}}$ by the filling factor, i. e. the middle of the 36th and 37th band in the $\Gamma$ point of the 60 bands (2 layers $\times$ 5 atoms $\times$ $p$ orbitals $\times$ spin). The obtained Wannier functions and matrix elements are also used in the calculations of the Wilson loop spectra with the WannierTools package \cite{wanniertools}. 
\blue{We also calculate the Sb$_2$Te$_3$ under the same condition (see Appendix \ref{sec:SM_Sedope}).}

%The detailed theory of the model derivation and implementation in numerical calculations are described in Appendix.X.

\begin{figure*}
    \centering
    \includegraphics[width=0.99\textwidth]{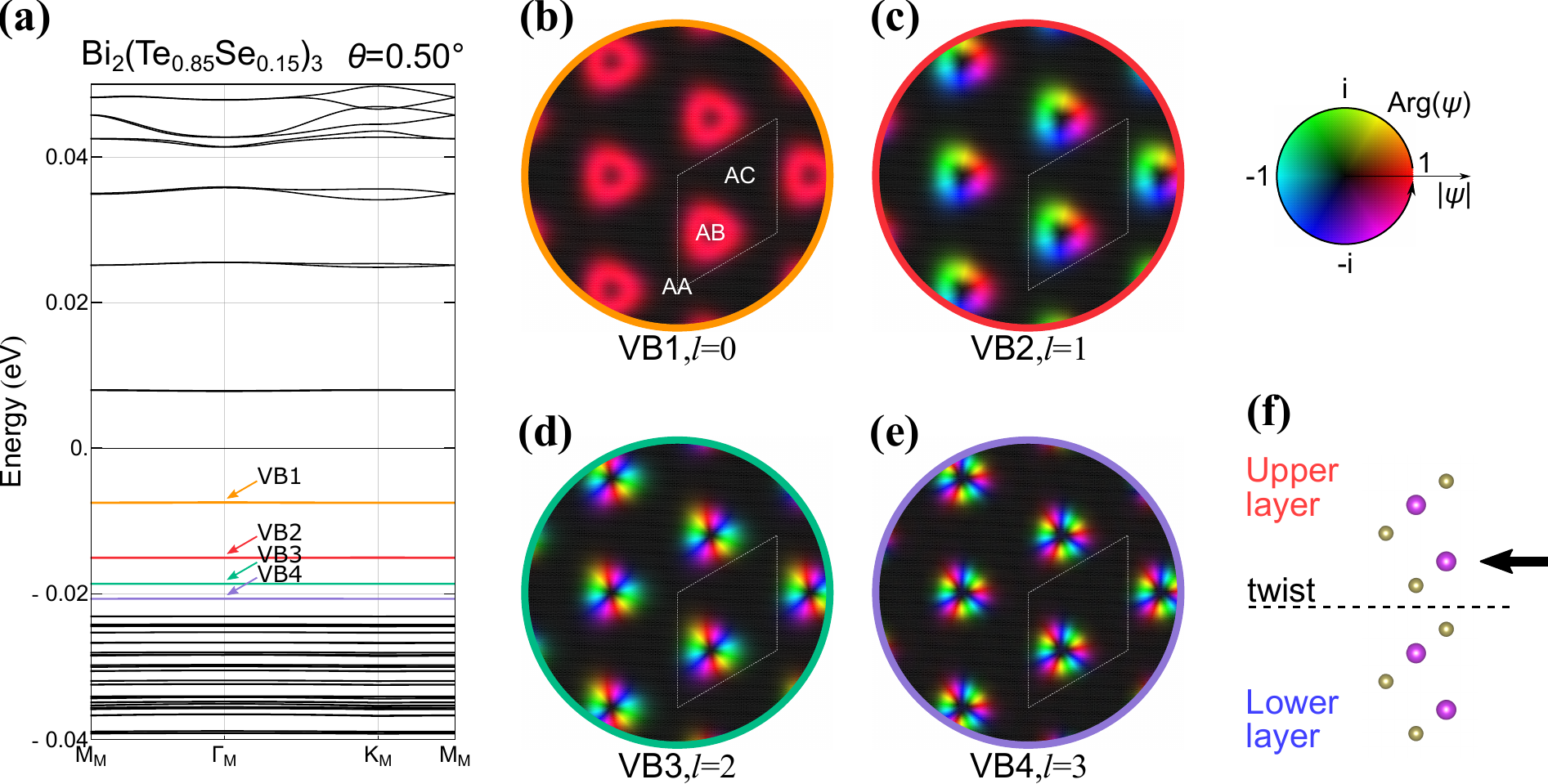}
    \caption{Moir\'e band structure of twisted bilayer Bi$_2$(Te$_{1-x}$Se$_x$)$_3$. (a) Moir\'e band dispersion of Bi$_2$(Te$_{0.85}$Se$_{0.15}$)$_3$ with a twist angle $\theta = 0.50^\circ$. The top (VB1), second (VB2), third (VB3), and fourth (VB4) valence band pairs are shown as orange, red, green, and purple lines, respectively. The symbols of the high-symmetry points are defined in Fig.\ref{fig:main1_1}(b). (b)-(e) Real space plot of the wave functions (upper layer, lower Bi, $p_z$ orbital, spin up component, as shown in (f)) of VB1-4 in the $\Gamma_M$ point. The brightness and color indicate the absolute value (normalized to have the maximum value of 1) and phase of the wave function as shown in the right of (c). The Moir\'e unit cell is shown as a white dashed rhombus. The angular momentum $l$ is also shown in each figure. (f) Bi atom that is focused in the plots. }
    \label{fig:main2_1}
\end{figure*}

\begin{figure}
    \centering
    \includegraphics[width=0.45\textwidth]{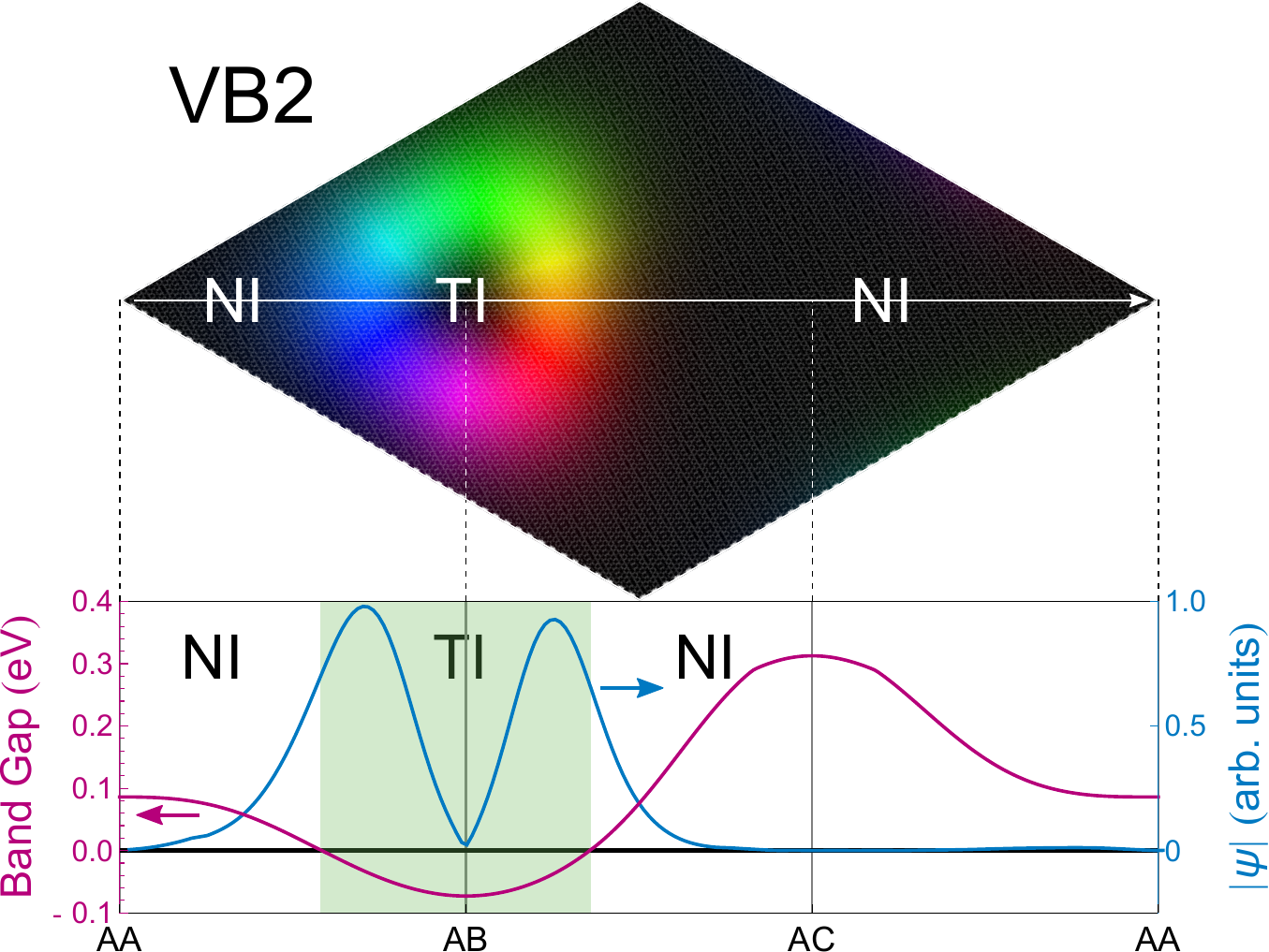}
    \caption{Moir\'e wave function and local electronic state. The rhombus is a wave function plot of VB2 in the Moir\'e unit cell (same as Fig.\ref{fig:main2_1}(c)). The lower plot shows the real space dependence of the absolute value of the wave function of VB2 (cyan line, right vertical axis) and the bandgap calculated in the interpolated untwisted Hamiltonian (violet line, left vertical axis). The horizontal axis is the real space position along the AA-AB-AC-AA-stacking line, which is the longer diagonal of the Moir\'e unit cell as shown with a white arrow in the top figure. The negative bandgap means that the bands are inverted. The two gapless points around the AB-stacking area are the domain boundary, and the topological insulator domain is shown as a green-shaded range.}
    \label{fig:main2_2}
\end{figure}

\begin{figure*}
    \centering
    \includegraphics[width=\textwidth]{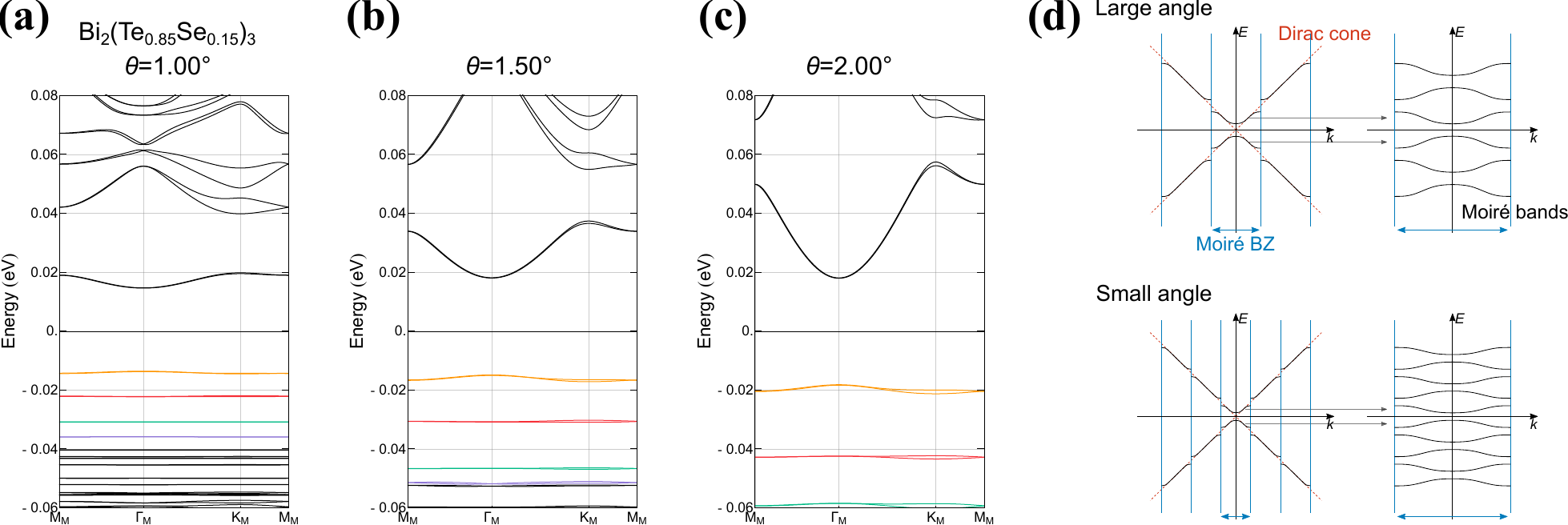}
    \caption{Twist angle dependence of Moir\'e band dispersion of Bi$_2$(Te$_{0.85}$Se$_{0.15}$)$_3$. (a), (b), and (c) are the cases with a twist angle $\theta = 1.00^\circ$, $1.50^\circ$, and $2.00^\circ$, respectively. VB1-4 are shown with the same color as Fig.\ref{fig:main2_1}(a). (d) Schematic picture to explain the twist angle dependence of the flatness and gap of the edge-state-originated nearly flat bands. }
    \label{fig:main2_3}
\end{figure*}

\section{Bilayer Bi$_2$(Te$_{1-x}$Se$_x$)$_3$ \label{sec:main_untwist}}
First, we show the result of calculations on the untwisted atomic lattices of bilayer Bi$_2$Te$_3$ and Bi$_2$Se$_3$ for the three stacking orders.
The atomic positions are shown in Fig.~\ref{fig:main1_1}(c). All these untwisted bilayer lattices belong to the layer group \#72 (or the space group \#166 with infinitely long c-axis). The optimized in-plane lattice constant $a$ and interlayer distance $D$, which is defined as the vertical distance between the two Te/Se atoms in the twist face (See Fig.\ref{fig:main1_1}(c), right), are listed in Table \ref{tab:opt_lat}. We neglect the stacking order dependence in the in-plane lattice constant in each material and use an averaged value in the effective model calculations. For Te/Se doping, the in-plane lattice constant is linearly interpolated.
%$D_{\mathrm{Bi_2Te_3,AA}}=3.18 \mathrm{\AA}$, $D_{\mathrm{Bi_2Te_3,AA}}=2.73 \mathrm{\AA}$, and $D_{\mathrm{Bi_2Te_3,AA}}=3.98 \mathrm{\AA}$ for the Bi$_2$Te$_3$ cases and $D_{\mathrm{Bi_2Se_3,AA}}=3.05 \mathrm{\AA}$, $D_{\mathrm{Bi_2Se_3,AA}}=2.75 \mathrm{\AA}$, and $D_{\mathrm{Bi_2Se_3,AA}}=3.78 \mathrm{\AA}$ for the Bi$_2$Se$_3$ cases, \blue{respectively}. The obtained in-plane lattice constants are $a_{\mathrm{Bi_2Te_3,AA}}= 4.399 \mathrm{\AA}$, $a_{\mathrm{Bi_2Te_3,AB}}=4.410$, $a_{\mathrm{Bi_2Te_3,AC}}=4.389$, $a_{\mathrm{Bi_2Se_3,AA}}=4.150$, $a_{\mathrm{Bi_2Se_3,AB}}=4.161$, and $a_{\mathrm{Bi_2Se_3,AC}}=4.150$. We neglect the difference in the in-plane lattice constant and use an averaged value in the effective model calculations.
In both materials, the AB-stacking has the smallest interlayer distance and the AC-stacking has the largest. The obtained electronic band structures are shown in Figs.\ref{fig:main1_2}(a)-(f) for both of Bi$_2$Te$_3$ and Bi$_2$Se$_3$, where the Fermi level is determined by the filling factor. The magenta and yellow dots represent the projected weight on the $p_z$ orbitals of Bi and (Te,Se) atoms, respectively. In these materials, the overlap of the $p_z$ orbitals contributes to a topological phase transition. Therefore, the smaller the interlayer distance, the more likely it is to be a topological insulator. To evaluate their topological invariants, we make Wannier functions for them and calculate the Wilson loop spectra as shown in the right panel of each figure of Figs.\ref{fig:main1_2}(a)-(f).  We can see only AB-stacking Bi$_2$Te$_3$ is a topological insulator and all of the others are normal insulators. These results indicate that a twisted bilayer Bi$_2$Te$_3$ system has a topological insulator domain around the AB-stacking region, and a normal insulator domain in the other region. Although Bi$_2$Se$_3$ is a normal insulator in the three stacking orders, we consider Se doping in Bi$_2$Te$_3$ by the linear interpolation to tune the system parameters. However, note that the topological non-triviality of the AB-stacking Bi$_2$Te$_3$ plays an essential role even in doped cases.

\section{Moir\'e band and quantized edge state in twisted bilayer Bi$_2$(Te$_{1-x}$Se$_x$)$_3$ \label{sec:main_twist}}
Next, we show the Moir\'e band dispersion of twisted bilayer Bi$_2$(Te$_{1-x}$Se$_x$)$_3$. Due to the twisting, the inversion symmetry is broken and twisted bilayer Bi$_2$(Te$_{1-x}$Se$_x$)$_3$ belongs to the layer group \#67 (or the space group \#149 with infinitely long c-axis). The in-plane $C_2$ axis exists along the AA-AB-AC-stacking line \blue{(see Fig. \ref{fig:main1_1}(a)), which corresponds to the $\Gamma_M$-M$_M$ line in the reciprocal space.} The symbols of the high symmetry points in the Moir\'e BZ are defined as Fig. \ref{fig:main1_1}(b). Figure \ref{fig:main2_1}(a) shows the band dispersion of the twisted bilayer Bi$_2$(Te$_{0.85}$Se$_{0.15}$)$_3$ with a twist angle $\theta=0.50^\circ$. Here, to obtain a clear domain boundary, the amount of Se ($x$) is determined so that the gap in the AA- and AB-stacking would be roughly the same (See Appendix \ref{sec:SM_Sedope} for the detail of Se-doping dependence of the gap and Moir\'e band dispersion).
%Due to the time-reversal symmetry, all bands are doubly degenerate \blue{at} the $\Gamma_M$ and M$_M$ points, which are the time-reversal invariant momenta (TRIM).
\blue{All bands are doubly degenerate at time-reversal invariant momenta (TRIM), the $\Gamma_M$ and M$_M$ points.}
%Because of the absence of the inversion symmetry, the double degeneracy splits in a general momentum. 
Because of the absence of the inversion symmetry, the Kramers degeneracy in the untwisted bilayer splits at a general momentum.
The split is easy to see in the conduction bands above $0.020$ eV in the K$_M$ point, while it is too small to see in the valence bands. It is worth noting that there are nearly flat bands around the Fermi level. For the nearly flat valence bands (from the top to fourth valence band pairs VB1, VB2, VB3, and VB4 in the Fig. \ref{fig:main2_1}(a)), real space plots of the wave functions (upper layer, lower Bi, $p_z$ orbital, spin up component. See Fig.\ref{fig:main2_1}(f)) at the $\Gamma_M$ point are shown in Fig.\ref{fig:main2_1}(b)-(e). The brightness and color indicate the absolute value (normalized to have the maximum value of 1) and phase of the wave function, respectively (as shown in the right of Fig.\ref{fig:main2_1}(c)). The wave function has a ring-shaped density distribution, clearly indicating that these nearly flat bands originate from the edge state corresponding to the topological insulator domain around the AB-stacking area. To compare with the wave functions of the flat bands, we calculate the real space dependence of the bandgap between the valence top and conduction bottom bands in the $\Gamma$ point $\Delta E (\Gamma)$ in the interpolated untwisted Hamiltonian obtained by the discrete Fourier transform in Eq. (\ref{eq.main.t}) (See Appendix \ref{sec:SM_Sedope} for more detail). In Fig.\ref{fig:main2_2}, the real space dependence of the band bap (violet line) and the absolute value of the wave function of VB2 (cyan line) along the AA-AB-AC-AA-stacking line (the longer diagonal of the Moir\'e unit cell shown as a white arrow) are shown. The negative bandgap means that the bands are inverted. The gapless points are the domain boundary and the topological insulator domain is shown as a green-shaded range. We can confirm that the wave function has a large amplitude around the domain boundary. Next, we focus on the difference between the wave functions of VB1-4 (Fig.\ref{fig:main2_1}(b)-(e)). They show ring-shaped density distributions in common, but their phase structures are different from each other. We can see a lower energy band has a larger angular momentum $l$, which is calculated as a winding number of the phase along the ring. Note that because these states are coupled with spin, the spin-down component has one different $l$ and the total angular momentum is a half-integer (See Appendix \ref{sec:SM_supinfo_twist} for more detail). The phase structure indicates that the nearly flat bands are formed by quantization of topological edge states due to the finite size effect of the domain boundary. The flatness of these bands and the gap size between them depend on the twist angle. Figures \ref{fig:main2_3}(a)-(c) show the twist angle dependence of the Moir\'e band dispersion. It can be seen that as the twist angle increases, the nearly flat bands get more dispersive and the energy gaps between them get larger. This tendency is due to the fact that the length of the Moir\'e reciprocal lattice vector, i.e. the width of the Moir\'e BZ, is proportional to $\theta$. When the original Dirac cone on the domain boundary is folded in Moir\'e BZ and quantized, folding with large Moir\'e BZ results in the folding of bands with widely different energies (Fig.\ref{fig:main2_3}(d)). Further, folding with large Moir\'e BZ relatively keeps the original band dispersion, and thus the obtained Moir\'e bands tend to have larger bandwidth (See Appendix \ref{sec:SM_supinfo_twist} for more detail). With these points of view, the twist angle dependence of the flatness and energy gap of the edge-state-originated nearly flat bands is roughly explained by the size of Moir\'e BZ, but in more detail, the effect of the Rashba splitting and hybridization with other bands must be taken into account.
Finally, we mention the conduction bands. In Bi$_2$(Te$_{1-x}$Se$_x$)$_3$ case, since the conduction bands are more dispersive than the valence bands in the band structures of untwisted lattices, the Moir\'e conduction bands are also more likely to have dispersion in the higher energy region. For the same reason, a bandgap between Moir\'e conduction bands tends to be larger than that of Moir\'e valence bands.

\begin{figure*}
    \centering
    \includegraphics[width=\textwidth]{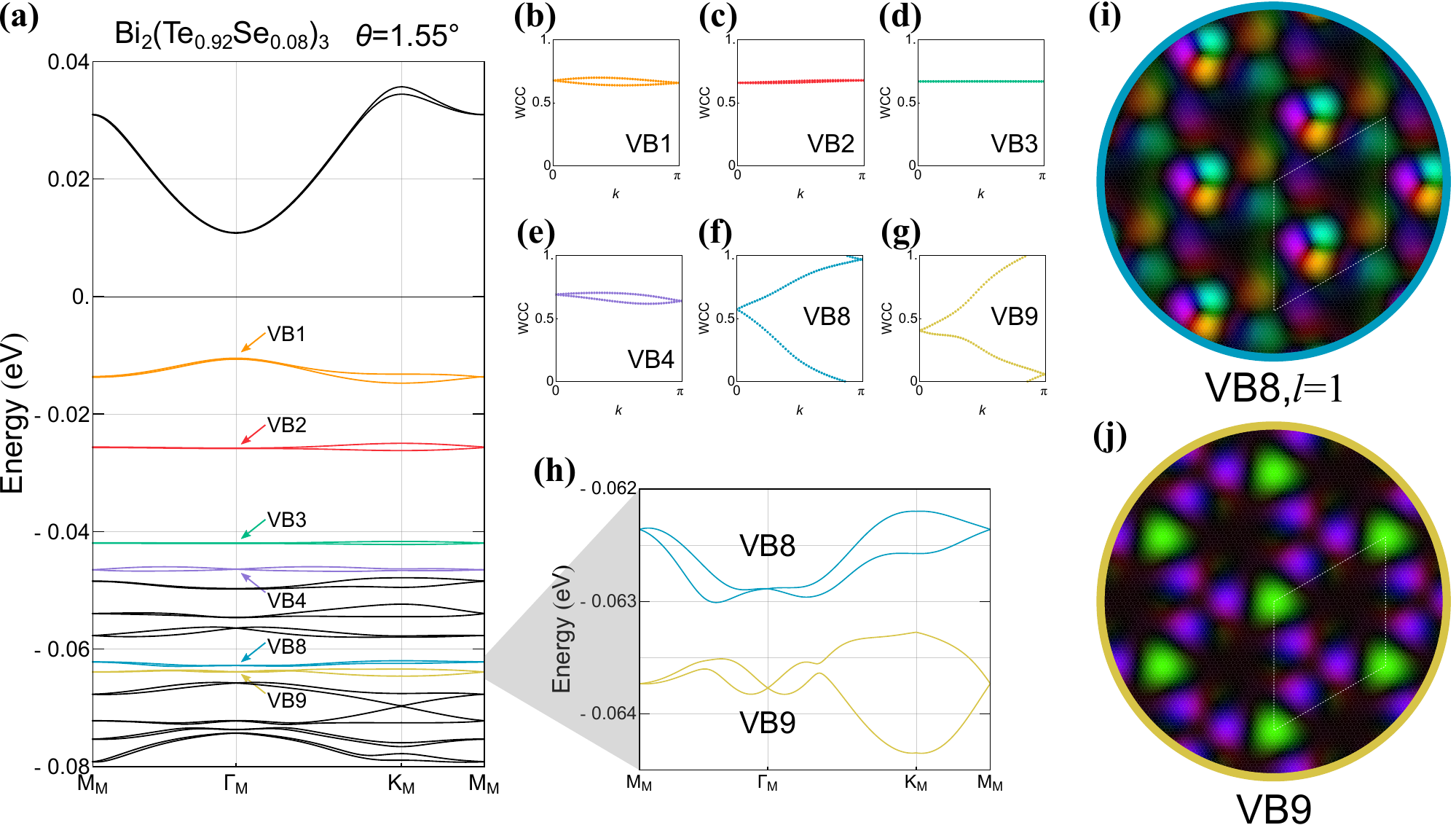}
    \caption{Quantum spin Hall effect in twisted bilayer Bi$_2$(Te$_{1-x}$Se$_x$)$_3$. (a) Moir\'e band dispersion of Bi$_2$(Te$_{0.92}$Se$_{0.08}$)$_3$ with a twist angle $\theta = 1.55^\circ$. The top (VB1), second (VB2), third (VB3), fourth (VB4), eighth (VB8), and ninth (VB9)  valence band pairs are shown as orange, red, green, purple, blue, and yellow lines, respectively. (b)-(g) Wilson loop spectra calculated on VB1-4, VB8, and VB9. While (b)-(e) VB1-4 are topologically trivial, (f) VB8 and (g) VB9 are topologically nontrivial. (h) Magnified view of the VB8 and VB9. In this figure. the numerical error due to the first-principles calculation is corrected to recover the exact time-reversal symmetry. (i),(j) Real space plot of the wave functions (upper layer, lower Bi, $p_z$ orbital, spin up component) of \blue{VB8 and VB9 in the $\Gamma_M$ point, respectively.} VB8 has a ring-shaped density distribution around the AB-stacking area, but VB9 has no longer ring-shaped one but localized density distribution around the AA-stacking area.}
    \label{fig:main3}
\end{figure*}

\begin{figure}
    \centering
    \includegraphics[width=0.45\textwidth]{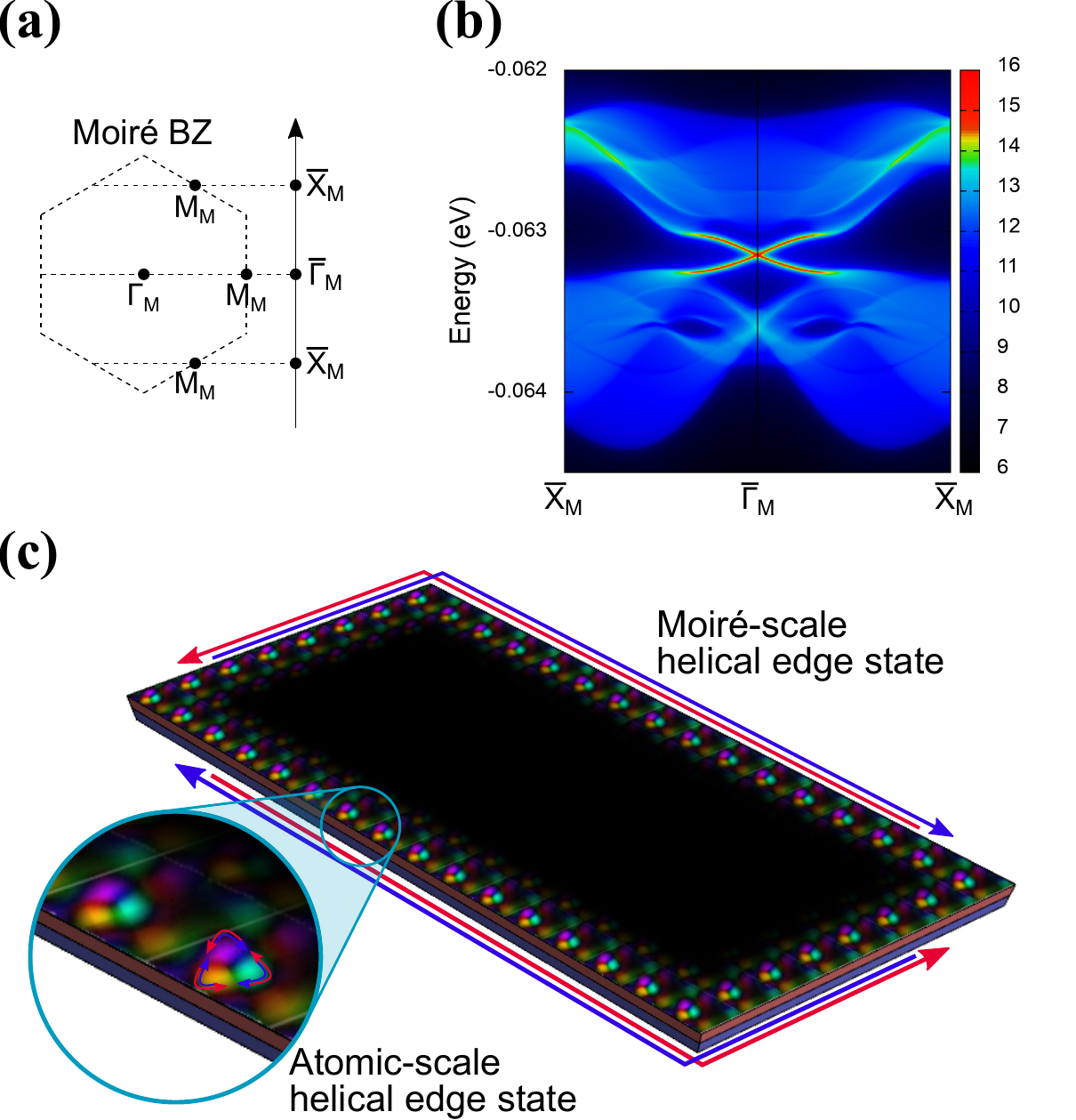}
    \caption{Topological edge state in twisted bilayer Bi$_2$(Te$_{1-x}$Se$_x$)$_3$. (a) Moir\'e-scale edge BZ that is used in the edge calculation. (b) Edge band spectrum around VB8 and VB9. The symbols of the high-symmetry points are defined in (a). Dirac cones of the Moir\'e-scale helical edge state can be seen appearing around the $\bar{\Gamma}_M$ point. (c) Schematic picture of ``edge state from edge state".
    \blue{The red and blue arrows correspond to spin currents having opposite spin directions to each other.}
    The helical edge state obtained in Moir\'e bands (Moir\'e-scale helical edge state) is running along the edge of the Moir\'e-scale lattice system. The Moir\'e-scale edge state is made from the ring-shaped edge state on the domain boundary around the AB-stacking area (atomic-scale helical edge state).}
    \label{fig:main4}
\end{figure}

\section{Moir\'e-scale topological band and edge state \label{sec:main_topo}}
We move on to a discussion of the emergence of topological bands in the Moir\'e bands of twisted bilayer Bi$_2$(Te$_{1-x}$Se$_x$)$_3$. As we discussed above, the formation of the edge-state-originated nearly flat bands are explained by a simple theory and they are ordered with angular momentum $l$. Therefore, we could not find a case to have a band inversion between the edge-state-originated bands. However, in the region farther from the Fermi level, a bulk-state-originated band appears and it can hybridize with an edge-state-originated to form topologically nontrivial bands. For example, the band dispersion of Bi$_2$(Te$_{0.92}$Se$_{0.08}$)$_3$ with a twist angle $\theta=1.55^\circ$ is shown in Fig.\ref{fig:main3}(a). Here, the amount of Se ($x$) is determined so that the gap in the AA-stacking area would be smaller than that of the AB-stacking area to obtain the bulk and edge states in different areas. To evaluate the topological invariants for each band, the Wilson loop spectra are calculated and given in Fig.\ref{fig:main3}(b)-(g). While the nearly flat band around the Fermi level (VB1-4, Fig\ref{fig:main3}(b)-(e)) are evaluated as topologically trivial, the bands around $-0.065$ eV (the eighth and ninth valence band pairs VB8 and VB9, Fig. \ref{fig:main3}(f),(g)) are evaluated as topologically nontrivial. Figure \ref{fig:main3}(h) is a magnified view of the VB8 and VB9. Here, we correct the numerical error due to the first-principles calculation to recover the exact time-reversal symmetry, because the Moir\'e system generally deals with a small energy scale and thus the numerical errors get more noticeable. It can be seen that the two bands are gapped and thus if the Fermi level is tuned and placed between them the system becomes a topological insulator protected by the time-reversal symmetry. The real space plots of the wave functions in the $\Gamma_M$ point of the two bands are shown in Fig.\ref{fig:main3}(i),(j). The wave function of VB8 (Fig.\ref{fig:main3}(i)) has a ring-shaped density distribution as well as the nearly flat bands around the Fermi level. On the other hand, VB9 (Fig.\ref{fig:main3}(j)) does not have a ring-shaped one, but has a density distribution localized around the AA-stacking area (the corner of the Moir\'e unit cell). Considering that the AA-stacking Bi$_2$Te$_3$ has the smallest gap among all stacking orders (Fig.\ref{fig:main1_2}), this state is understood as a state originated from the bulk state of the AA-stacking area. Due to a symmetry restriction of the hybridization, the simple bulk-state-originated band can hybridize with a band with a small $l$. That is why the VB8 has $l=1$ (as we explained, the spin-down component has one different $l$). To obtain edge states corresponding to these topological bands, we make Wannier functions for these bands and calculate the edge band spectra with Green's function method (technical details are described in Appendix.\ref{sec:SM_numerical}.). The Moir\'e-scale edge BZ that is used in the edge calculation is shown in Fig.\ref{fig:main4}(a). The obtained edge band spectrum is shown in Fig.\ref{fig:main4}(b). As in a typical topological insulator, a Dirac-cone of the helical edge state can be seen appearing around the $\bar{\Gamma}_M$ point (See Appendix \ref{sec:SM_supinfo_twist} for a symmetry restriction on the Dirac-cone). These topological bands are obtained in the Moir\'e band dispersion, and thus the corresponding helical edge state is running along the edge of the Moir\'e-scale lattice system (Fig.\ref{fig:main4}(c)). As we have seen, the Moir\'e-scale edge state is partially made from the ring-shaped edge state on the domain boundary around the AB-stacking area (let us say atomic-scale helical edge state). Therefore, we can say the Moir\'e edge state is ``edge state from edge state" (Fig.\ref{fig:main4}(c)), and this is a novel and unique phenomenon that is observed in twisted bilayer Bi$_2$(Te$_{1-x}$Se$_x$)$_3$.
These topological properties also have twist angle dependence. In the small-angle limit, more edge-state-originated nearly flat bands appear around the Fermi level due to the band folding with the small Moir\'e BZ, as seen in Fig.\ref{fig:main2_3}. Therefore, as twist angle is decreased, topological phase transitions occur on VB8 and VB9 and eventually they become topologically trivial bands. The topological phase transition goes through a gap-closing at general momenta as described in a general theory of topological phase transitions in 2D systems \cite{PhysRevB.76.205304,murakami2007phase} (See Appendix \ref{sec:SM_supinfo_twist} and \ref{sec:SM_noncentroTI} for more detail). From this perspective, when the twist angle gets smaller, topological bands tend to appear in lower (or higher) energy bands.

The edge-state-originated nearly flat bands, Moir\'e-scale edge states, and other properties obtained in twisted bilayer Bi$_2$(Te$_{1-x}$Se$_x$)$_3$ are also reproduced in a more simplified model that we propose, twisted Bernevig-Hughes-Zhang model (See Appendix \ref{sec:SM_tBHZ}). This fact not only allows us to easily explore the properties of these systems, but also indicates that our essential strategy can be applied to other topological materials.

\section{Discussion \label{sec:main_discussion}}
%\noindent{\bf Conclusion and discussion}\\
In conclusion, we have theoretically studied the electronic structure of twisted bilayer Bi$_2$(Te$_{1-x}$Se$_x$)$_3$. It is revealed that twisted bilayer Bi$_2$(Te$_{1-x}$Se$_x$)$_3$ has a topological insulator domain and a normal insulator domain in the Moir\'e unit cell due to the stacking modulation of the Moir\'e superlattice structure. 
%The Moir\'e band structure has nearly flat bands around the Fermi level, and their ring-shaped wave functions indicate that they are formed by quantization of the topological edge state on the domain boundary.
%An angular momentum sequence has also been found in the flat bands.
%In the smaller energy region, we have found Moir\'e bands with nontrivial topological invariants that are formed by hybridization between an edge-state-originated band and a bulk-state-originated band.
%We have also obtained corresponding Moir\'e-scale helical edge states. The Moir\'e-scale edge state is made from the atomic-scale edge state on the domain boundary, 
We have obtained a Moir\'e-scale band inversion and corresponding edge states that are made from the atomic-scale edge state on the domain boundary, thus it can be called ``edge state from edge state". 
\rev{
With these results, we propose twisted bilayer Bi$_2$(Te$_{1-x}$Se$_x$)$_3$ as a new topological Moir\'e material that hosts novel low-energy states and unique multi-scale band inversion. Not only as a proposal of new material, but this proposal also provides a new platform to study topological phases both in atomic- and Moir\'e-scale. The atomic-scale helical edge states are expected to be used as an ideal platform to realize the Chalker-Coddington network model \cite{chalker1988percolation,PhysRevB.76.075301,PhysRevB.78.115301,PhysRevB.89.155315,ryu2010network} without an external field. The Moir\'e-scale topological states also propose a realization of Moir\'e topological spin Hall materials that do not require external fields. Further, because the wave function of the Moir\'e-scale helical edge state has characteristic density distribution and far larger real space size than that of helical edge states in previous topological insulator crystals, it allows us to observe the nontriviality of wave function with a real space observation such as scanning tunneling microscope (STM). While those phenomena proposed by some previous studies require a strong external field \cite{sanchez2017helical,tong2017topological,rickhaus2018transport,PhysRevB.88.121408,PhysRevB.98.035404,PhysRevLett.121.037702,PhysRevLett.121.146801} or a particular symmetry-broken valley setup \cite{PhysRevLett.122.086402}, the role is replaced by the intrinsic strong spin-orbit coupling in twisted bilayer Bi$_2$(Te$_{1-x}$Se$_x$)$_3$. Moreover, twisted bilayer Bi$_2$(Te$_{1-x}$Se$_x$)$_3$ can be a new platform to investigate correlated phases. As shown in the Moir\'e band dispersion, large Rashba splits are predicted around K$_M$ points in our calculation. It indicates that the ratio between bandwidth and the size of the Rashba split can be tuned by changing the twist angle. 
}
%Furthermore, because the ring-shaped edge-originated-states are protected by the atomic-scale topology, their existences are expected to be robust against perturbations including a degree of freedom in the edge truncation.

Although we have studied a specific material twisted bilayer Bi$_2$(Te$_{1-x}$Se$_x$)$_3$ here, the strategy underlying this study --- making a Moir\'e material with a stacking sensitive topological material --- is quite versatile. \rev{There is an abundance of stacking-sensitive topological materials, and it should be possible to design Moir\'e materials in which multiple topological phase domains coexist with this strategy.} The combination of a novel quantum system with several different topological phase domains and the high tunability of the Moir\'e materials provides an avenue in the search for new quantum materials and devices.

%\noindent{\bf Moir\'e edge calculation}\\
%To obtain the edge state spectra of the Moir\'e topological bands, we construct Wannier functions from the Moir\'e bands.
%\blue{To obtain the edge state spectra of the Moir\'e topological bands, we construct maximally localized Wannier functions from the Moir\'e bands by WANNIER90~\cite{wannier90}.}
%We use the option of the WANNIER90 to set the Bloch functions as the initial guess for the projections.
%The input data, energy eigenvalues and overlaps $\braket{\psi_{\bm{k}} | \psi_{\bm{k}+\delta\bm{k}}}$, are given from the effective model. The $k$-point mesh is set to $8\times8\times1$. Because the Moir\'e system generally deals with a small energy scale, numerical errors get more noticeable. Therefore, we correct the numerical errors in the obtained matrix elements that are slightly breaking the time-reversal symmetry. The correction is done by replacing $t_{\bm{a}_{nm}}^{\alpha\sigma,\beta'\sigma'}$ with $\left( t_{\bm{a}_{nm}}^{\alpha\sigma,\beta'\sigma'} + (-1)^{1-\delta_{\sigma\sigma'}} \left. t_{\bm{a}_{nm}}^{\alpha\bar{\sigma},\beta'\bar{\sigma}'} \right.^* \right)/2$, where $\bar{\sigma}$ is the opposite spin of $\sigma$. The edge state spectra are calculated by the Green's function method implemented in the WannierTools package~\cite{wanniertools}.

\section*{Acknowledgements \label{sec:main_ack}}
We are grateful to Akira Furusaki for fruitful discussion.
This work was supported by JST CREST (Grants No. JPMJCR19T2).
M.H. was supported by PRESTO, JST (JPMJPR21Q6) and JSPS KAKENHI Grants No. 20K14390.

%\noindent{\bf Author contributions}\\
%All authors contributed to the main contents of this work.
%I.T. constructed the theory for the Moir\'e system, and performed the effective model calculation.
%M.H. conceived and supervised the project, and performed the \blue{first-principles} calculation.

%\noindent{\bf Data availability}\\
%The authors declare that the data supporting the findings of this study are available within the paper and its supplementary information file.

%\textcolor{red}{ab initio or first-principles? We should choose one.}

\appendix

\section{Moir\'e pattern \label{sec:SM_moire}}

To make a 2D Moir\'e superlattice structure, there are two fundamental methods, lattice constant mismatch and twisting identical lattices. General Moir\'e superlattice systems are made by combining the two methods. In this paper, we focus on the cases of twisting identical lattices. 

\subsection{General cases \label{sec:SM_moire_general}}
\begin{figure}
    \centering
    \includegraphics[width=0.45\textwidth]{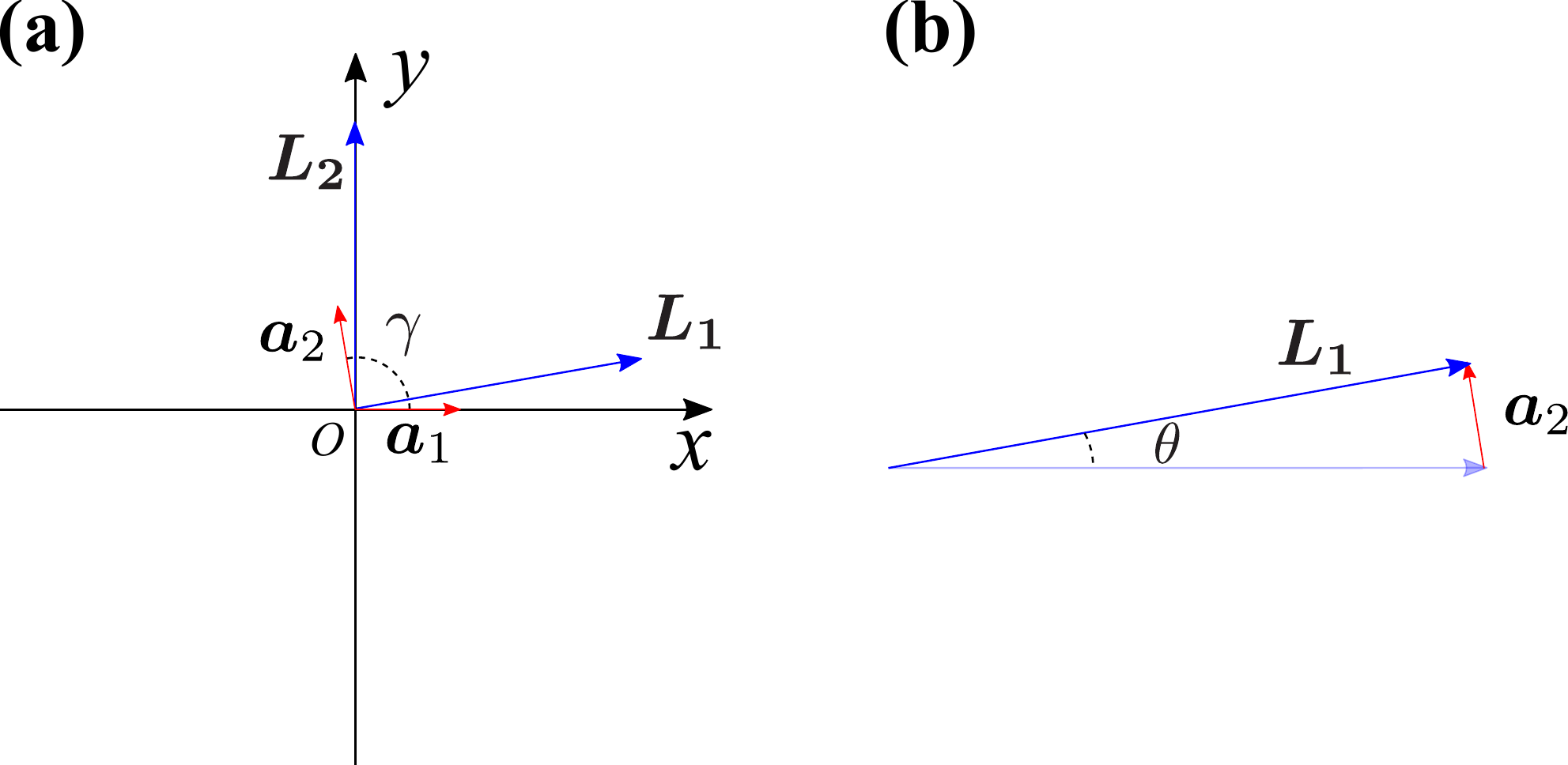}
    \caption{(a) Atomic lattice vectors $\bm{a}_1, \bm{a}_2$ and Moir\'e lattice venctors $\bm{L}_1,\bm{L}_2$. (b) Relation between the atomic lattice vector, the Moir\'e lattice vector, and twist angle $\theta$.}
    \label{fig:SM_AtomToMoire}
\end{figure}

Generally, given atomic lattice vectors $\bm{a}_1$, $\bm{a}_2$, and twist angle $\theta$, the Moir\'e lattice vectors $\bm{L}_1$ and $\bm{L}_2$ (Fig.\ref{fig:SM_AtomToMoire}(a)) satisfy
\begin{equation}
    \begin{split}
        2 \sin \frac{\theta}{2} \bm{e}_z \times \bm{L}_1 &= -\bm{a}_2 , \\
        2 \sin \frac{\theta}{2} \bm{L}_2 \times \bm{e}_z &= \bm{a}_1 ,
    \end{split}
\end{equation}
within the small angle approximation (Fig.\ref{fig:SM_AtomToMoire}(b)), where $\bm{e}_z$ is a unit vector along the $z$ axis.
By solving these equations, the Moire lattice vectors are obtained as
\begin{equation}
    \begin{split}
        \bm{L}_1 &= \bm{a}_2 \times \frac{1}{2 \sin \frac{\theta}{2} } \bm{e}_z  \approx \bm{a}_2 \times \frac{1}{\theta} \bm{e}_z , \\
        \bm{L}_2 &= \frac{1}{2 \sin \frac{\theta}{2} } \bm{e}_z \times \bm{a}_1 \approx \frac{1}{\theta} \bm{e}_z \times \bm{a}_1 . \\
    \end{split}
\end{equation}

It is also proved that $\bm{a}_l$ and $\bm{L}_m$ satisfies
\begin{equation}
    \begin{split}
        \bm{a}_l \cdot \bm{L}_m = \delta_{lm} \frac{|\bm{a}_1||\bm{a}_2|}{2 \sin \frac{\theta}{2}} \sin \gamma \approx \delta_{lm} \frac{|\bm{a}_1||\bm{a}_2|}{\theta} \sin \gamma ,
    \end{split}
\end{equation}
where $\gamma$ is the angle between $\bm{a}_1$ and $\bm{a}_2$.

This relation indicates that the Moir\'e lattice (Moir\'e reciprocal lattice) of the twisted system is similar to the atomic reciprocal lattice (atomic lattice) in the original untwisted system.

\subsection{Trigonal lattice \label{sec:SM_moire_tri}}
For a trigonal lattice system, atomic lattice vectors are given with the lattice constant $a_0$ as

\begin{equation}
    \begin{split}
        \bm{a}_1 &= a_0 \left( 1, 0 \right) , \\
        \bm{a}_2 &= a_0 \left( -1/2, \sqrt{3}/2 \right) .
    \end{split}
\end{equation}
Moir\'e lattice vectors are given as
\begin{equation}
    \begin{split}
        \bm{L}_1 &= L \left( \sqrt{3}/2, 1/2 \right) , \\
        \bm{L}_2 &= L \left( 0 , 1 \right) ,
    \end{split}
\end{equation}
where the Moir\'e lattice constant $L$ is written as
\begin{equation}
    L=\frac{a_0}{2 \sin \frac{\theta}{2}} .
\end{equation}

\section{Effective model of Moir\'e superlattice systems \label{sec:SM_model}} 
In this section, we derive a model of a Moir\'e superlattice system. In the following, the atomic lattice constant is set to $a_0=1$. In subsection \ref{sec:SM_model_general}, we describe the general and exact part of the model derivation that is independent of the detail of the target Moir\'e system. The description in this subsection is consistent with previous studies \cite{bistritzer2011moire,jung2014ab}. In subsection \ref{sec:SM_model_smallangle}, we introduce the small angle approximation, which is generally used in a Moir\'e system. In this subsection, we improve the approximation method of the previous studies to make them more efficient. In subsection \ref{sec:SM_model_specific}, we describe the system dependent part of the model derivation. In this subsection, we make a correction to make the model satisfy symmetry restrictions of the twisted bilayer Bi$_2$(Te$_{1-x}$Se$_x$)$_3$ system.

\subsection{General formula of model derivation \label{sec:SM_model_general}}

The exact Hamiltonian of a Moir\'e superlattice system is given by the following with a microscopic picture.

\begin{equation}
    \begin{split}
    H = & \sum_{\bm{r}_j} \sum_{\alpha\sigma,\beta'\sigma'} \sum_{pq} t_{pq}^{\alpha\sigma,\beta'\sigma'}\left( \bm{r}_j \right) \\
    & \times c^\dagger_{\alpha\sigma} \left( \bm{r}_j + \bm{d}_{\alpha} \left( \bm{r}_j \right) \right) c^{\phantom{\dagger}}_{\beta'\sigma'} \left( \bm{r}_j + \bm{d}_{\beta'} \left( \bm{r}_j \right) + \bm{a}_{pq} \right) .
    \end{split}
    \label{eq.Model.Hbegin}
\end{equation}

\begin{equation}
    \begin{array}{ccl}
        \bm{r}_j & : & \mathrm{atomic~unit~cell~position} \\
        \alpha\sigma, \beta'\sigma' & : & \mathrm{orbital,~spin~index} \\
        \bm{d}_{\alpha,\beta'} \left( \bm{r}_j \right) & : & \mathrm{site~position~of~} \alpha,\beta' \mathrm{~in~the~unit~cell~} \bm{r}_j \\
        pq & : & \mathrm{hopping~lattice~index~}(\bm{a}_{pq}) \\
        c^\dagger,c & : & \mathrm{creation/annihilation~operator} \\
    \end{array}
\end{equation}

\begin{figure}
    \centering
    \includegraphics[width=0.45\textwidth]{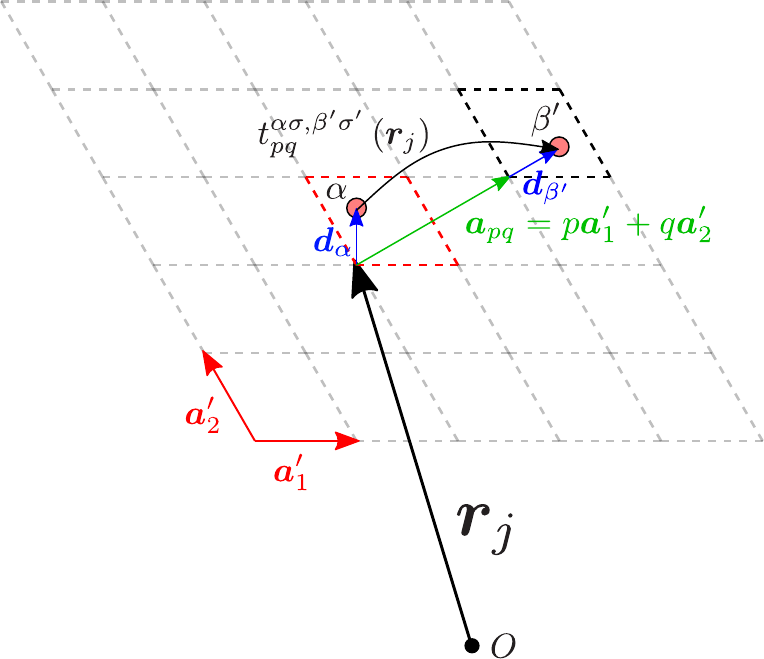}
    \caption{Definition of the hopping $t_{pq}^{\alpha\sigma,\beta'\sigma'}\left( \bm{r}_j \right)$. $\bm{r}_j$ is a position that indicates atomic unit cell index. The $t_{pq}^{\alpha\sigma,\beta'\sigma'}\left( \bm{r}_j \right)$ is a hopping between orbital $\alpha$ spin $\sigma$ in the atomic unit cell $\bm{r}_j$ (red dashed cell) and prbital $\beta'$ spin $\sigma'$ in the atomic unit cell $\bm{r}_j+\bm{a}_{pq}$ (black dashed cell).}
    \label{fig:SM_HoppingDef}
\end{figure}

Here, $\bm{r}_j$ is a position of an atomic unit cell, and $\alpha\sigma,\beta'\sigma'$ are indices to specify orbital and spin. Now $\bm{r}_j$ is defined as the corner of the atomic unit cell, and thus we define $\bm{d}_\alpha(\bm{r}_j)$ as the site position of the orbital $\alpha$ measured from the corner of the atomic unit cell $\bm{r}_j$. Due to the effect of twisting, $\bm{d}_\alpha(\bm{r}_j)$ generally depends on the atomic unit cell position $\bm{r}_j$. $c^\dagger_{\alpha \sigma}(\bm{r}_j + \bm{d}_\alpha (\bm{r}_j) )$ is a creation operator of a fermion on the orbital $\alpha$ spin $\sigma$ in the atomic unit cell $\bm{r}_j$, and $c$ is an annihilation operator. $t_{pq}^{\alpha\sigma,\beta'\sigma'}\left( \bm{r}_j \right)$ is a hopping parameter which is defined as a hopping between  ``orbital $\alpha$ spin $\sigma$ in the atomic unit cell $\bm{r}_j$" and ``orbital $\beta'$ spin $\sigma'$ in the atomic unit cell $\bm{r}_j + \bm{a}_{pq}$" (Fig.\ref{fig:SM_HoppingDef}). The $\bm{a}_{pq}$ are defined with integers $p,q$ as
\begin{equation}
    \bm{a}_{pq} = p \bm{a}'_1 + q \bm{a}'_2 ,
\end{equation}
where $\bm{a}'_1$, $\bm{a}'_2$ are atomic lattice vectors defined differently in intralayer and interlayer cases. The definitions of $\bm{a}'_1$ and $\bm{a}'_2$ are
\begin{equation}
    \begin{array}{ccl}
        \bm{a}'_1 = C_{\theta/2} \bm{a}_1 & \bm{a}'_2 = C_{\theta/2} \bm{a}_2 & \mathrm{(intralayer,~upper~layer)} \\
        \bm{a}'_1 = C_{-\theta/2} \bm{a}_1 & \bm{a}'_2 = C_{-\theta/2} \bm{a}_2 & \mathrm{(intralayer,~lower~layer)} \\ 
        \bm{a}'_1 = \bm{a}_1 & \bm{a}'_2 = \bm{a}_2 & \mathrm{(interlayer)} \\
        \label{eq.Model.a'def}
    \end{array} .
\end{equation}
We assume that long range hoppings are negligible and $\sum_{pq}$ is a finite sum. There are two points to note about $\bm{a}_{pq}$ and $\sum_{pq}$. First, in the interlayer case, $(\bm{a}_1,\bm{a}_2)$ does not match either the atomic lattice period of the upper layer or that of the lower layer. As a result, it can be possible that there is no $\beta'$ site (or there are two $\beta'$ sites) in the atomic unit cell $\bm{r}_j+\bm{a}_{pq}$. However, because the small angle limit is interpreted as a continuous limit, those inconvenient cases are neglected. Second, because $\bm{d}_{\alpha}(\bm{r}_j)$ depends on the position in the Moir\'e unit cell $\bm{r}_j$, a site can disappear on one end of the summation range of $\sum_{pq}$ and appear on the other end when $\bm{r}_j$ is changed. This will essentially make a ``discontinuity" in the Hamiltonian. However, as long as the range of $\sum_{pq}$ is large enough, the disappearing and appearing hoppings have small absolute values and thus the discontinuity is negligible.

We perform a Fourier transform of Eq. (\ref{eq.Model.Hbegin}). We assume that the wannier functions are independent of position in the Moir\'e unit cell, and thus the annihilation operator is transformed as
\begin{equation}
    c_{\alpha\sigma}(\bm{r}_j) = \int d\bm{k}~ e^{i \bm{k} \cdot \bm{r}_j} c_{\alpha\sigma,\bm{k}} .
\end{equation}

The Fourier transform of Eq. (\ref{eq.Model.Hbegin}) is obtained as
\begin{equation}
    \begin{split}
        H = & \int d\bm{k} d\bm{k}' \sum_{\bm{r}_j} \sum_{\alpha\sigma,\beta'\sigma'} \\
        & \left[ \sum_{pq} t_{pq}^{\alpha\sigma,\beta'\sigma'} \left( \bm{r}_j \right)  e^{-i \bm{k} \cdot \bm{d}_{\alpha} \left( \bm{r}_j \right) } e^{i \bm{k}' \cdot \left( \bm{d}_{\beta'}\left( \bm{r}_j \right) + \bm{a}_{pq} \right) } \right] \\
        & \times e^{i \left( -\bm{k}'+\bm{k} \right) \bm{r}_j} ~ c_{\alpha\sigma,\bm{k}}^\dagger c_{\beta'\sigma', \bm{k}'}^{\phantom{\dagger}} .
\end{split}
\end{equation}

We focus on the bracket part. The $\bm{d}_\alpha(\bm{r}_j)$ and $t_{pq}^{\alpha\sigma,\beta'\sigma'} \left( \bm{r}_j \right)$ are Moir\'e periodic. Therefore, the bracket part is Moir\'e periodic and can be decomposed into components of the Moir\'e reciprocal lattice vectors $\bm{G}_l$ as
\begin{equation}
    \begin{split}
        & \left[ \sum_{pq} t_{pq}^{\alpha\sigma,\beta'\sigma'} \left( \bm{r}_j \right)  e^{-i \bm{k} \cdot \bm{d}_{\alpha} \left( \bm{r}_j \right) } e^{i \bm{k}' \cdot \left( \bm{d}_{\beta'}\left( \bm{r}_j \right) + \bm{a}_{pq} \right) } \right] \\
        & = \sum_{\bm{G}_l} e^{i \bm{G}_{l} \cdot \bm{r}_j} ~t^{\alpha\sigma,\beta'\sigma'}_{\bm{k},\bm{k}',\bm{G}_{l}} ,
    \end{split}
    \label{eq.Model.tFT1}
\end{equation}
where $t^{\alpha\sigma,\beta'\sigma'}_{\bm{k},\bm{k}',\bm{G}_{l}}$ is the Fourier transform of the bracket part. Note that the $t^{\alpha\sigma,\beta'\sigma'}_{\bm{k},\bm{k}',\bm{G}_{l}}$ is not the Fourier transform of $t_{pq}^{\alpha\sigma,\beta'\sigma'} \left( \bm{r}_j \right)$ because the bracket part includes a sum on $p,q$ and phase factors.

By using Eq. (\ref{eq.Model.tFT1}), the Hamiltonian is writetn as
\begin{equation}
    \begin{split}
        H &= \int d\bm{k} d\bm{k}' \sum_{\bm{r}_j} \sum_{\alpha\sigma,\beta'\sigma'} \sum_{\bm{G}_l} ~ t_{\bm{k},\bm{k}',\bm{G}_l}^{\alpha\sigma,\beta'\sigma'} ~ e^{i \left( -\bm{k} + \bm{k}' + \bm{G}_l \right) \cdot \bm{r}_{j} } \\
        & ~~~~~~~~~~~~~~~~~~~~~~~~~~~~~~~~~ \times c_{\alpha\sigma,\bm{k}}^\dagger c_{\beta'\sigma', \bm{k}'}^{\phantom{\dagger}} \\
        &=  \int d\bm{k} \sum_{\alpha\sigma,\beta'\sigma'} \sum_{\bm{G}_l} t_{\bm{k},\bm{k}-\bm{G}_l,\bm{G}_l}^{\alpha\sigma,\beta'\sigma'}~ c_{\alpha\sigma,\bm{k}}^\dagger c_{\beta'\sigma', \bm{k}-\bm{G}_l}^{\phantom{\dagger}} .
        \label{eq.Model.Hsimple}
\end{split}
\end{equation}
Note that $\bm{k}'$ is restricted to $\bm{k}-\bm{G}_l$.

So far, the model derivation is exact and general. However, in the Eq. (\ref{eq.Model.tFT1}), all $t_{pq}^{\alpha\sigma,\beta'\sigma'} \left( \bm{r}_j \right)$ in the Moir\'e unit cell are needed to obtain exact value of $t^{\alpha\sigma,\beta'\sigma'}_{\bm{k},\bm{k}',\bm{G}_{l}}$. Because it is usually unrealistic, we use a small angle approximations to obtain $t^{\alpha\sigma,\beta'\sigma'}_{\bm{k},\bm{k}',\bm{G}_{l}}$, as described in the following.

\subsection{Small angle approximation \label{sec:SM_model_smallangle}}
In a small twist angle case, $t^{\alpha\sigma,\beta'\sigma'}_{\bm{k},\bm{k}',\bm{G}_{l}}$ can be approximately obtained as following.

First, we obtain $t_{pq}^{\alpha\sigma,\beta'\sigma'} \left( \bm{r}_j \right)$ at finite number of sampling points $\{ \bm{r}_j \}$ in the Moir\'e unit cell. In a small twist angle case, the local atomic lattice structures around each sampling points can be approximated by untwisted atomic lattices with proper layer stacking orders. We perform first-principles calculation for each untwisted structure and use the results as approximate value of $t_{pq}^{\alpha\sigma,\beta'\sigma'} \left( \bm{r}_j \right)$ at the sampling points. With the approximated $t_{pq}^{\alpha\sigma,\beta'\sigma'} \left( \bm{r}_j \right)$, $t^{\alpha\sigma,\beta'\sigma'}_{\bm{k},\bm{k}',\bm{G}_{l}}$ are obtained by the discrete Fourier transform Eq. (\ref{eq.Model.tFT1}).

\begin{figure*}
    \centering
    \includegraphics[width=0.7\textwidth]{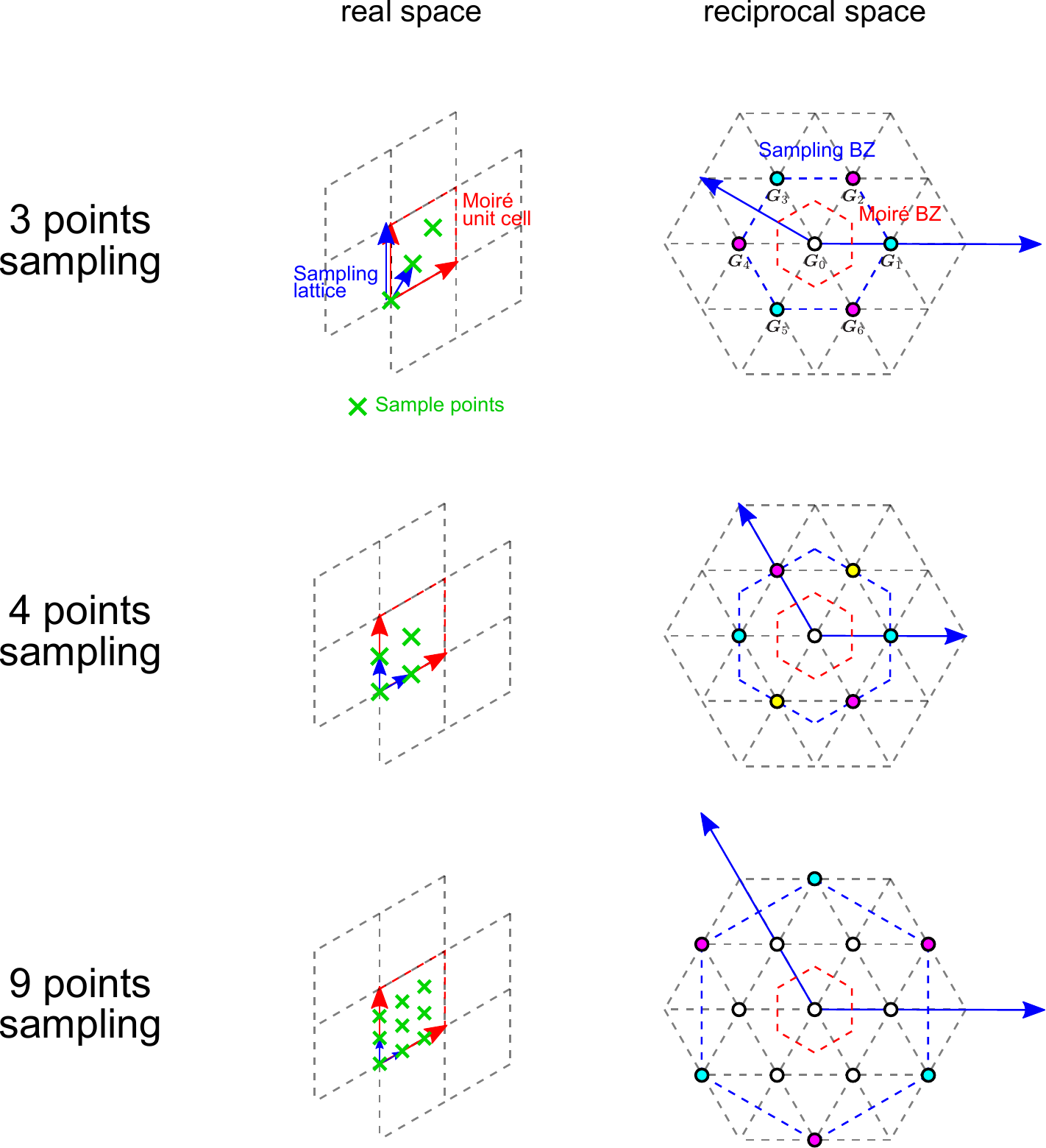}
    \caption{Examples of parameter sampling and $\bm{G}_l$ adopted in the discrete Fourier transform in a trigonal lattice system. The left figure shows the Moir\'e unit cell (red), sampling points, and sampling lattice vectors (blue arrows) in the real space. The right figure shows the Moir\'e BZ (red dashed), reciprocal vectors of the sampling lattice (blue arrows), and $\bm{G}_l$ adopted in the discrete Fourier transform (white and colored dots). For $\bm{G}_l$ shown with the same colored dot, $t^{\alpha\sigma,\beta'\sigma'}_{\bm{k},\bm{k}',\bm{G}_{l}}$ have the same value.}
    \label{fig:SM_sampling_C3}
\end{figure*}

\begin{figure*}
    \centering
    \includegraphics[width=0.7\textwidth]{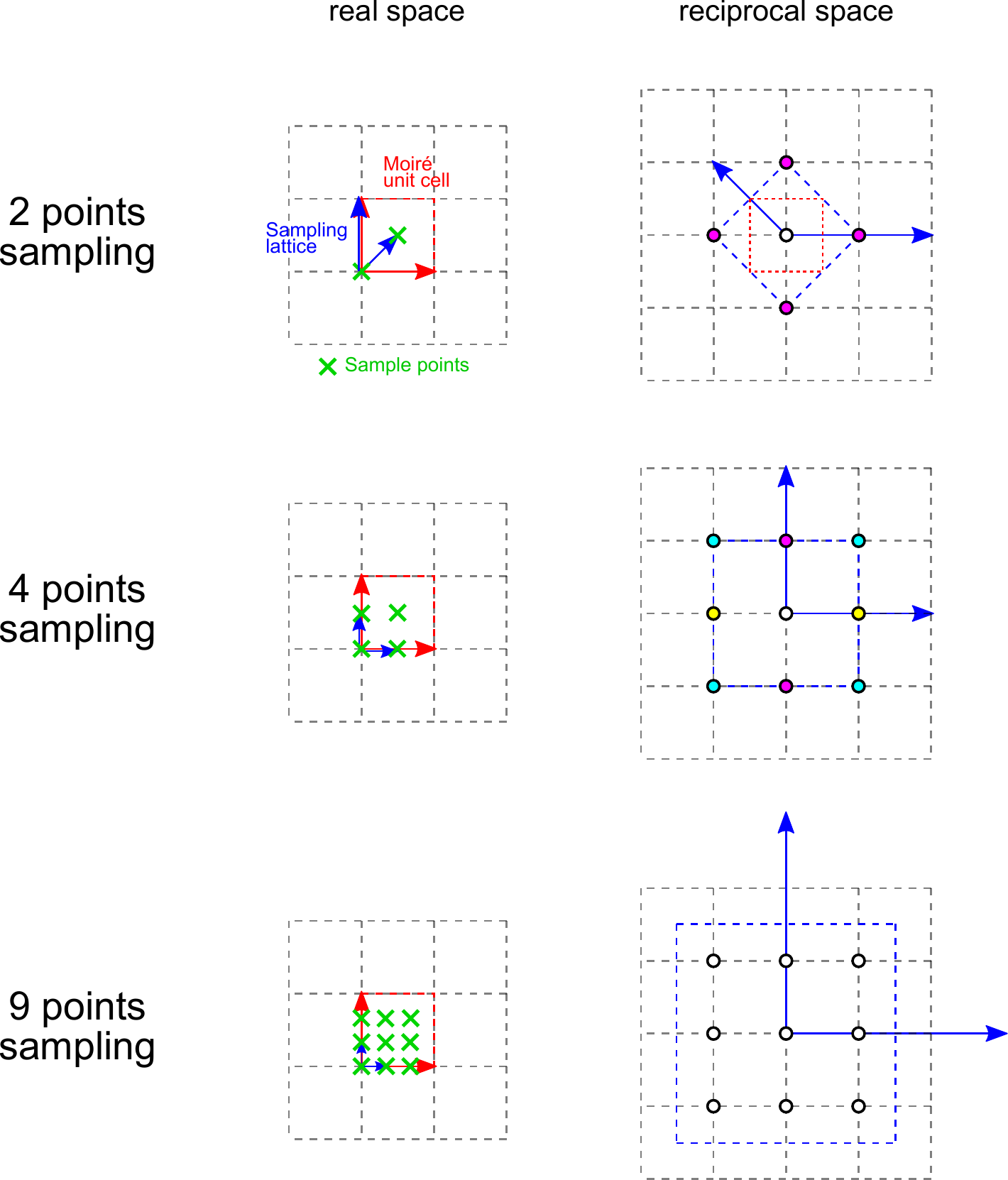}
    \caption{Examples of parameter sampling and $\bm{G}_l$ adopted in the discrete Fourier transform in a square lattice system. The left figure shows the Moir\'e unit cell (red), sampling points, and sampling lattice vectors (blue arrows) in the real space. The right figure shows the Moir\'e BZ (red dashed), reciprocal vectors of the sampling lattice (blue arrows), and $\bm{G}_l$ adopted in the discrete Fourier transform (white and colored dots). For $\bm{G}_l$ shown with the same colored dot, $t^{\alpha\sigma,\beta'\sigma'}_{\bm{k},\bm{k}',\bm{G}_{l}}$ have the same value.}
    \label{fig:SM_sampling_C4}
\end{figure*}

Generally, the discrete Fourier transform guarantees that the parameters in the sampling points are reproduced, but those in the intermediate area are not necessarily estimated in a physically reasonable way. To obtain a reasonable estimation for the whole Moir\'e unit cell by discrete Fourier transform, we need to adopt appropriate $\bm{G}_l$ and sampling points must be determined to capture the ``feature values", for example, the maximum and minimum values of $t_{pq}^{\alpha\sigma,\beta'\sigma'} \left( \bm{r}_j \right)$. Therefore, the sampling points should be determined from the atomic positions in the untwisted atomic lattice. Here, we explain general way to decide the sampling points and how to choose $\bm{G}_l$ in detail, and then we show some examples in Figs.\ref{fig:SM_sampling_C3} and \ref{fig:SM_sampling_C4}. As the first step, we decide untwisted bilayer structures with interlayer in-plane shift $\{ \Delta \bm{r}_j \}$ that should be sampled. For the convenience of the discrete Fourier transform, the list $\{ \Delta \bm{r}_j \}$ should be taken as a mesh in the atomic unit cell. As explained above, the $\{ \Delta \bm{r}_j \}$ should include the cases where feature values are obtained. Typically, the $\{ \Delta \bm{r}_j \}$ include the cases where the atoms on the upper layer and the lower layer are closest or farthest. We translate the $\{ \Delta \bm{r}_j \}$ to a position in the Moir\'e unit cell $\{ \bm{r}_j \}$ by using the equation $\Delta \bm{r}_j = \bm{r}_j \times \theta \bm{e}_z$ (easily obtained from Fig.\ref{fig:SM_AtomToMoire}(b)). As a result, a sampling lattice (green crosses and blue arrows in left figures in Figs.\ref{fig:SM_sampling_C3} and \ref{fig:SM_sampling_C4}) is obtained in the Moir\'e unit cell and a sampling BZ is also defined in the reciprocal space (blue hexagons in the right figures). The number of sampling points (the number of $\Delta \bm{r}_j$, originally) decides a number of $\bm{G}_l$ that is taken into account in Eq. (\ref{eq.Model.tFT1}) and Eq. (\ref{eq.Model.Hsimple}). Moir\'e reciprocal lattice points included in the sampling BZ (white dots) are adopted as $\bm{G}_l$. Additionally, Moir\'e reciprocal lattice points on the boundary of the sampling BZ (Cyan, magenta, and yellow dots) are also adopted as $\bm{G}_l$, but we assume that $t^{\alpha\sigma,\beta'\sigma'}_{\bm{k},\bm{k}',\bm{G}_{l}}$ have same value in the points that are connected by the reciprocal vectors of the sampling lattice (blue arrows in the right figures). $\bm{G}_l$ outside of the sampling BZ are neglected, i. e. components with frequencies higher than the sampling points are neglected. By using this method, the number of the sampling points and $\bm{G}_l$ are equal, and thus $t^{\alpha\sigma,\beta'\sigma'}_{\bm{k},\bm{k}',\bm{G}_{l}}$ are uniquely determined by the discrete Fourier transform Eq. (\ref{eq.Model.tFT1}). Examples of the way to choose $\bm{G}_l$ are shown in Figs.\ref{fig:SM_sampling_C3} and \ref{fig:SM_sampling_C4} (for trigonal and square cases, respectively). In each row, the left figure is the Moir\'e lattice in the real space, and the right figure is the Moir\'e reciprocal lattice in the reciprocal space. In the left figure, the Moir\'e unit cell (red) and sampling points (green crosses) are shown. Blue arrows are basic lattice vectors of the sampling lattice. In the right figure, the Moir\'e BZ (red dashed) and the sampling BZ (blue dashed), which is defined by the sampling lattice vectors, are shown. Moir\'e reciprocal lattice points that is adopted as $\bm{G}_l$ are shown with dots. If it is included in the sampling BZ, the dot is white. If it is on the boundary of the sampling BZ, the dot is colored (cyan, magenta, or yellow) and the points that is connected by the reciprocal vectors of the sampling lattice (blue arrows) have the same color. $t^{\alpha\sigma,\beta'\sigma'}_{\bm{k},\bm{k}',\bm{G}_{l}}$ on those $\bm{G}_l$ have the same value. For example, in the case of the three points sampling in Fig.\ref{fig:SM_sampling_C3},
\begin{equation}
    t^{\alpha\sigma,\beta'\sigma'}_{\bm{k},\bm{k}',\bm{G}_{1}} = t^{\alpha\sigma,\beta'\sigma'}_{\bm{k},\bm{k}',\bm{G}_{3}} = t^{\alpha\sigma,\beta'\sigma'}_{\bm{k},\bm{k}',\bm{G}_{5}} .
\end{equation}
Note that the number of the sampling points in the left figure equals to the number of $\bm{G}_l$ in the right figure (dots with same color on the boundary should be counted as one) in each column.

Note that depending on the symmetry of the system and the center of momentum of the cutoff (See Sec.\ref{sec:SM_numerical_cutoff}), hopping parameters in some stacking orders can be fixed to be 0 \cite{PhysRevResearch.1.033076}.

\subsection{Specific formula for Bi$_2$(Te$_{1-x}$Se$_x$)$_3$ \label{sec:SM_model_specific}}
In this subsection, we apply the method given in the previous subsection to Bi$_2$(Te$_{1-x}$Se$_x$)$_3$.
In the case of Bi$_2$(Te$_{1-x}$Se$_x$)$_3$, the in-plane atomic positions are only three in the atomic unit cell ($C_3$ symmetry centers). Therefore, the simplest choice is to calculate three untwisted stacking structures (the three points sampling in Fig.\ref{fig:SM_sampling_C3}). Here, we define the three stacking orders as AA, AB, and AC stacking. In this case, we take components of seven $\bm{G}_l$,
\begin{equation}
    \bm{G}_l = \left\{ \bm{G}_0, \bm{G}_1, \bm{G}_2, \bm{G}_3, \bm{G}_4, \bm{G}_5, \bm{G}_6 \right\} ,
\end{equation}
and assume equivalences between them as
\begin{equation}
    \begin{split}
        t^{\alpha\sigma,\beta'\sigma'}_{\bm{k},\bm{k}',\bm{G}_{1}} &= t^{\alpha\sigma,\beta'\sigma'}_{\bm{k},\bm{k}',\bm{G}_{3}} = t^{\alpha\sigma,\beta'\sigma'}_{\bm{k},\bm{k}',\bm{G}_{5}} , \\
        t^{\alpha\sigma,\beta'\sigma'}_{\bm{k},\bm{k}',\bm{G}_{2}} &= t^{\alpha\sigma,\beta'\sigma'}_{\bm{k},\bm{k}',\bm{G}_{4}} = t^{\alpha\sigma,\beta'\sigma'}_{\bm{k},\bm{k}',\bm{G}_{6}} .
    \end{split}
    \label{eq.Model.tGrelation}
\end{equation}
Including a completely independent component
\begin{equation}
    t^{\alpha\sigma,\beta'\sigma'}_{\bm{k},\bm{k}',\bm{G}_{0}} ,
\end{equation}
the degree of freedom is three, which is the same as the number of the sampling points. Therefore, using the relation Eq. (\ref{eq.Model.tGrelation}), we can determine the seven $t^{\alpha\sigma,\beta'\sigma'}_{\bm{k},\bm{k}',\bm{G}_{l}}$ from the three sampling points with the discrete Fourier transform.

The hopping parameters in the three sampling points, $t_{pq}^{\alpha\sigma,\beta'\sigma'} \left( \bm{r}_{AA} \right)$, $t_{pq}^{\alpha\sigma,\beta'\sigma'} \left( \bm{r}_{AB} \right)$, and $t_{pq}^{\alpha\sigma,\beta'\sigma'} \left( \bm{r}_{AC} \right)$, are calculated with the first-principle calculation of untwisted structures. However, when the hopping parameters of the twisted structure are estimated from untwisted structures, the Hamiltonian Eq. (\ref{eq.Model.Hsimple}) will break the Hermitian and $C_{3z}$ symmetry slightly. Therefore, we add a correction factor in Eq. (\ref{eq.Model.tFT1}).

The restrictions by each symmetry are given as following. The detail of the derivation of the restrictions is given in later subsection.
The Hermitian symmetry restriction is
\begin{equation}
    \left. t_{\bm{k}+\bm{G}_l, \bm{k},\bm{G}_l}^{\alpha\sigma,\beta'\sigma'} \right.^* = t_{\bm{k},\bm{k}+\bm{G}_l,-\bm{G}_l}^{\beta'\sigma',\alpha\sigma}
    \label{eq.Model.HeCondition}
\end{equation}
and the $C_{3z}$ symmetry restriction is
\begin{equation}
    t^{\alpha\sigma,\beta'\sigma'}_{\bm{k},\bm{k}-\bm{G}_l,\bm{G}_l} = t^{\widetilde{\alpha\sigma},\widetilde{\beta'\sigma'}}_{C_3\bm{k},C_3\bm{k}-C_3\bm{G}_l,C_3\bm{G}_l} ,
    \label{eq.Model.C3Condition}
\end{equation}
where the orbital-spin index $\widetilde{\alpha\sigma}$ means the $C_{3z}$ rotated orbital-spin.
The twisted structure also have time-reversal symmetry and its restriction is
\begin{equation}
    (-1)^{1-\delta_{\sigma\sigma'}} \left. t_{-\bm{k},-\bm{k}-\bm{G}_l,\bm{G}_l}^{\alpha\sigma,\beta'\sigma'} \right.^* = t_{\bm{k},\bm{k}+\bm{G}_l,-\bm{G}_l}^{\alpha\bar{\sigma},\beta'\bar{\sigma}'} ,
    \label{eq.Model.TRCondition}
\end{equation}
where $\bar{\sigma}$ means the opposite spin, but this equation is satisfied when the three untwisted structures have the time-reversal symmetry.

To satisfy these equation, we add a correction factor $e^{i \frac{\bm{G}_m}{2} \cdot \left( \bm{d}_{\alpha}(\bm{r}_j) + \bm{d}_{\beta'}(\bm{r}_j) + \bm{a}_{pq} \right) }$ in the Eq. (\ref{eq.Model.tFT1}) as
\begin{equation}
    \begin{split}
        & \sum_{\bm{G}_l} e^{i \bm{G}_{l} \cdot \bm{r}_j} ~t^{\alpha\sigma,\beta'\sigma'}_{\bm{k},\bm{k}-\bm{G}_{m},\bm{G}_{l}} \\
        & = \left[ \sum_{pq} t_{pq}^{\alpha\sigma,\beta'\sigma'} \left( \bm{r}_j \right)  e^{-i \bm{k} \cdot \bm{d}_{\alpha} \left( \bm{r}_j \right) } e^{i \left( \bm{k}-\bm{G}_{m} \right) \cdot \left( \bm{d}_{\beta'}\left( \bm{r}_j \right) + \bm{a}_{pq} \right) } \right. \\
        & \left. \phantom{~= \sum_{pq} t_{pq}^{\alpha\sigma,\beta'\sigma'}\left( \bm{r}_j \right) } \times e^{i \frac{\bm{G}_m}{2} \cdot \left( \bm{d}_{\alpha}(\bm{r}_j) + \bm{d}_{\beta'}(\bm{r}_j) + \bm{a}_{pq} \right) } \right] \\
        & = \left[ \sum_{pq} t_{pq}^{\alpha\sigma,\beta'\sigma'} \left( \bm{r}_j \right)  e^{i \left( \bm{k}- \frac{\bm{G}_{m}}{2} \right) \cdot \left( -\bm{d}_{\alpha}(\bm{r}_j) + \bm{d}_{\beta'}\left( \bm{r}_j \right) + \bm{a}_{pq} \right) } \right] \\
        & = ~ h_{\bm{r}_j}^{\alpha\sigma,\beta'\sigma'} ( \bm{k}-\bm{G}_m/2 ) .
    \end{split}
    \label{eq.Model.Hdftcorrected}
\end{equation}
Because the correction factor tends to 1 as the twist angle get small, this correction is reasonable.
Note that the right-hand-side $h_{\bm{r}_j}^{\alpha\sigma,\beta'\sigma'} ( \bm{k}-\bm{G}_m/2 )$ is basically the matrix element of the Hamiltonian with momentum $\bm{k}-\bm{G}_m/2$ given by the parameters of untwisted structure around $\bm{r}_j$. The only difference is the momentum twisting due to the different definition of $\bm{a}'_{1,2}$ in Eq. (\ref{eq.Model.a'def}). The explicit definition is given as, for $\alpha\sigma, \beta'\sigma'\in$ Upper layer,
\begin{equation}
    \begin{split}
        & h_{\bm{r}_j}^{\alpha\sigma,\beta'\sigma'} ( \bm{k}-\bm{G}_m/2 ) \\
        & = \sum_{pq} t_{\mathrm{DFT}, pq}^{\alpha\sigma, \beta'\sigma'}(\bm{r}_j)~ e^{i C_{
        -\theta/2} \left( \bm{k}-\bm{G}_m/2 \right) \cdot \left( - \bm{d}_{\alpha}(\bm{r}_j) + \bm{d}_{\beta'}(\bm{r}_j) + \bm{a}_{pq} \right) } ,
    \end{split}
    \label{eq.Model.Uk}
\end{equation}
for $\alpha\sigma, \beta'\sigma'\in$ Lower layer,
\begin{equation}
    \begin{split}
        & h_{\bm{r}_j}^{\alpha\sigma,\beta'\sigma'} ( \bm{k}-\bm{G}_m/2 ) \\
        & = \sum_{pq} t_{\mathrm{DFT}, pq}^{\alpha\sigma, \beta'\sigma'}(\bm{r}_j)~ e^{i C_{
        \theta/2} \left( \bm{k}-\bm{G}_m/2 \right) \cdot \left( - \bm{d}_{\alpha}(\bm{r}_j) + \bm{d}_{\beta'}(\bm{r}_j) + \bm{a}_{pq} \right) } ,
    \end{split}
    \label{eq.Model.Lk}
\end{equation}
for $\alpha\sigma\in$ Upper Layer and $\beta'\sigma'\in$ Lower layer,
\begin{equation}
    \begin{split}
        & h_{\bm{r}_j}^{\alpha\sigma,\beta'\sigma'} ( \bm{k}-\bm{G}_m/2 ) \\
        & = \sum_{pq} t_{\mathrm{DFT}, pq}^{\alpha\sigma, \beta'\sigma'}(\bm{r}_j)~ e^{i \left( \bm{k}-\bm{G}_m/2 \right) \cdot \left( - \bm{d}_{\alpha}(\bm{r}_j) + \bm{d}_{\beta'}(\bm{r}_j) + \bm{a}_{pq} \right) } , \\
        & h_{\bm{r}_j}^{\beta'\sigma',\alpha\sigma} ( \bm{k}-\bm{G}_m/2 ) = { h_{\bm{r}_j}^{\alpha\sigma,\beta'\sigma'} }( \bm{k}-\bm{G}_m/2 )^\dagger ,
    \end{split}
    \label{eq.Model.Tk}
\end{equation}
where $t_{\mathrm{DFT}, pq}^{\alpha\sigma, \beta'\sigma'}(\bm{r}_j)$ is the hopping parameter of the untwisted structure around $\bm{r}_j$ calculated by the first-principles calculation.

The discrete Fourier transform is performed as
\begin{equation}
    \begin{split}
        t^{\alpha\sigma,\beta'\sigma'}_{\bm{k},\bm{k},\bm{G}_{0}} =& \frac{1}{3} \left[ h_{\bm{r}_{AA}}^{\alpha\sigma,\beta'\sigma'}(\bm{k}) + h_{\bm{r}_{AB}}^{\alpha\sigma,\beta'\sigma'}(\bm{k}) + h_{\bm{r}_{AC}}^{\alpha\sigma,\beta'\sigma'}(\bm{k}) \right] , \\
        t^{\alpha\sigma,\beta'\sigma'}_{\bm{k},\bm{k}-\bm{G}_{l},\bm{G}_{l}} =& \frac{1}{9} \left[  h_{\bm{r}_{AA}}^{\alpha\sigma,\beta'\sigma'}(\bm{k}-\bm{G}_{l}/2) \right. \\
        &~~~~ + e^{-i \frac{2\pi}{3} } h_{\bm{r}_{AB}}^{\alpha\sigma,\beta'\sigma'}(\bm{k}-\bm{G}_{l}/2) \\
        &~~~~ \left. + e^{i \frac{2\pi}{3} } h_{\bm{r}_{AC}}^{\alpha\sigma,\beta'\sigma'}(\bm{k}-\bm{G}_{l}/2) \right] ~~ (l=1,3,5), \\
        t^{\alpha\sigma,\beta'\sigma'}_{\bm{k},\bm{k}-\bm{G}_{l},\bm{G}_{l}} =& \frac{1}{9} \left[  h_{\bm{r}_{AA}}^{\alpha\sigma,\beta'\sigma'}(\bm{k}-\bm{G}_{l}/2) \right. \\
        &~~~~ + e^{i \frac{2\pi}{3} } h_{\bm{r}_{AB}}^{\alpha\sigma,\beta'\sigma'}(\bm{k}-\bm{G}_{l}/2) \\
        &~~~~ \left. + e^{-i \frac{2\pi}{3} } h_{\bm{r}_{AC}}^{\alpha\sigma,\beta'\sigma'}(\bm{k}-\bm{G}_{l}/2) \right] ~~ (l=2,4,6) .
        \label{eq.Model.tFT2}
    \end{split}
\end{equation}
With these definitions, Eqs. (\ref{eq.Model.tGrelation}), (\ref{eq.Model.HeCondition}), (\ref{eq.Model.C3Condition}) and (\ref{eq.Model.TRCondition}) are satisfied. The $t^{\alpha\sigma,\beta'\sigma'}_{\bm{k},\bm{k},\bm{G}_{0}}$ term can be understood as an averaged term, and the others as Moir\'e modulation terms.

\subsubsection{Gauge selection}
In Eqs.(\ref{eq.Model.Uk})(\ref{eq.Model.Lk})(\ref{eq.Model.Tk}), the phase factors are calculated with the explicit difference of the orbital positions. There is another standard way to calculate the phase factor, where a phase of the basic translations is given only to inter-cell hopping terms. In an untwisted system, these two are connected by a gauge transform, and thus have no physical difference. However, in a Moir\'e system, they are physically different in a strict sense due to the interpolation of the Hamiltonian. Although the difference is negligible in the small angle limit with a small cutoff momentum, the former gives a better interpolation for the large $\bm{k}$ region. Therefore, the former definition is recommended, as we used.

\begin{widetext}
\subsubsection{Derivation of symmetry restrictions}
We start from the Hamiltonian Eq.(\ref{eq.Model.Hbegin}).
\begin{equation}
    \begin{split}
        H =& \sum_{\bm{r}_j} \sum_{\alpha\sigma,\beta'\sigma'} \sum_{pq} t_{pq}^{\alpha\sigma,\beta'\sigma'}\left( \bm{r}_j \right)~ c^\dagger_{\alpha\sigma} \left( \bm{r}_j + \bm{d}_{\alpha} \left( \bm{r}_j \right) \right) c^{\phantom{\dagger}}_{\beta'\sigma'} \left( \bm{r}_j + \bm{d}_{\beta'} \left( \bm{r}_j \right) + \bm{a}_{pq} \right)
    \end{split}
\end{equation}

\begin{itemize}

\item{\bf Hermitian symmetry}

The Hermite conjugate of the Hamiltonian is given as
\begin{equation}
    \begin{split}
        H^\dagger =& \sum_{\bm{r}_j} \sum_{\alpha\sigma,\beta'\sigma'} \sum_{pq} \left. t_{pq}^{\alpha\sigma,\beta'\sigma'} \right.^* \left( \bm{r}_i \right)~ c^{\dagger}_{\beta'\sigma'} \left( \bm{r}_j + \bm{d}_{\beta'} \left( \bm{r}_j \right) + \bm{a}_{pq} \right) c^{\phantom{\dagger}}_{\alpha\sigma} \left( \bm{r}_j + \bm{d}_{\alpha} \left( \bm{r}_j \right) \right) \\
         =& \int d\bm{k} d\bm{k}' \sum_{\bm{r}_j} \sum_{\alpha\sigma,\beta'\sigma'} \left[ \sum_{pq} t_{pq}^{\alpha\sigma,\beta'\sigma'}(\bm{r}_j) e^{-i \bm{k}' \cdot \bm{d}_{\alpha}(\bm{r}_j) } e^{i \bm{k} \cdot \left( \bm{d}_{\beta'}(\bm{r}_j) + \bm{a}_{pq} \right) } \right]^* e^{i (- \bm{k} + \bm{k}') \cdot \bm{r}_j } ~ c^{\dagger}_{\beta'\sigma',\bm{k}} c^{\phantom{\dagger}}_{\alpha\sigma,\bm{k}'} \\
         =& \int d\bm{k} d\bm{k}' \sum_{\bm{r}_j} \sum_{\alpha\sigma,\beta'\sigma'} \sum_{\bm{G}_l} e^{-i \bm{G}_l \cdot \bm{r}_j} \left. t_{\bm{k}',\bm{k},\bm{G}_l}^{\alpha\sigma,\beta'\sigma'} \right.^* e^{i (- \bm{k} + \bm{k}') \cdot \bm{r}_j } ~ c^{\dagger}_{\beta'\sigma',\bm{k}} c^{\phantom{\dagger}}_{\alpha\sigma,\bm{k}'} \\
         =& \int d\bm{k} \sum_{\alpha\sigma,\beta'\sigma'} \sum_{\bm{G}_l} \left. t_{\bm{k}+\bm{G}_l,\bm{k},\bm{G}_l}^{\alpha\sigma,\beta'\sigma'} \right.^* c^{\dagger}_{\beta'\sigma',\bm{k}} c^{\phantom{\dagger}}_{\alpha\sigma,\bm{k}+\bm{G}_l} .
    \end{split}
\end{equation}

The conditions for this to be Hermite (Eq. (\ref{eq.Model.HeCondition})) is
\begin{equation}
    \left. t_{\bm{k}+\bm{G}_l, \bm{k},\bm{G}_l}^{\alpha\sigma,\beta'\sigma'} \right.^* = t_{\bm{k},\bm{k}+\bm{G}_l,-\bm{G}_l}^{\beta'\sigma',\alpha\sigma} .
\end{equation}

When the Hamiltonian of the untwisted structures are Hermite,
\begin{equation}
    \left. h_{\bm{r}}^{\alpha\sigma,\beta'\sigma'}(\bm{k}) \right.^* = h_{\bm{r}}^{\beta'\sigma',\alpha\sigma}(\bm{k})
\end{equation}
is satisfied and the Eq. (\ref{eq.Model.HeCondition}) is satisfied by using Eq. (\ref{eq.Model.tFT2}).

\item{\bf $C_{3z}$ symmetry}

The $C_3$ rotated Hamiltonian is
\begin{equation}
    \begin{split}
        C_3 H C_3^\dagger =& \sum_{\bm{r}_j} \sum_{\alpha\sigma,\beta'\sigma'} \sum_{pq} t_{pq}^{\alpha\sigma,\beta'\sigma'}\left( \bm{r}_j \right)~ c^\dagger_{\widetilde{ \alpha\sigma }} \left( C_3 \bm{r}_j + C_3 \bm{d}_{\alpha} (\bm{r}_j) \right) c^{\phantom{\dagger}}_{\widetilde{ \beta'\sigma' }} \left( C_3 \bm{r}_j + C_3 \bm{a}_{pq} + C_3 \bm{d}_{\beta'} (\bm{r}_j) \right) \\
        &= \int d\bm{k} d\bm{k'} \sum_{\bm{r}_j} \sum_{\alpha\sigma,\beta'\sigma'} \left[ \sum_{pq} t_{pq}^{\alpha\sigma,\beta'\sigma'}(\bm{r}_j) e^{-i C_3^{-1} \bm{k} \cdot \bm{d}_{\alpha}(\bm{r}_j)} e^{i C_3^{-1} \bm{k}' \cdot \left( \bm{d}_{\beta'}(\bm{r}_j) + \bm{a}_{pq} \right) } \right] e^{i C_3^{-1}(-\bm{k}+\bm{k}') \cdot \bm{r}_j } c^\dagger_{\widetilde{ \alpha\sigma }, \bm{k}} c_{\widetilde{\beta'\sigma'} , \bm{k}' } \\
        &= \int d\bm{k} d\bm{k}' \sum_{\bm{r}_j} \sum_{\alpha\sigma,\beta'\sigma'} \left[ \sum_{\bm{G}_l} e^{i \bm{G}_l \cdot \bm{r}_j} t^{\alpha\sigma,\beta'\sigma'}_{C_3^{-1}\bm{k},C_3^{-1}\bm{k}',\bm{G}_l} \right] e^{i C_3^{-1}(-\bm{k}+\bm{k}') \cdot \bm{r}_j } c^\dagger_{\widetilde{ \alpha\sigma }, \bm{k}} c_{\widetilde{\beta'\sigma'} , \bm{k}' } \\
        &= \int d\bm{k} \sum_{\alpha\sigma,\beta'\sigma'} \sum_{\bm{G}_l} t^{\alpha\sigma,\beta'\sigma'}_{C_3^{-1}\bm{k},C_3^{-1}\bm{k}-\bm{G}_l,\bm{G}_l}~ c^\dagger_{\widetilde{ \alpha\sigma }, \bm{k}} c_{\widetilde{\beta'\sigma'} , \bm{k}-C_3 \bm{G}_l } \\
        &= \int d\bm{k} \sum_{\alpha\sigma,\beta'\sigma'} \sum_{\bm{G}_l} t^{\alpha\sigma,\beta'\sigma'}_{\bm{k},\bm{k}-\bm{G}_l,\bm{G}_l}~ c^\dagger_{\widetilde{ \alpha\sigma }, C_3 \bm{k}} c_{\widetilde{\beta'\sigma'} , C_3 \bm{k}-C_3 \bm{G}_l } ,
    \end{split}
\end{equation}
where the orbital-spin index $\widetilde{\alpha\sigma}$ means the $C_{3z}$ rotated orbital-spin.

The conditions for this to be $C_3$ symmetric (Eq. (\ref{eq.Model.C3Condition})) is
\begin{equation}
    t^{\alpha\sigma,\beta'\sigma'}_{\bm{k},\bm{k}-\bm{G}_l,\bm{G}_l} = t^{\widetilde{\alpha\sigma},\widetilde{\beta'\sigma'}}_{C_3\bm{k},C_3\bm{k}-C_3\bm{G}_l,C_3\bm{G}_l} .
\end{equation}

When the Hamiltonian of the untwisted structures are $C_3$ symmetric,
\begin{equation}
    h_{\bm{r}}^{\alpha\sigma,\beta'\sigma'}(\bm{k}) = h_{\bm{r}}^{\widetilde{\alpha\sigma},\widetilde{\beta'\sigma'}}(C_3 \bm{k})
\end{equation}
is satisfied and the Eq. (\ref{eq.Model.C3Condition}) is satisfied by using Eq. (\ref{eq.Model.tFT2}).

\item{\bf Time-reversal symmetry}

The time-reversed Hamiltonian is
\begin{equation}
    \begin{split}
        {\cal T} H {\cal T}^\dagger =& \sum_{\bm{r}_j} \sum_{\alpha\sigma,\beta'\sigma'} \sum_{pq} \left. t_{pq}^{\alpha\sigma,\beta'\sigma'} \right.^* \left( \bm{r}_i \right)~ c^{\dagger}_{\alpha \bar{\sigma}} \left( \bm{r}_j + \bm{d}_{\alpha} \left( \bm{r}_j \right) \right) c^{\phantom{\dagger}}_{\beta' \bar{\sigma}'} \left( \bm{r}_j + \bm{d}_{\beta'} \left( \bm{r}_j \right) + \bm{a}_{pq} \right) (-1)^{1-\delta_{\sigma \sigma'}} \\
         =& \int d\bm{k} d\bm{k}' \sum_{\bm{r}_j} \sum_{\alpha\sigma,\beta'\sigma'} \left[ \sum_{pq} t_{pq}^{\alpha\sigma,\beta'\sigma'}(\bm{r}_j) e^{i \bm{k} \cdot \bm{d}_{\alpha}(\bm{r}_j) } e^{i \bm{k}' \cdot \left( \bm{d}_{\beta'}(\bm{r}_j) + \bm{a}_{pq} \right) } \right]^* e^{i (- \bm{k} + \bm{k}') \cdot \bm{r}_j } ~ c^{\dagger}_{\alpha \bar{\sigma}, \bm{k} } c^{\phantom{\dagger}}_{\beta' \bar{\sigma}', \bm{k}' } (-1)^{1-\delta_{\sigma \sigma'}} \\
         =& \int d\bm{k} d\bm{k}' \sum_{\bm{r}_j} \sum_{\alpha\sigma,\beta'\sigma'} \sum_{\bm{G}_l} e^{-i \bm{G}_l \cdot \bm{r}_j} \left. t_{-\bm{k},-\bm{k}',\bm{G}_l}^{\alpha\sigma,\beta'\sigma'} \right.^* e^{i (- \bm{k} + \bm{k}') \cdot \bm{r}_j } ~ c^{\dagger}_{\alpha \bar{\sigma}, \bm{k} } c^{\phantom{\dagger}}_{\beta' \bar{\sigma}', \bm{k}' } (-1)^{1-\delta_{\sigma \sigma'}} \\
         =& \int d\bm{k} \sum_{\alpha\sigma,\beta'\sigma'} \sum_{\bm{G}_l} \left. t_{-\bm{k},-\bm{k}-\bm{G}_l,\bm{G}_l}^{\alpha\sigma,\beta'\sigma'} \right.^* ~ c^{\dagger}_{\alpha \bar{\sigma}, \bm{k} } c^{\phantom{\dagger}}_{\beta' \bar{\sigma}', \bm{k}+\bm{G}_l } (-1)^{1-\delta_{\sigma \sigma'}} ,
    \end{split}
\end{equation}
where $\bar{\sigma}$ means the opposite spin.

The conditions for this to be time-reversal symmetric (Eq. (\ref{eq.Model.TRCondition})) is
\begin{equation}
    (-1)^{1-\delta_{\sigma\sigma'}} \left. t_{-\bm{k},-\bm{k}-\bm{G}_l,\bm{G}_l}^{\alpha\sigma,\beta'\sigma'} \right.^* = t_{\bm{k},\bm{k}+\bm{G}_l,-\bm{G}_l}^{\alpha\bar{\sigma},\beta'\bar{\sigma}'} .
\end{equation}

When the Hamiltonian of the untwisted structures are time-reversal symmetric,
\begin{equation}
    \left. h_{\bm{r}}^{\alpha\sigma,\beta'\sigma'} (\bm{k}) \right.^* = (-1)^{1-\delta_{\sigma\sigma'}} ~ h_{\bm{r}}^{\alpha\bar{\sigma},\beta'\bar{\sigma}'} (-\bm{k})
\end{equation}
is satisfied and the Eq. (\ref{eq.Model.TRCondition}) is satisfied by using Eq. (\ref{eq.Model.tFT2}).

\end{itemize}

\end{widetext}

\section{Numerical calculation methods \label{sec:SM_numerical}}
In this section, we describe methods used in our numerical calculations.

\subsection{Matrix representation \label{sec:SM_numerical_matrixrep}}
We have obtained the Hamiltonian of the Moir\'e system,
\begin{equation}
    \begin{split}
        H &= \int d\bm{k} \sum_{\alpha\sigma,\beta'\sigma'} \sum_{\bm{G}_l} t_{\bm{k},\bm{k}-\bm{G}_l,\bm{G}_l}^{\alpha\sigma,\beta'\sigma'}~ c_{\alpha\sigma,\bm{k}}^\dagger c_{\beta'\sigma', \bm{k}-\bm{G}_l}^{\phantom{\dagger}} ,
\end{split}
\end{equation}
and here we give the matrix representation of this Hamiltonian so that it can be implemented in numerical calculations.

Because all hopping processes have momentum difference $\bm{G}_l$, the basis is given for $\bm{k} \in$ Moir\'e~BZ as
\begin{equation}
    \left\{ c_{\alpha\sigma,\bm{k}+\bm{v}_{mn}} \left| \bm{v}_{mn}=m \bm{G}_1 + n \bm{G}_2 , ~ m,n \in \mathbb{Z} \right. \right\} .
\end{equation}
To describe low-energy physics with a finite dimension matrix, we take a momentum cutoff $k_c$ as an approximation.  In the Bi$_2$Te$_3$ system, the low-energy physics are described around the $\Gamma$ point. Therefore, we take a finite base set as
\begin{equation}
    \left\{ c_{\alpha\sigma,\bm{k}+\bm{v}_{mn}} \left| \bm{v}_{mn}=m \bm{G}_1 + n \bm{G}_2 , ~ m,n \in \mathbb{Z},~|\bm{v}_{mn}|<k_c \right. \right\} .
    \label{eq:SM_effectivebasis}
\end{equation} 

With this basis, the Matrix representation of the Hamiltonian is
\tiny
\begin{equation}
    \begin{split}
        H(\bm{k}) = \left(
        \begin{array}{cccccc}
            t_{\bm{k},\bm{k},\bm{0}} & & & & \mathrm{if}~\bm{v}_{mn}-\bm{v}_{m'n'}=\bm{G}_l & \\
             & \ddots & & & \downarrow & \\
             & & t_{\bm{k}+\bm{v}_{mn},\bm{k}+\bm{v}_{mn},\bm{0}} & \cdots & t_{\bm{k}+\bm{v}_{mn},\bm{k}+\bm{v}_{m'n'},\bm{G}_l} & \\
             & & & \ddots & \vdots & \\
             & & & & t_{\bm{k}+\bm{v}_{m'n'},\bm{k}+\bm{v}_{m'n'},\bm{0}} & \\
             & & & & & \ddots \\
        \end{array}
        \right) ,
    \end{split}
\end{equation}
\normalsize
where $\bm{k} \in$ Moir\'e BZ.

Because we consider only seven $\bm{G}_l$, an off-diagonal block is non-zero only when $\bm{v}_{mn}-\bm{v}_{m'n'}$ matches one of the seven $\bm{G}_l$.

\subsection{Validity of the cutoff $k_c$ \label{sec:SM_numerical_cutoff}}
\begin{figure}
    \centering
    \includegraphics[width=0.45\textwidth]{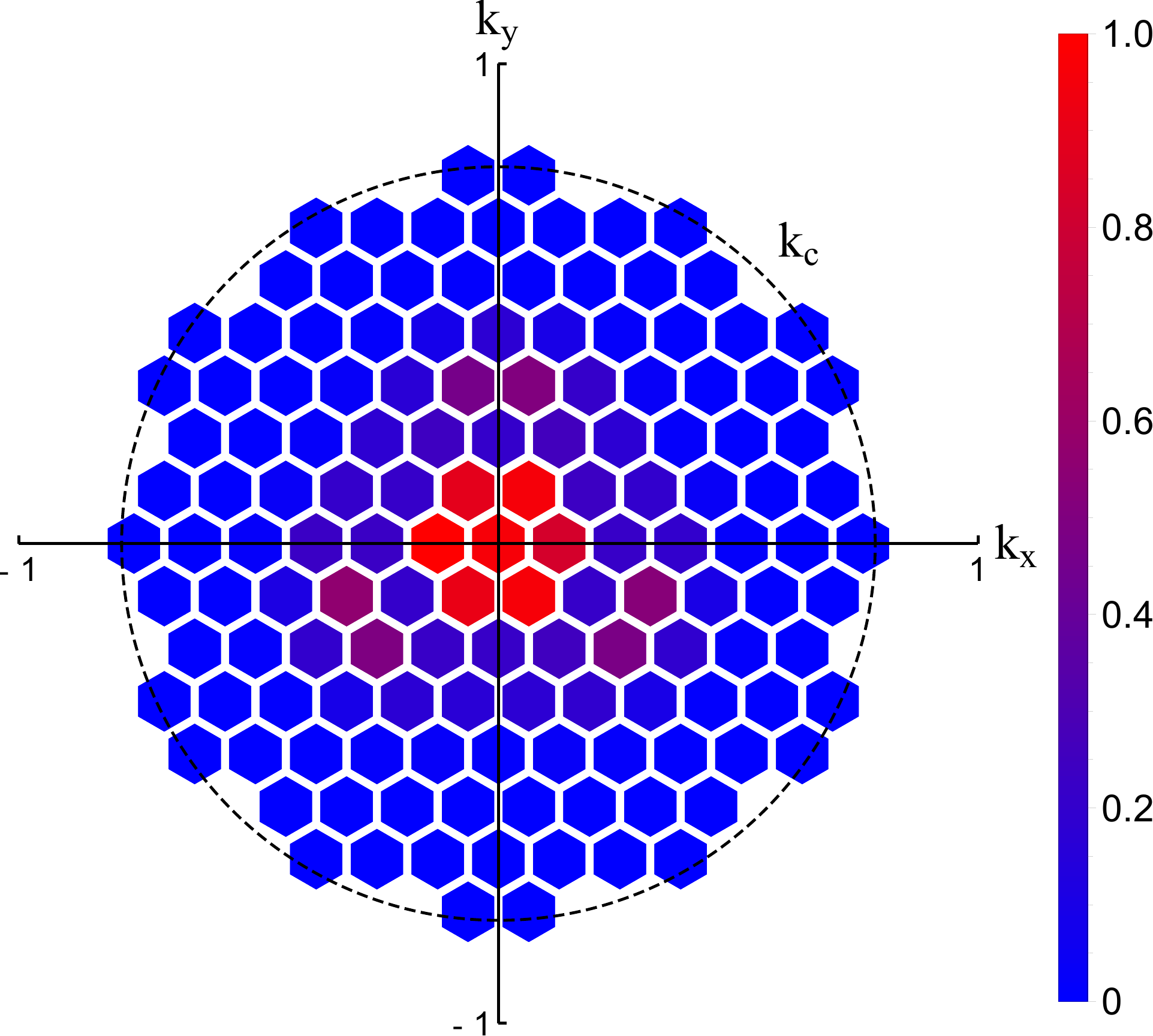}
    \caption{Weight mapping of the 17th valence band at the $\Gamma$ point of twisted bilayer Bi$_2$(Te$_{0.92}$Se$_{0.08}$)$_3$ with a twist angle $\theta=1.0^\circ$ and cutoff $k_c=\pi/4$. The hexagons are Moir\'e BZs and their colors represent the weight on them (normalized with the maximum value). In this case, it can be seen that the cutoff is large enough.}
    \label{fig:SM_kcmap}
\end{figure}

The validity of the cutoff $k_c$ is evaluated as follows. Once we calculate a wave function $\psi_{\bm{k}}$, and make a weight mapping onto Moir\'e BZs within $k_c$ as
\begin{equation}
    \begin{split}
        \psi_{\bm{k}} = & \left(
    \begin{array}{c}
         \phi_1 \\
         \phi_2 \\
         \vdots
    \end{array}
    \right) \\
    & \mapsto
    \left(
    \begin{array}{c}
         \sum_{l=1}^{60} |\phi_l|^2 \\
         \sum_{l=61}^{120} |\phi_l|^2 \\
         \vdots
    \end{array}
    \right) ~ 
    \begin{array}{l}
         \leftarrow \mathrm{weight~on~Moir\acute{e}~BZ\#1} \\
         \leftarrow \mathrm{weight~on~Moir\acute{e}~BZ\#2} \\
         \vdots
    \end{array} .
    \end{split}
\end{equation}
By plotting the weight mapping, we check the weight on the BZs around $k_c$. If the BZs around $k_c$ have small weights, the cutoff $k_c$ is large enough and the effect of $k_c$ is negligible. If the BZs around $k_c$ have significant weight compared to other BZs, larger $k_c$ should be taken. Generally, wave functions whose energies are far from Fermi level requires larger $k_c$. Therefore, evaluating with wave functions that are distant from the Fermi level ensures the validity of $k_c$ for wave functions that are closer to the Fermi level. Figure \ref{fig:SM_kcmap} is an example of the weight mapping for the 17th valence band of Bi$_2$(Te$_{0.92}$Se$_{0.08}$)$_3$ with a twist angle $\theta = 1.0 ^\circ$, cutoff $k_c = \pi/4$. Each hexagon represents a Moir\'e BZ, and its color represents the weight (normalized with the maximum value). It can be seen that the Moir\'e BZs around $k_c$ have almost no weights and thus the cutoff is large enough at least for the 1st $\sim$ 17th valence bands in this case. The validity of $k_c$ has been confirmed in all calculations in this paper.

\subsection{Periodicity in the momentum space \label{sec:SM_numerical_periodink}}
From the Bloch theorem, a wave function $\psi_{\bm{k}+\bm{G}_l}$ is physically identical with $\psi_{\bm{k}}$, and thus $| \braket{\psi_{\bm{k}+\bm{G}_l} | \psi_{\bm{k}} } | = 1$. This fact is used in the numerical evaluation of the Wilson loop spectra.

In the effective model of a Moir\'e superlattice system with basis (\ref{eq:SM_effectivebasis}), the momentum shift $\bm{k} \to \bm{k}+\bm{G}_l$ effectively result in a shift in $\bm{v}_{mn}$ indices. Therefore, components of a numerically obtained eigenvector $\psi_{\bm{k}+\bm{G}_l}$ are shifted from those of $\psi_{\bm{k}}$, and $| \braket{\psi_{\bm{k}+\bm{G}_l} | \psi_{\bm{k}} } | \neq 1$ due to the $\bm{k}$ dependence of the basis. Note that, as long as the cutoff $k_c$ is large enough, i.e. no significant component is put outside of the cutoff by the shift, $\psi_{\bm{k}+\bm{G}_l}$ and $\psi_{\bm{k}}$ are physically identical states. To evaluate the inner product of wave functions $\psi_{\bm{k}+\bm{G}_l}$ and $\psi_{\bm{k}}$ correctly with the numerically obtained eigenvectors, we introduce a base-shift matrix $T$, which is defined as
\begin{equation}
    T = \{ T_{(mn)(m'n')} \} = \left\{
    \begin{array}{ll}
        I_{\eta} & \bm{v}_{mn}+\bm{G}_l=\bm{v}_{m'n'} \\
        O_{\eta} & \mathrm{others} \\
    \end{array}
    \right. ,
\end{equation}
where $\eta$ is the orbital-spin degree of freedom ($\eta=60$ in twisted bilayer Bi$_2$(Te$_{1-x}$Se$_x$)$_3$), and $I_\eta$ and $O_\eta$ are the $\eta$ dimension identity matrix and zero matrix, respectively. By using this $T$, the inner product of wave functions $\psi_{\bm{k}+\bm{G}_l}$ and $\psi_{\bm{k}}$ is evaluated as $ \braket{\psi_{\bm{k}+\bm{G}_l} |T| \psi_{\bm{k}} } $, which approximately satisfies $ | \braket{\psi_{\bm{k}+\bm{G}_l} |T| \psi_{\bm{k}} } | =1 $ when the cutoff is large enough.

\subsection{Wannierization and surface state calculation \label{sec:SM_numerical_wannierization}}
To calculate surface states, we construct Wannier functions from wave functions and a band dispersion obtained from the effective model. We use the option \texttt{use\_bloch\_phases=T} of the WANNIER90 \cite{wannier90} to set the Bloch functions as the initial guess for the projections. In calculations with this option, we need a list of eigenenergies $E_{n\bm{k}}$ in file \texttt{prefix.eig} and a list of wave function overlaps $ \braket{\psi_{\bm{k} + \delta \bm{k}} | \psi_{\bm{k}} } $ in file \texttt{prefix.mmn} for the target bands. These lists are numerically obtained from the effective model calculation. Both of them should be calculated on a finite $\bm{k}$ mesh. Note that in the calculation of overlaps, the base-shift matrix given in the previous subsection should be used for overlaps that cross the end of the $\bm{k}$ mesh. With this method, Wannier functions and hoppings between them are obtained for Moir\'e bands. With the obatined parameters, edge state spectra are calculated by the Green's function method implemented in the WannierTools package \cite{wanniertools}.

For the parameters obtained by the wannierization, we make a correction to recover the time-reversal symmetry. It is because the energy scale of a Moir\'e superlattice system is so small that numerical errors in the first-principles calculations in untwisted systems get more noticeable. The correction is done by replacing the hopping parameters as
\begin{equation}
    t_{\bm{a}_{nm}}^{\alpha\sigma,\beta'\sigma'} \to \frac{1}{2} \left( t_{\bm{a}_{nm}}^{\alpha\sigma,\beta'\sigma'} + (-1)^{1-\delta_{\sigma\sigma'}} \left. t_{\bm{a}_{nm}}^{\alpha\bar{\sigma},\beta'\bar{\sigma}'} \right.^* \right) ,
\end{equation}
where $\bar{\sigma}$ is the opposite spin of $\sigma$. The $\delta_{\sigma\sigma'}$ is the Kronecker delta for spin $\sigma$ and $\sigma'$, which is essentially comes from ${\cal T}^2=-1$ in a spin-$1/2$ system. We have confirmed that this correction works only as a minor correction in the edge state spectra.

\subsection{Symmetry in wave function plot \label{sec:SM_numerical_symmetry}}
Twisted bilayer Bi$_2$(Te$_{1-x}$Se$_x$)$_3$ has the time-reversal symmetry and the layer group symmetry \#67, which include $C_{3z}$ rotation symmetry. Due to the time-reversal symmetry, all bands at the $\Gamma$ point appear as doubly degenerate Kramers' pairs. Because the numerical diagonalization does not necessarily give simultaneous eigenvectors of the Hamiltonian and the $C_{3z}$ rotation, we need to symmetrize the Kramers' pair to obtain a $C_{3z}$ symmetric wave function plot at the $\Gamma$ point.

For a given Kramers' pair $\ket{\psi_1},\ket{\psi_2}$ and $C_{3z}$ operator, a matrix representation of $\Gamma_{C_{3z}}$ is given as
\begin{equation}
    \Gamma_{C_{3z}} = \left(
    \begin{array}{cc}
         \braket{\psi_1|C_{3z}|\psi_1} & \braket{\psi_1|C_{3z}|\psi_2} \\
         \braket{\psi_2|C_{3z}|\psi_1} & \braket{\psi_2|C_{3z}|\psi_2}
    \end{array}
    \right).
\end{equation}
With a Unitary transform $U$ that diagonalize $\Gamma_{C_{3z}}$ as $U^\dagger \Gamma_{C_{3z}} U = \mathrm{diag}(\lambda_1,\lambda_2)$, symmetrized wave functions $\ket{\tilde{\psi}_1},\ket{\tilde{\psi}_2}$ are obtained as
\begin{equation}
    \left( \begin{array}{cc} \ket{\tilde{\psi}_1} & \ket{\tilde{\psi}_2} \end{array} \right) = \left( \begin{array}{cc} \ket{\psi_1} & \ket{\psi_2} \end{array} \right) U .
\end{equation}

\section{Effect of Se-doping on twisted bilayer Bi$_2$(Te$_{1-x}$Se$_x$)$_3$ \label{sec:SM_Sedope}}
In this section, we discuss the effect of Se-doping in in twisted bilayer Bi$_2$(Te$_{1-x}$Se$_x$)$_3$. First, we see the Se-doping effect in parameters of untwisted Bi$_2$(Te$_{1-x}$Se$_x$)$_3$ systems. Next, we see the Moir\'e band dispersions for various Se-doping values.

\subsection{Effect of Se-doping on parameters in untwisted bilayer Bi$_2$(Te$_{1-x}$Se$_x$)$_3$ \label{sec:SM_Sedope_untwisted}}
\begin{figure*}
    \centering
    \includegraphics[width=\textwidth]{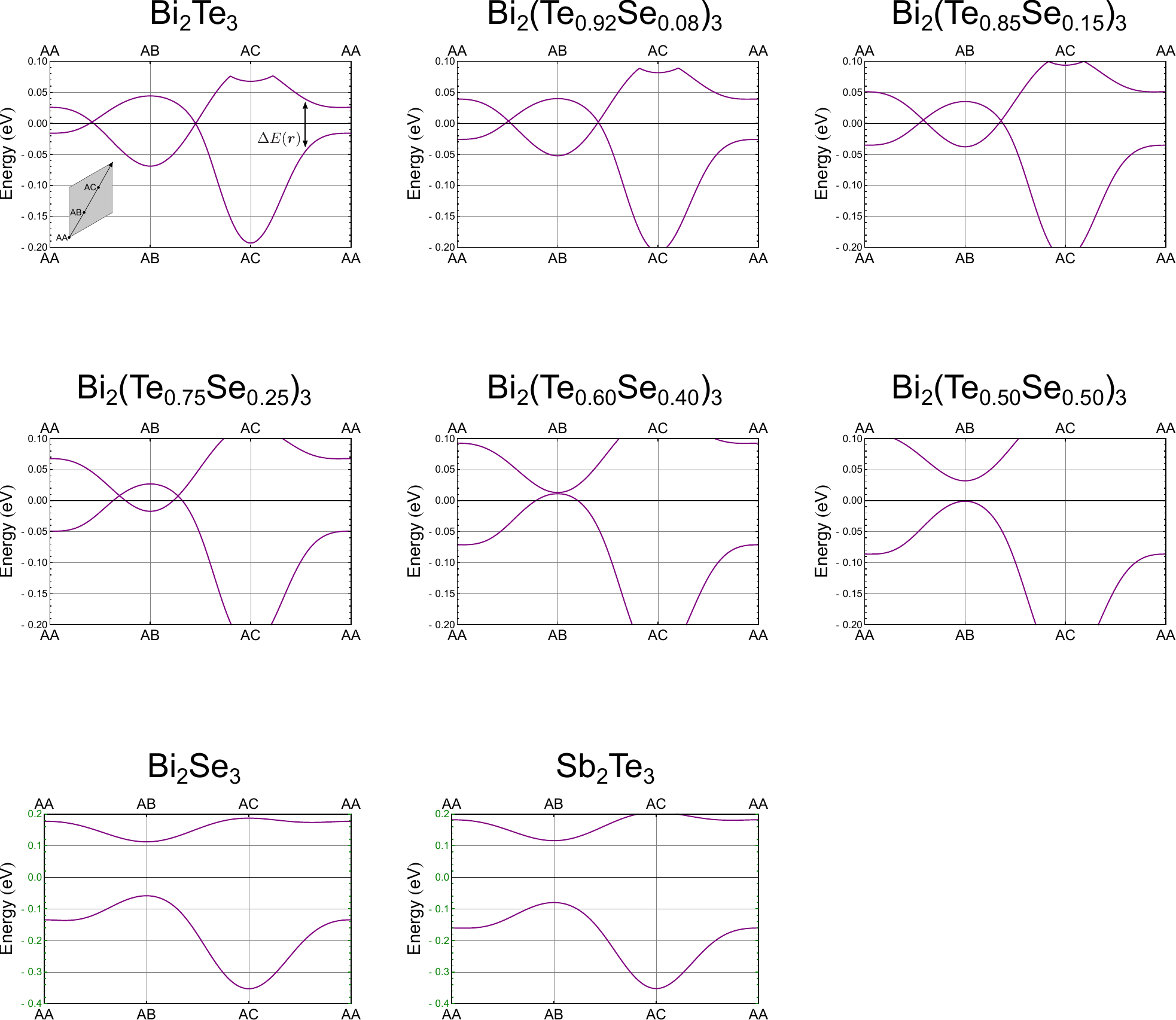}
    \caption{Energies of the valence top band and the conduction bottom band at the $\Gamma$ point plotted along the AA-AB-AC-AA stacking line for Bi$_2$Te$_3$, Bi$_2$(Te$_{1-x}$Se$_x$)$_3$ with various $x$, Bi$_2$Se$_3$, and Sb$_2$Te$_3$. Note that the Bi$_2$Se$_3$ and Sb$_2$Te$_3$ cases are plotted in a different energy range.}
    \label{fig:SM_stackgap}
\end{figure*}

The discrete Fourier transform Eq. (\ref{eq.Model.Hdftcorrected}) interpolates the untwisted Hamiltonians to the intermediate area. Therefore, we can define ``bandgap in each real space position $\bm{r}$" in the Moir\'e unit cell, and it can be used to estimate where the topological domain boundary is. To evaluate the bandgap in the $\Gamma$ point in the small twist angle limit, we consider $\bm{k}=\bm{0}$ in Eq. (\ref{eq.Model.Hdftcorrected}) as
\begin{equation}
    \sum_{\bm{G}_l} e^{i \bm{G}_l \cdot \bm{r}} ~ t_{\bm{0},-\bm{G}_l,\bm{G}_l}^{\alpha\sigma,\beta'\sigma'} = h_{\bm{r}}^{\alpha\sigma,\beta'\sigma'}(\bm{0}) ,
    \label{eq:rdependentH}
\end{equation}
where $t_{\bm{0},-\bm{G}_l,\bm{G}_l}^{\alpha\sigma,\beta'\sigma'}$ are given in Eq. (\ref{eq.Model.tFT2}). Note that $\bm{G}_l$ are negligible compared to the atomic BZ in the small twist angle limit. By diagonalizing the right-hand-side, we obtain a bandgap $\Delta E (\bm{r})$. The position where $\Delta E (\bm{r}) = 0 $ is satisfied is the topological domain boundary. In Fig.\ref{fig:SM_stackgap}, the energies of the valence top band and the conduction bottom band at the $\Gamma$ point calculated with Eq.(\ref{eq:rdependentH}) are plotted along the AA-AB-AC-AA stacking line for Bi$_2$Te$_3$, Bi$_2$(Te$_{1-x}$Se$_x$)$_3$ with various $x$, Bi$_2$Se$_3$, and Sb$_2$Te$_3$. 
\blue{Here, we calculate the electronic band structure of Sb$_2$Te$_3$ under the same condition as Bi$_2$Te$_3$ and Bi$_2$Se$_3$.}
Note that Bi$_2$Se$_3$ and Sb$_2$Te$_3$ cases are plotted in a different energy range. $\Delta E (\bm{r}) = 0 $ is the gap between the two band in $\bm{r}$ (horizontal axis).

As shown in the main article, the valence and conduction bands in Bi$_2$Te$_3$ are inverted only around the AB-stacking case. As Se is doped ($x$ gets larger), the (inverted) bandgap in the AB-stacking area gets smaller, and the bandgap in the AA- and AC-stacking are get larger. Around Bi$_2$(Te$_{0.60}$Se$_{0.40}$)$_3$ ($x=0.40$), a band-touching occurs at the AB-stacking point, and the band structure in the AB-stacking area become trivial in the $x>0.40$ region. Because the plots of Bi$_2$Se$_3$ and Sb$_2$Te$_3$ are almost identical, it can be seen that replacing Bi with Sb has a similar effect as replacing Te with Se.

\subsection{Effect of Se-doping on Moir\'e band dispersion \label{sec:SM_Sedope_twisted}}
\begin{figure*}
    \centering
    \includegraphics[width=0.8\textwidth]{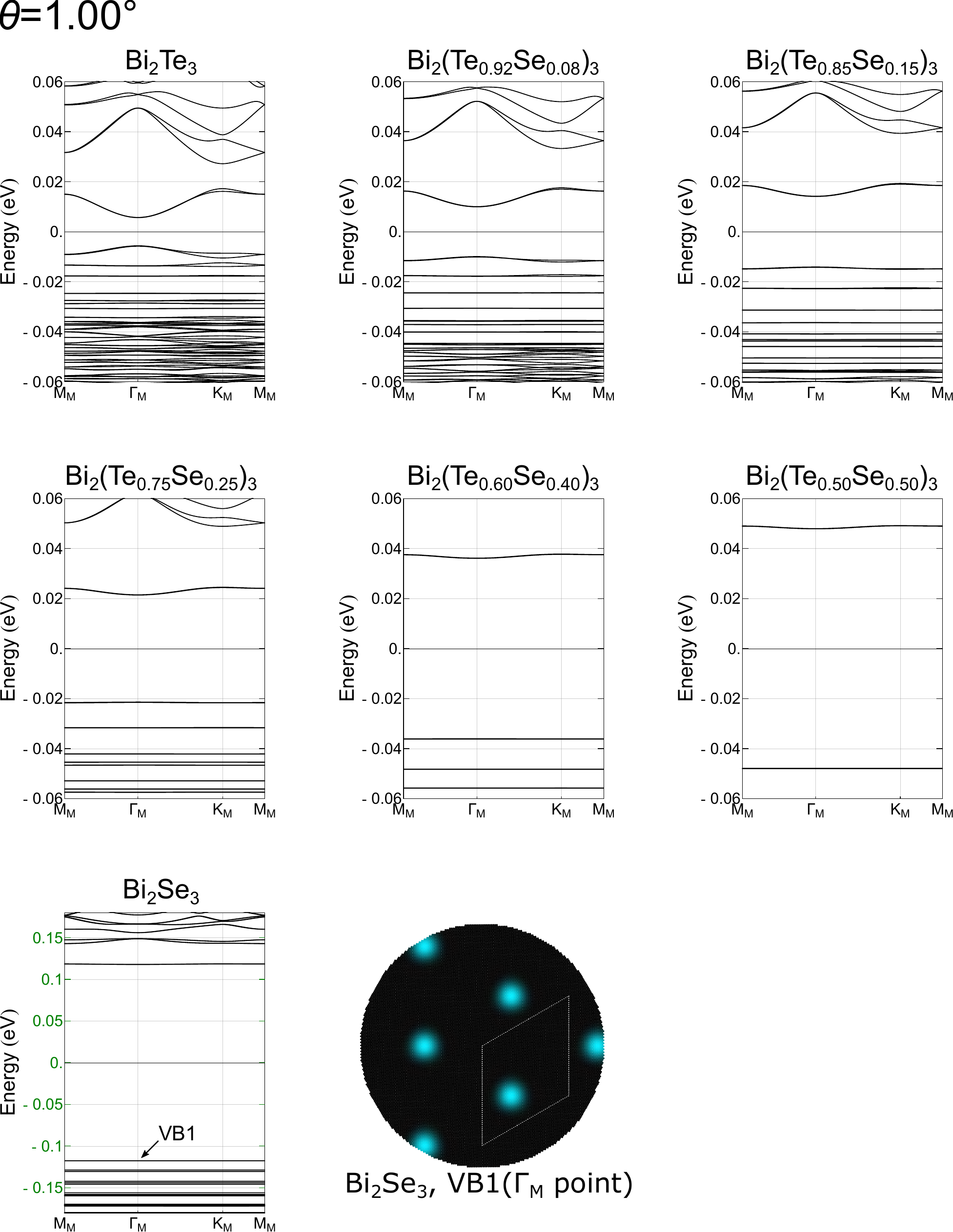}
    \caption{Moir\'e band dispersions of twisted bilayer Bi$_2$(Te$_{1-x}$Se$_x$)$_3$ for various $x$. The twist angle is fixed at $\theta=1.00^\circ$. Note that the Bi$_2$Se$_3$ case is plotted in a different energy range. A wave function of the valence top band at the $\Gamma_M$ point (upper layer, lower Bi atom, $p_z$ orbital, spin-up component) is also shown.}
    \label{fig:SM_Sedope}
\end{figure*}

Figure \ref{fig:SM_Sedope} shows Moir\'e band dispersions of twisted bilayer Bi$_2$(Te$_{1-x}$Se$_x$)$_3$ for various $x$. The twist angle is fixed at $\theta=1.00^\circ$. The values of $x$ are the same as those in Fig.\ref{fig:SM_stackgap}. After the gap-closing at the AB-stacking occurs ($x>0.40$), the edge-state-originated bands disappear and the band gap around the Fermi level get larger. For twisted bilayer Bi$_2$Se$_3$, a wave function of the valence top band at the $\Gamma_M$ point is plotted. The wave function is found to be a bulk-originated state around the AB-stacking area.

\section{Supplemental Information of twisted bilayer Bi$_2$(Te$_{1-x}$Se$_x$)$_3$ \label{sec:SM_supinfo_twist}}
In this section, we give supplemental information of twisted bilayer Bi$_2$(Te$_{1-x}$Se$_x$)$_3$.

\subsection{Angular momentum of edge-state-originated bands \label{sec:SM_supinfo_twist_AM}}

\begin{figure}
    \centering
    \includegraphics[width=0.45\textwidth]{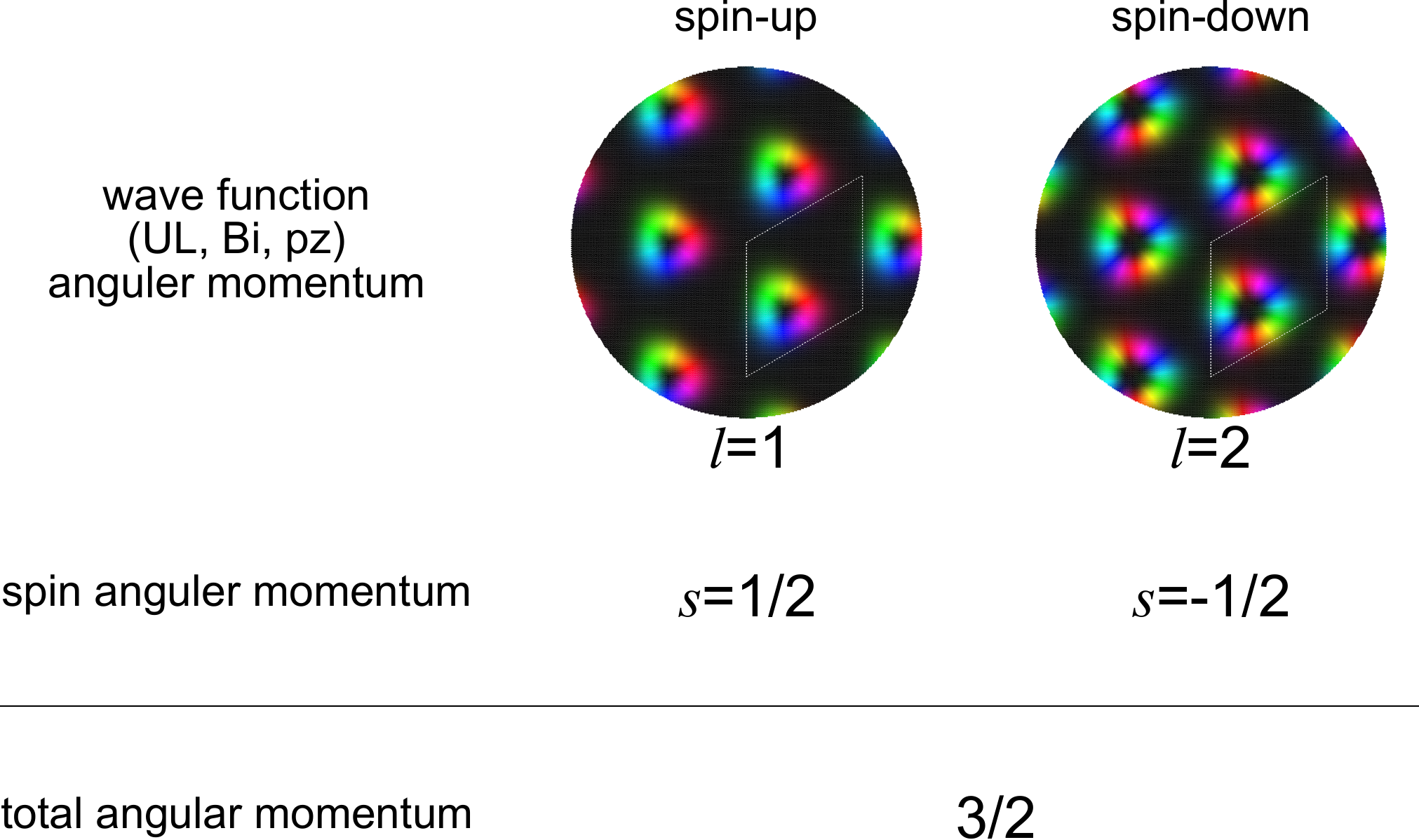}
    \caption{Example of the wave function and angular momentum in the VB2 of twisted bilayer Bi$_2$(Te$_{0.85}$Se$_{0.15}$)$_3$ with $\theta=0.50^\circ$. The wave function plots shows the components of spin-up and spin-down on (upper layer, lower Bi atom, $p_z$ orbital). The total angular momentum of this wave function is calculated as 3/2.}
    \label{fig:SM_totalAM}
\end{figure}

When a wave function is plotted for an edge-state-originated band, it appears to have an integer angular momentum. However, the electronic state is coupled with spin degree-of-freedom and thus the total angular momentum is a half-integer. Actually, when both of spin-up and spin-down components are plotted for a wave function, the phase winding number ($\approx$ orbital angular momentum) is different by 1 (Fig.\ref{fig:SM_totalAM}). By coupling the orbital and spin angular momentum, a total angular momentum (or a rotation eigenvalue of the wave function) is uniquely defined.

\subsection{Twist angle dependence of the bandwidth \label{sec:SM_supinfo_twist_bandwidth}}

\begin{figure}
    \centering
    \includegraphics[width=0.45\textwidth]{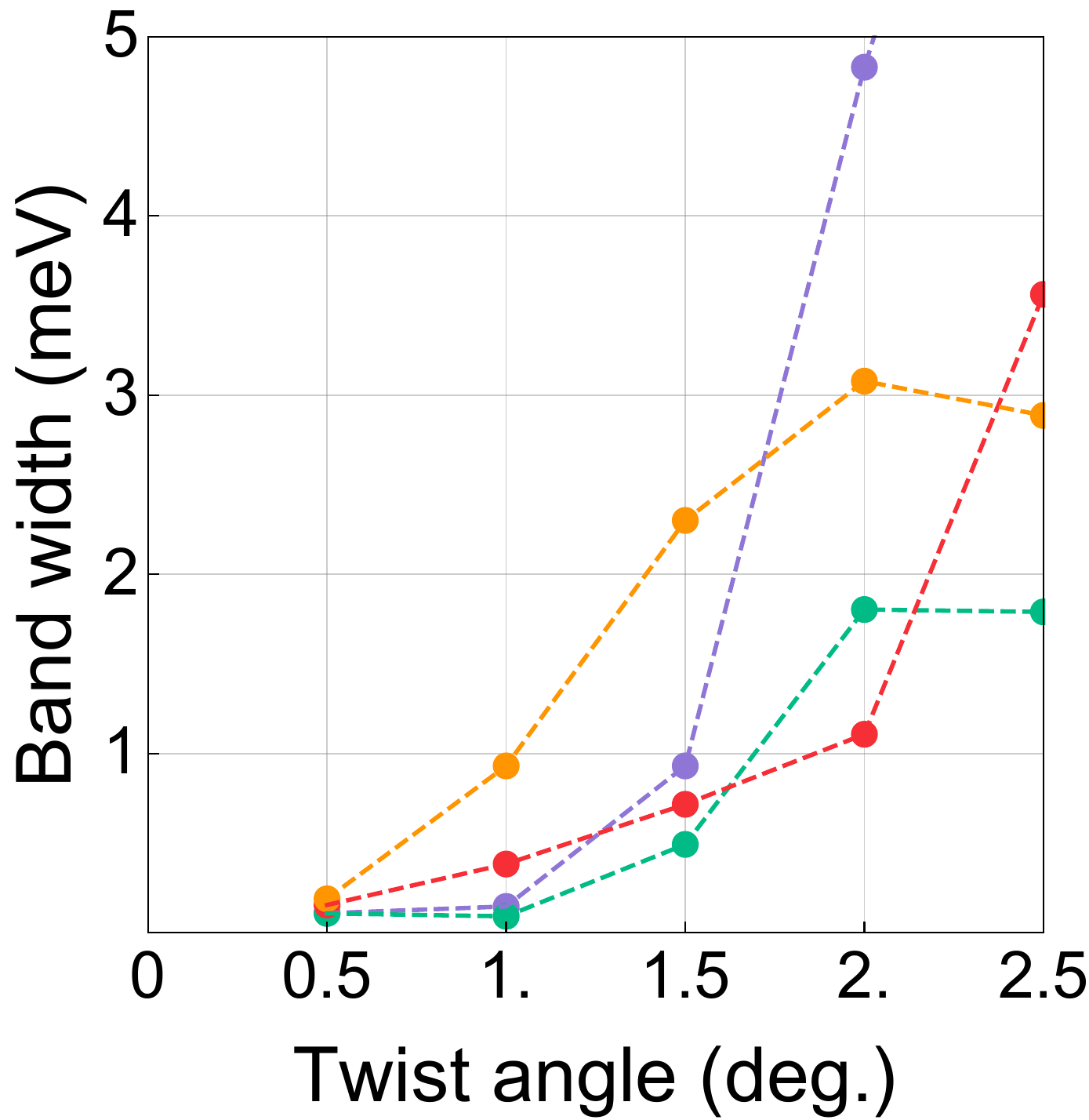}
    \caption{Twist angle dependence of the bandwidth of VB1 (orange), VB2 (red), VB3 (green), and VB4 (purple) in twisted bilayer Bi$_2$(Te$_{0.85}$Se$_{0.15}$)$_3$. The band width is estimated with the band dispersion along the M$_M$-$\Gamma_M$-K$_M$-M$_M$ line.}
    \label{fig:SM_bwangle}
\end{figure}

Figure \ref{fig:SM_bwangle} shows the twist angle dependence of the bandwidth of VB1-4 in twisted bilayer Bi$_2$(Te$_{0.85}$Se$_{0.15}$)$_3$. The bandwidth is estimated as $\mathrm{Max}(E(\bm{k})) - \mathrm{Min}(E(\bm{k})$) ($\bm{k}\in\{$M$_M$-$\Gamma_M$-K$_M$-M$_M$ line$\}$). As a general trend, the bandwidth tends to increase as the twist angle increases. This tendency is explained by the band folding in the Moir\'e BZ, as explained in the main article. More detailed behavior is determined by a combination of multiple factors such as Rashba splitting and hybridization with other bands.

\subsection{Edge dependence of Moir\'e edge state spectra \label{sec:SM_supinfo_twist_LRedge}}

\begin{figure*}
    \centering
    \includegraphics[width=\textwidth]{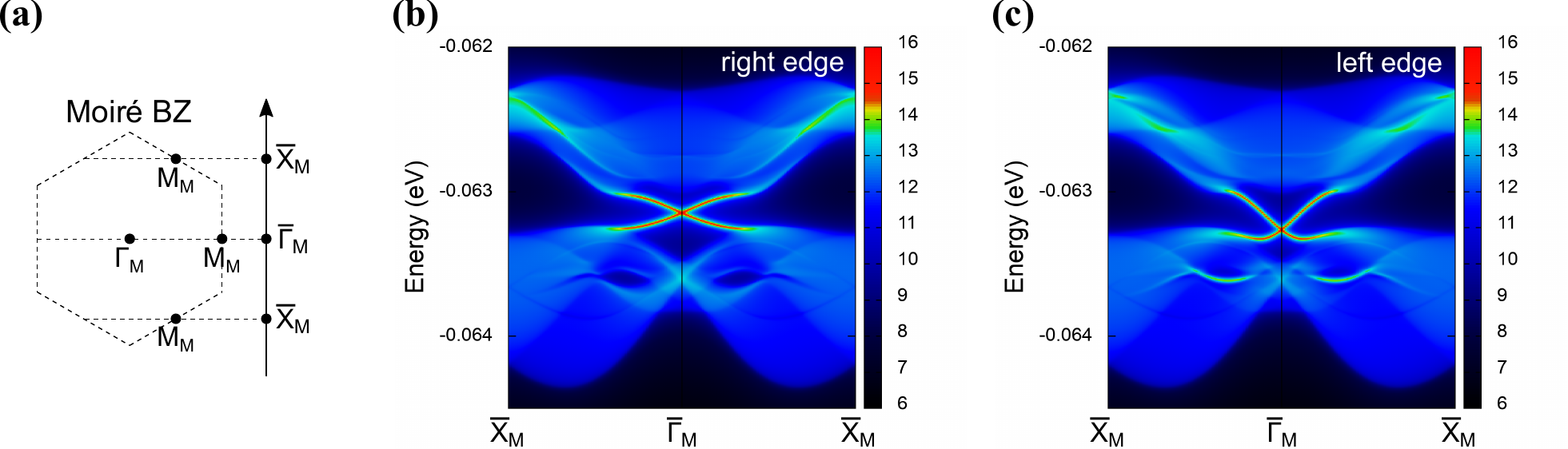}
    \caption{Edge dependence of edge state spectra. (a) The edge BZ used in the edge state calculations. (b),(c), Edge state spectra of the right edge and left edge in twisted bilayer Bi$_2$(Te$_{0.92}$Se$_{0.08}$)$_3$.}
    \label{fig:SM_lredge}
\end{figure*}

Here, we compare the edge state spectra for the left and right edge. Note that the left and right edges are defined by using the Moir\'e unit cell as a unit to build the half-infinite plane, i.e. the the left edge (right edge) of the Moir\'e unit cell appears on the left edge (right edge) of the half-infinite plane. Figure \ref{fig:SM_lredge} shows the edge BZ (a) and edge state spectra of the right edge (b) and left edge (c) in twisted bilayer Bi$_2$(Te$_{0.92}$Se$_{0.08}$)$_3$. The edge state spectra are topologically identical for the both edge, but the energies of the Dirac cones of the edge states are different. It is because there is no symmetry to guarantee the equivalence of their energies in twisted bilayer Bi$_2$(Te$_{0.92}$Se$_{0.08}$)$_3$. Although the Moir\'e superlattice of twisted bilayer Bi$_2$(Te$_{1-x}$Se$_x$)$_3$ has in-plane $C_2$ rotation symmetries along the $\Gamma_M$-M$_M$ lines, they do not exchange the two edges. Therefore, each edge state does not necessarily have the same spectrum. This edge dependence is not Moir\'e-specific but generally found in a noncentrosymmetric 2D topological insulator. See Sec.\ref{sec:SM_noncentroTI} for general and more detailed discussion. Note that in more realistic conditions, the energy difference between the two edge states may be suppressed by the charge neutral condition.

\subsection{Topological phase transition in Moir\'e bands \label{sec:SM_supinfo_twist_tpt}}

\begin{figure*}
    \centering
    \includegraphics[width=0.8\textwidth]{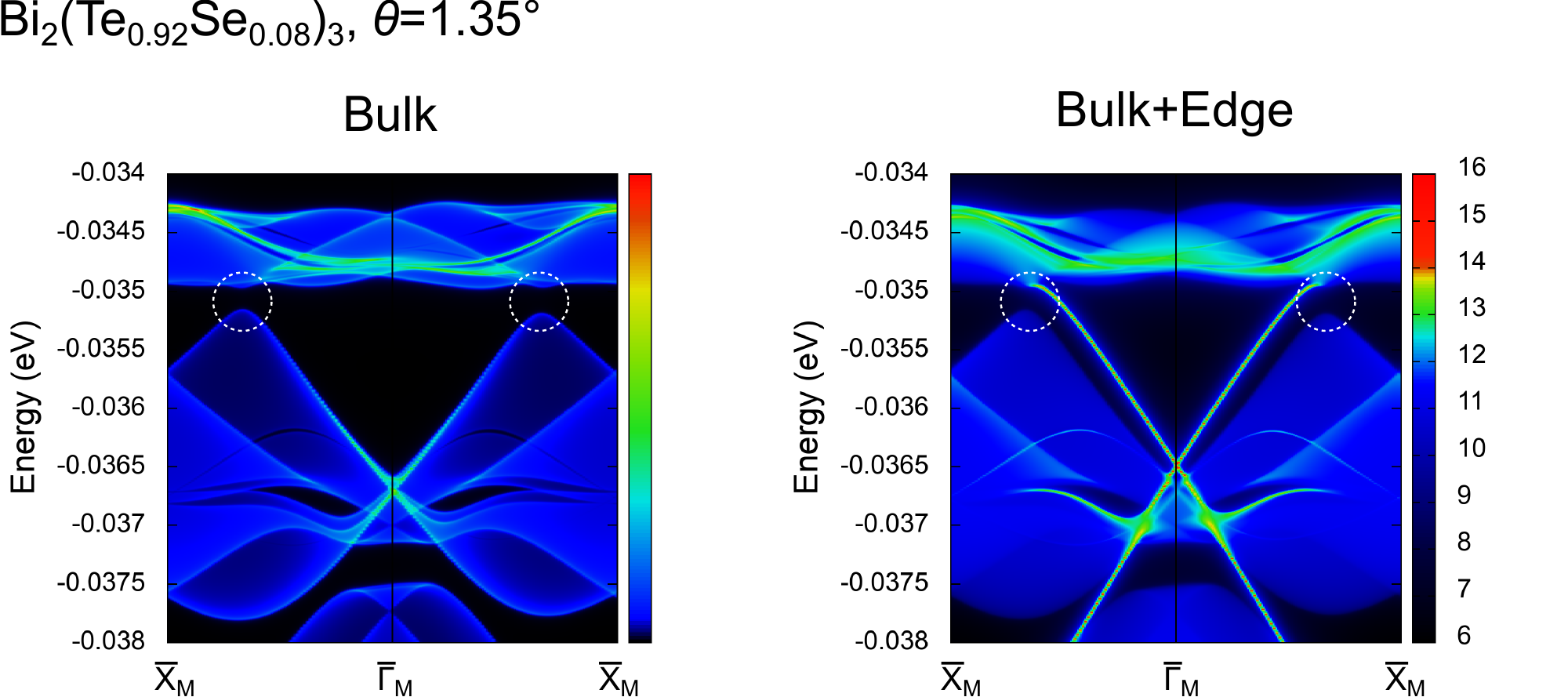}
    \caption{Bulk and edge band spectra of twisted bilayer Bi$_2$(Te$_{0.92}$Se$_{0.08}$)$_3$ with $\theta=1.35^\circ$. The gap-closing of the bulk band spectra occurs at general momenta (white dashed circles).}
    \label{fig:SM_topotrans}
\end{figure*}

Here, we show how a gap-closing and a topological phase transition occur in Moir\'e bands.
As the twist angle get smaller, more edge-state-originated nearly flat bands appear around the Fermi level due to the band folding with the small Moir\'e BZ. For example in the case of twisted bilayer Bi$_2$(Te$_{0.92}$Se$_{0.08}$)$_3$ with $\theta \sim 1.55^\circ$, topological phase transitions occur on VB8 and VB9 and eventually they become topologically trivial bands as the twist angle is decreased. Generally for the topological phase transition in twisted bilayer Bi$_2$(Te$_{1-x}$Se$_x$)$_3$, the gap-closing occurs on the $\Gamma_M$-K$_M$ lines due to the in-plane $C_2$ rotation symmetry \cite{yu2020piezoelectricity}. Note that the $\Gamma_M$-K$_M$ lines are not high-symmetry lines, although the position of the gap-closing is determined by the symmetry restriction. When edges are made so that there is no symmetry between the two edges, the gap-closing is also seen in the 1D edge BZ as a gap-closing of bulk-band spectra at general momenta. Figure \ref{fig:SM_topotrans} shows a bulk band spectra and an edge band spectra of twisted bilayer Bi$_2$(Te$_{0.92}$Se$_{0.08}$)$_3$ with the twist angle $\theta=1.35^\circ$, which is close to a phase transition point. Gap-closing points are indicated with white dashed circles. Note that the edge states appear from the points where the gap-closing will occur and make a Dirac-cone that is close to the bulk-band crossing of a Kramers pair at the $\bar{\Gamma}_M$ point.
The topological phase transition that occurs as the twist angle is decreased is not simple as a pair annihilation between VB8 and VB9. As the twist angle is decreased, the gap between edge-state-originated bands gets smaller and thus they shift relatively upward, while the bulk-state-originated band stays. Therefore, the nontrivial structure moves to the next and next band as the twist angle is decreased. Actually, another gap-closing occurs between VB9 and VB10 and there is midgap edge states in the lower energy region VB9 in Fig.\ref{fig:SM_topotrans}.

\section{Noncentrosymmetric 2D topological phase transition \label{sec:SM_noncentroTI}}
A Moir\'e system does not have the inversion symmetry and we have found that a topological phase transition occurs in twisted Bi$_2$(Te$_{1-x}$Se$_x$)$_3$.
In this section, we explain how a topological phase transition occurs in a general noncentrosymmetric 2D system. We consider a topological phase transition between a topological insulator (TI) phase and a normal insulator (NI) phase in time-reversal ${\cal T}$ symmetric systems. Generally, when a topological phase transition occurs, a gap-closing must occurs in the bulk band dispersion \cite{PhysRevB.76.205304,murakami2007phase}.
First, we briefly review the case of a centrosymmetric TI. In a phase transition occurs from a NI to a TI in a centrosymmetric system, the bulk gap-closing occurs at one of (or an odd number of) the TRIM. When a surface band spectrum is drawn, the gap-closing is seen as a gap-closing of the bulk continuous spectra at one of the TRIM of the surface BZ, because TRIM of 3D BZ are projected on TRIM of the surface BZ. In the TI phase, a corresponding surface state emerges around the TRIM in which the gap-closing occurred.
In noncentrosymmetric systems, the bulk gap-closing occurs at a generic $\bm{k}$ point. In 3D case, the gapless phase has finite width in the parameter space, which is Weyl semimetal (WSM) phase. In 2D case, the gap-closing occurs only on the critical point in the parameter space. Because the bulk gap-closing occurs at a generic $\bm{k}$ point, a gap-closing in the surface band spectrum should also be seen at a generic $\bm{k}$ in the surface BZ. Ref.\cite{yu2020piezoelectricity} clarified that, in noncentrosymmetric 2D systems, $\bm{k}$ points where the gap-closing can occur are restricted by crystalline symmetries. However, it has not been clearly described how a corresponding surface state emerges from the gap-closing point in the surface band spectrum. Therefore, here we demonstrate the emergence of the surface state in topological phase transition in noncentrosymmetric 2D systems, and explain the relation between the surface state and the crystalline symmetries.

As explained above, in 3D system, a topological phase transition in a noncentrosymmetric system is described as a NI-WSM-TI transition. Therefore, we start from a 3D model that describe a NI-WSM-TI transition, and by focusing a particular 2D momentum plane in the model we discuss the bulk and surface band spectra in a noncentrosymmetric 2D system.
We use the inversion broken TI/NI multilayer model given in Refs.\cite{PhysRevLett.107.127205} and \cite{PhysRevB.85.035103}, and it is a four by four model written as
\begin{equation}
    \begin{split}
        H = & v \tau_z \left( k_y \sigma_x - k_x \sigma_y \right) + V \tau_z \\
        & + \Delta_N(\bm{k}) \tau_x + \Delta_T(\bm{k}) \left( e^{ik_z} \tau_+ + e^{-ik_z} \tau_- \right) .
    \end{split}
\end{equation}
The four components consist of surface states on the top surface of the TI layer with spin up and down ($t_{\uparrow \bm{k}} , t_{\downarrow \bm{k}}$), and those on the bottom surface ($b_{\uparrow \bm{k}} , b_{\downarrow \bm{k}}$). The $\sigma_{x,y,z}$ and $\tau_{x,y,z}$ are Pauli matrices for spin and top/bottom surfaces, respectively, where $\tau_{\pm} = \left( \tau_x \pm i \tau_y \right) / \sqrt{2} $. We assume a two-fold rotation symmetry about the $z$-axis ($C_{2z}$), and thus the $\Delta_N(\bm{k})$ and $\Delta_T(\bm{k})$ are written as
\begin{equation}
    \begin{split}
        \Delta_T \left( \bm{k} \right) &= \Delta_T^0 + \delta_T^x k_x^2 + \delta_T^y k_y^2 , \\
        \Delta_N \left( \bm{k} \right) &= \Delta_N^0 + \delta_N^x k_x^2 + \delta_N^y k_y^2 . \\
    \end{split}
\end{equation}

\begin{figure*}
    \centering
    \includegraphics[width=0.9\textwidth]{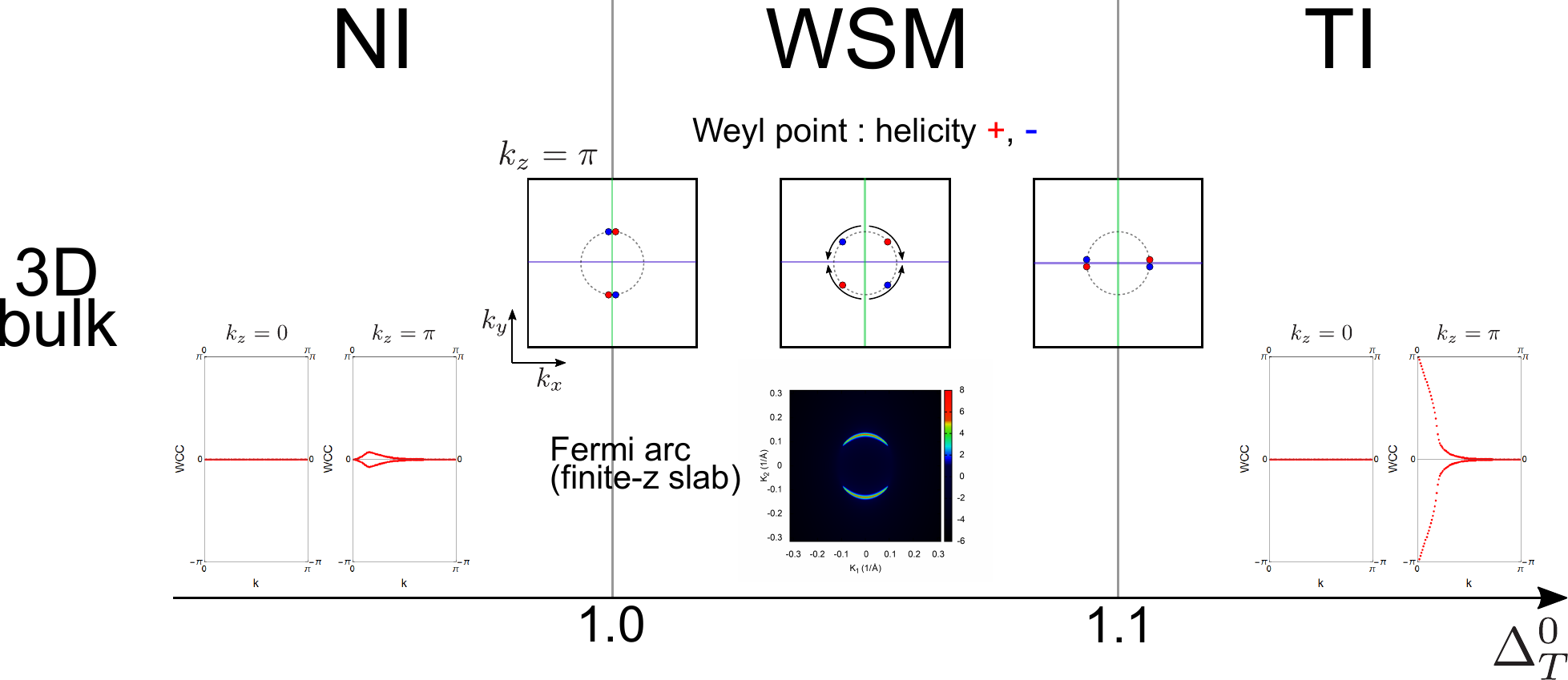}
    \caption{Phase diagram of 3D WSM model (\ref{eq:SM_3DWeylmodel}). A WSM phase appears between $\Delta_T^0=1.0$ and $\Delta_T^0=1.1$ as a gapless phase between NI and TI. At $\Delta_T^0=1.0$, Weyl points are created on the $(k_x=0,k_z=\pi)$ line. As $\Delta_T^0$ increases, the Weyl points move in the $k_z=\pi$ plane and annihilate on the $(k_y=0,k_z=\pi)$ line at $\Delta_T^0=1.1$.}
    \label{fig:SM_WSMphase}
\end{figure*}

To draw the surface band spectrum, we transform the model into an equivalent lattice periodic model. The lattice periodic model is
\begin{equation}
    \begin{split}
        H = & v \tau_z \left( \sin k_y \sigma_x - \sin k_x \sigma_y \right) + V \tau_z \\
        & + \Delta_T(\bm{k}) \tau_x + \Delta_N(\bm{k}) \left( e^{ik_z} \tau_+ + e^{-ik_z} \tau_- \right),
    \end{split}
    \label{eq:SM_3DWeylmodel}
\end{equation}
where $\Delta_N(\bm{k})$ and $\Delta_T(\bm{k})$ are written as
\begin{equation}
    \begin{split}
        \Delta_T \left( \bm{k} \right) &= \Delta_T^0 + 2 \delta_T^x (1-\cos k_x) + 2 \delta_T^y (1-\cos k_y) , \\
        \Delta_N \left( \bm{k} \right) &= \Delta_N^0 + 2 \delta_N^x (1-\cos k_x) + 2 \delta_N^y (1-\cos k_y) . \\
    \end{split}
\end{equation}
By taking $\Delta_T^0$ as a tuning parameter, this model shows the NI-WSM-TI transition.
We fix the other parameters as
\begin{equation}
    \begin{split}
        & V = 0.2 ~,~ v = 0.4 ~,~ \delta_N^x = 0.60 ~,~ \delta_T^x = 0.20 , \\
        & \delta_N^y = \delta_N^x \left( 1 + \frac{0.1}{\delta_T^x-\delta_N^x} \frac{1/2}{1-\sqrt{1-V^2/v^2}} \right), \\
        & \delta_T^y = \delta_T^x \left( 1 + \frac{0.1}{\delta_T^x-\delta_N^x} \frac{1/2}{1-\sqrt{1-V^2/v^2}} \right), \\
        & \Delta_N^0 = 1.0 + 2 \left( \delta_T^y -\delta_N^y \right) \left( 1 - \sqrt{1-V^2/v^2} \right) .
    \end{split}
\end{equation}
Here, $\delta^y_N$, $\delta^y_T$, and $\Delta^0_N$ are determined to fix the transition points at $\Delta^0_T=1.0$ and $1.1$.
When $\Delta_T^0<1.0$, the model is a normal insulator (Fig.\ref{fig:SM_WSMphase}). When $\Delta_T^0=1.0$, Weyl points are created as pairs on the $(k_z=\pi,k_x=0)$ line (Fig.\ref{fig:SM_WSMphase}). Due to the time-reversal symmetry, a pair of Weyl points are created each of positive and negative $k_y$ region and thus four Weyl points are created in total. When $1.0<\Delta_T^0<1.1$, the Weyl points move on the $k_z=\pi$ plane, and when $\Delta_T^0=1.1$, they annihilate on the $(k_z=\pi,k_y=0)$ line. When $\Delta_T^0>1.1$, the model is a strong topological insulator.

By focusing on a particular 2D momentum space in the 3D BZ of this model, we can discuss a topological phase transition and surface states in a noncentrosymmetric 2D system. To consider a time-reversal symmetric system, the focused 2D plane must be a time-reversal invariant plane ($\forall \bm{k} \in\left\{ \mathrm{2D~plane} \right\},-\bm{k} \in\left\{ \mathrm{2D~plane} \right\}$). Here we focus on two planes, $k_x=0$ and $k_y=0$. To draw the surface band spectrum with Green's function method, we transform the lattice periodic model into a real space representation as
\begin{equation}
    \begin{split}
        H  = \sum_{\bm{r}} & \left[ i \frac{v}{2} \left( -t_{\uparrow \bm{r}}^\dagger t_{\downarrow \bm{r}+\bm{y}}^{\phantom{\dagger}} +t_{\uparrow \bm{r}}^\dagger t_{\downarrow \bm{r}-\bm{y}}^{\phantom{\dagger}} +b_{\uparrow \bm{r}}^\dagger b_{\downarrow \bm{r}+\bm{y}}^{\phantom{\dagger}} -b_{\uparrow \bm{r}}^\dagger b_{\downarrow \bm{r}-\bm{y}}^{\phantom{\dagger}} \right) \right. \\
        &~ + \frac{v}{2} \left( t_{\uparrow \bm{r}}^\dagger t_{\downarrow \bm{r}+\bm{x}}^{\phantom{\dagger}} -t_{\uparrow \bm{r}}^\dagger t_{\downarrow \bm{r}-\bm{x}}^{\phantom{\dagger}} -b_{\uparrow \bm{r}}^\dagger b_{\downarrow \bm{r}+\bm{x}}^{\phantom{\dagger}} +b_{\uparrow \bm{r}}^\dagger b_{\downarrow \bm{r}-\bm{x}}^{\phantom{\dagger}} \right) \\
        &~ + V \left( t_{\uparrow \bm{r}}^\dagger t_{\uparrow \bm{r}}^{\phantom{\dagger}} + t_{\downarrow \bm{r}}^\dagger t_{\downarrow \bm{r}}^{\phantom{\dagger}} - b_{\uparrow \bm{r}}^\dagger b_{\uparrow \bm{r}}^{\phantom{\dagger}} - b_{\downarrow \bm{r}}^\dagger b_{\downarrow \bm{r}}^{\phantom{\dagger}} \right) \\
        &~ + \left( \Delta_N^0 + 2\delta_N^x + 2\delta_N^y \right) \left( t_{\uparrow \bm{r}}^\dagger b_{\uparrow \bm{r}}^{\phantom{\dagger}} + t_{\downarrow \bm{r}}^\dagger b_{\downarrow \bm{r}}^{\phantom{\dagger}} \right) \\
        &~ - \delta_N^x \left( t_{\uparrow \bm{r}}^\dagger b_{\uparrow \bm{r}+\bm{x}}^{\phantom{\dagger}} + t_{\downarrow \bm{r}}^\dagger b_{\downarrow \bm{r}+\bm{x}}^{\phantom{\dagger}} + t_{\uparrow \bm{r}}^\dagger b_{\uparrow \bm{r}-\bm{x}}^{\phantom{\dagger}} + t_{\downarrow \bm{r}}^\dagger b_{\downarrow \bm{r}-\bm{x}}^{\phantom{\dagger}} \right) \\
        &~ - \delta_N^y \left( t_{\uparrow \bm{r}}^\dagger b_{\uparrow \bm{r}+\bm{y}}^{\phantom{\dagger}} + t_{\downarrow \bm{r}}^\dagger b_{\downarrow \bm{r}+\bm{y}}^{\phantom{\dagger}} + t_{\uparrow \bm{r}}^\dagger b_{\uparrow \bm{r}-\bm{y}}^{\phantom{\dagger}} + t_{\downarrow \bm{r}}^\dagger b_{\downarrow \bm{r}-\bm{y}}^{\phantom{\dagger}} \right) \\
        &~ + \left( \Delta_T^0 + 2\delta_T^x + 2\delta_T^y \right) \left( t_{\uparrow \bm{r}}^\dagger b_{\uparrow \bm{r}+\bm{z}}^{\phantom{\dagger}} + t_{\downarrow \bm{r}}^\dagger b_{\downarrow \bm{r}+\bm{z}}^{\phantom{\dagger}} \right) \\
        &~ - \delta_T^x \left( t_{\uparrow \bm{r}}^\dagger b_{\uparrow \bm{r}+\bm{x}+\bm{z}}^{\phantom{\dagger}} + t_{\downarrow \bm{r}}^\dagger b_{\downarrow \bm{r}+\bm{x}+\bm{z}}^{\phantom{\dagger}} \right. \\
        &~~~~~~~ \left. + t_{\uparrow \bm{r}}^\dagger b_{\uparrow \bm{r}-\bm{x}+\bm{z}}^{\phantom{\dagger}} + t_{\downarrow \bm{r}}^\dagger b_{\downarrow \bm{r}-\bm{x}+\bm{z}}^{\phantom{\dagger}} \right) \\
        &~ - \delta_T^y \left( t_{\uparrow \bm{r}}^\dagger b_{\uparrow \bm{r}+\bm{y}+\bm{z}}^{\phantom{\dagger}} + t_{\downarrow \bm{r}}^\dagger b_{\downarrow \bm{r}+\bm{y}+\bm{z}}^{\phantom{\dagger}} \right. \\
        &~~~~~~~ \left. \left. + t_{\uparrow \bm{r}}^\dagger b_{\uparrow \bm{r}-\bm{y}+\bm{z}}^{\phantom{\dagger}} + t_{\downarrow \bm{r}}^\dagger b_{\downarrow \bm{r}-\bm{y}+\bm{z}}^{\phantom{\dagger}} \right) \right] \\
        &~ + \mathrm{h.c.} ,
    \end{split}
\end{equation}
where $\bm{r}$ is the unit cell index and a cubic cell with a lattice constant 1 is assumed.

\begin{figure*}
    \centering
    \includegraphics[width=0.9\textwidth]{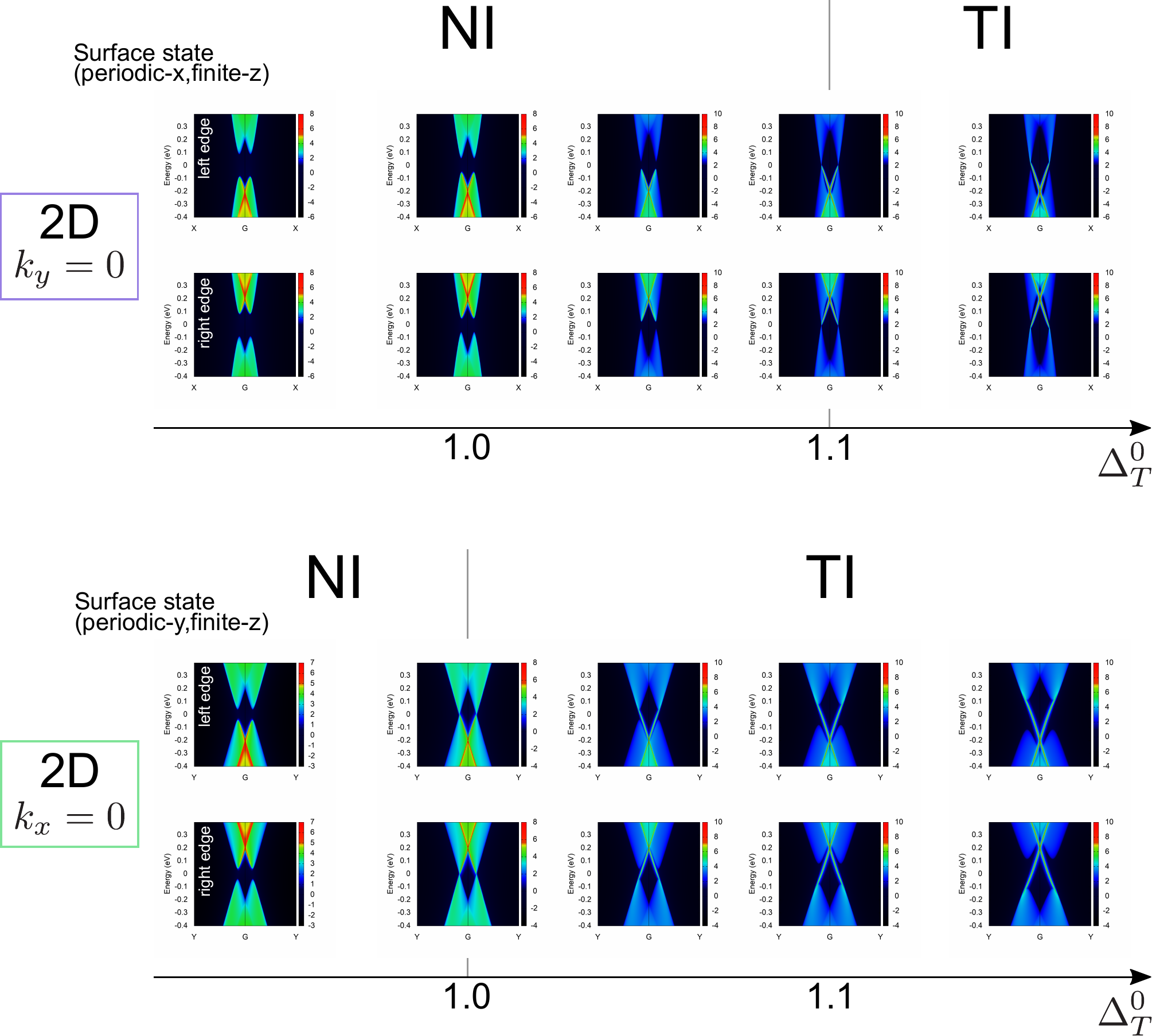}
    \caption{Phase diagrams of noncentrosymmetric 2D systems and surface band spectra obtained by focusing on the $k_y=0$ plane (top) and $k_x=0$ plane (bottom) of the 3D BZ of the WSM (\ref{eq:SM_3DWeylmodel}). In each phase diagram, the top and bottom density plots are the surface band spectra of the left and right edges, respectively.}
    \label{fig:SM_2DTIphase}
\end{figure*}

We obtain surface band spectrum with the real space lattice periodic Hamiltonian as shown in Fig.\ref{fig:SM_2DTIphase}. First, we focus on the $k_y=0$ plane. The $y$ direction is removed by fixing $k_y$, and the surface band spectrum is calculated in a ribbon with a periodic $x$ direction and a finite $z$ direction. The bulk gap closing occurs at $\Delta_T^0=1.1$, which is the point where the Weyl points in 3D BZ touch the $k_y=0$ plane. This is a topological phase transition point between 2D NI and TI, and when $\Delta_T^0>1.1$, we can see topological edge states that connect the valence and conduction band spectra. Next we focus on the $k_x=0$ plane. In this case, we consider a ribbon with a periodic $y$ direction and a finite $z$ direction. Also in this case, a topological phase transition occurs when the Weyl points in 3D BZ touch the $k_x=0$ plane, $\Delta_T^0>1.0$. These results correspond to looking a particular $\bm{k}$ slice of the surface state of the 3D WSM.
It should be noted that the energies of the surface states are different in the left and right edges in both cases. The emergence of the surface state with edge dependent energies is consistent with two natural requirement in properties of surface states: (i) edge state should emerges from a gap-closing point of the bulk band spectrum, (ii) a Dirac cone of a time-reversal protected surface state should be located on a TRIM. At the same time, this edge dependence indicates asymmetry of the left and right edges. If the gap-closing of the bulk band spectrum occurs on a generic point in the surface BZ, the bulk Hamiltonian is not allowed to have a symmetry that exchange the left and right edges. Consequently, the position in 2D BZ where a gap-closing point can appear is restricted by crystalline symmetries of the bulk Hamiltonian, which is consistent with Ref.\cite{yu2020piezoelectricity}. 

In conclusion of this section, we have shown the edge dependence of the emergence of the surface states and its relation with symmetries of the bulk Hamiltonian, and demonstrated it with an easy model.

\section{Twisted BHZ model \label{sec:SM_tBHZ}}
In this section, we propose a twisted Bernevig-Hughes-Zhang (BHZ) model as a simple model to describe the essential behavior of twisted bilayer Bi$_2$(Te$_{1-x}$Se$_x$)$_3$.

\begin{figure*}
    \centering
    \includegraphics[width=\textwidth]{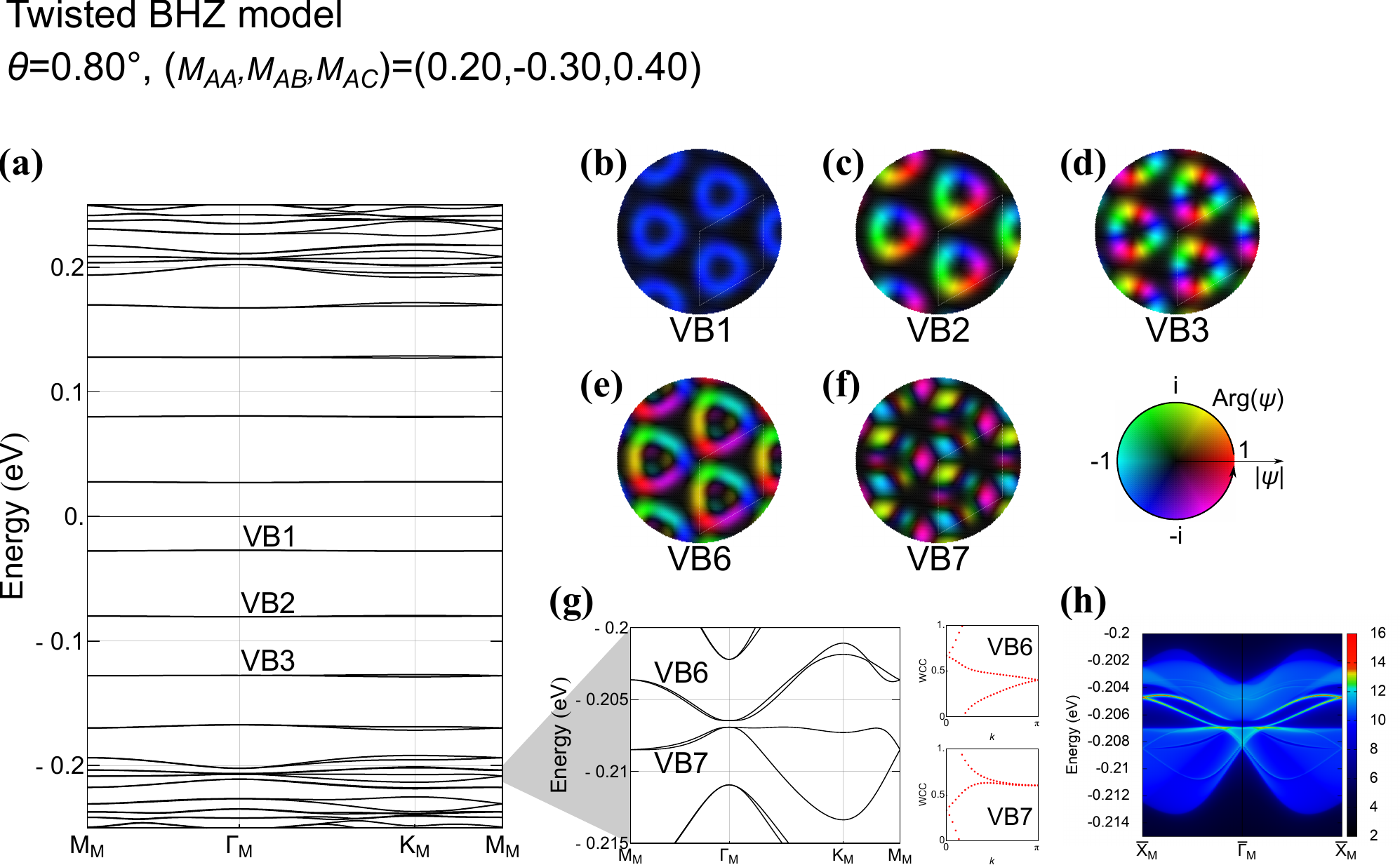}
    \caption{Example of the twisted BHZ model ($\theta=0.80^\circ,A=1,B=0.1$). The stacking dependent $M$ is set to $(M_{AA},M_{AB},M_{AC})=(0.2,-0.3,0.4)$. (a) Moir\'e band dispersion. (b)-(f) real space plot of the wave functions (Upper layer, spin-up) at the $\Gamma$ point for VB1, VB2, VB3, VB6, and VB7. (g) Magnified view of the VB6 and VB7 and Wilson loop spectra of them. (h) Moir\'e edge state spectrum.}
    \label{fig:SM_tBHZ}
\end{figure*}

The BHZ model is a well-known effective model of a 2D topological insulator, i.e. a quantum spin Hall insulator. The BHZ model is given with two orbitals and two spin, and the Hamiltonian is typically written as
\begin{equation}
    \begin{split}
        & H_{\mathrm{BHZ}}(\bm{k}) \\
        &= \left( M - B k^2 \right) \tau_z \sigma_0 + A k_x \tau_x \sigma_z + A k_y \tau_y \sigma_0 \\
        &= \left(
        \begin{array}{cccc}
            (M-Bk^2) & A k^- & & \\
            A k^- & -(M-Bk^2) & & \\
             & & (M-Bk^2) & -A k^+ \\
             & & -A k^- & -(M-Bk^2)
        \end{array}
        \right) ,
    \end{split}
\end{equation}
where $k^\pm = k_x \pm i k_y$, $k^2 = k_x^2+k_y^2$, and the basis is ordered as $\ket{1,\uparrow}$, $\ket{2,\downarrow}$, $\ket{1,\downarrow}$, $\ket{2,\uparrow}$. This model has the time-reversal symmetry ${\cal T}$ and a rotation symmetry around z-axis $C_\theta$ (perpendicular to the 2D system), and the operators are given as
\begin{equation}
    {\cal T} = i \sigma_y {\cal K} ~,~
    C_\theta = \left(
    \begin{array}{cccc}
         e^{i\frac{\theta}{2}} & & & \\
         & e^{-i\frac{\theta}{2}} & & \\
         & & e^{-i\frac{\theta}{2}} & \\
         & & & e^{i\frac{\theta}{2}} \\
    \end{array}
    \right)
\end{equation}

% inversion operator
%\begin{equation}
%    I = \tau_z \sigma_0 = \left(
%    \begin{array}{cccc}
%        1 & & & \\
%         & -1 & & \\
%         & & 1 & \\
%         & & & -1
%    \end{array}
%    \right)
%\end{equation}

This model describes a topological insulator when $MB>0$ and a normal insulator when $BM<0$.

By introducing a Moir\'e scale oscillation in $M$, we obtain a twisted BHZ model. To make the model realistic, we transform the basis. We assume that one of the two orbitals is located on the upper layer, and the other on the lower layer, and spin is located on each orbital. This assumption restricts the time-reversal and the rotation operators to be block-diagonal with intralayer blocks. We also assume that the Moir\'e oscillating term $M$ should be an interlayer element. To satisfy these assumptions, we transform the Hamiltonian with a unitary matrix 
\begin{equation}
    \begin{split}
        U &= \frac{1}{2\sqrt{2}} \left[ (\tau_0 + i \tau_y) (\sigma_0 -i \sigma_x) + (\tau_0 -i \tau_y) (\sigma_y + \sigma_z) \right] \\
        &= \frac{1}{\sqrt{2}} \left(
        \begin{array}{cccc}
            1 & 0 & -i & 0 \\
            0 & 1 & 0 & -i \\
            0 & -i & 0 & 1 \\
            i & 0 & -1 & 0 \\
        \end{array}
        \right),
    \end{split}
\end{equation}
% U dagger
%\begin{equation}
%    \begin{split}
%        U^\dagger &= \frac{1}{2\sqrt{2}} \left[ (\tau_0 - i \tau_y) (\sigma_0 +i \sigma_x) + (\tau_0 +i \tau_y) (\sigma_y + \sigma_z) \right] \\
%        &= \frac{1}{\sqrt{2}} \left(
%        \begin{array}{cccc}
%            1 & 0 & 0 & -i \\
%            0 & 1 & i & 0 \\
%            i & 0 & 0 & -1 \\
%            0 & i & 1 & 0 \\
%        \end{array}
%        \right)
%    \end{split}
%\end{equation}
and obtain a new representation of Hamiltonian
\footnotesize
\begin{equation}
    \begin{split}
        & \tilde{H}_{\mathrm{BHZ}}(\bm{k}) \\
        &= U^\dagger H_{\mathrm{BHZ}}(\bm{k}) U \\
        &= \left( M - B k^2 \right) \tau_z \sigma_y + A k_x \tau_x \sigma_0 + A k_y \tau_y \sigma_0 \\
        &= \left(
        \begin{array}{cccc}
             & A k^- & -i(M-Bk^2) & \\
            A k^+ & & & i(M-Bk^2) \\
            i(M-Bk^2) & & & A k^- \\
             & -i(M-Bk^2) & A k^+ & \\
        \end{array}
        \right) ,
    \end{split}
\end{equation}
\normalsize
where the new basis is written as $\ket{1,\uparrow}-i\ket{2,\uparrow}$, $\ket{2,\downarrow}+i\ket{1,\downarrow}$, $i\ket{1,\uparrow}-\ket{2,\uparrow}$, $i\ket{2,\downarrow}+\ket{1,\downarrow}$. In this representation, the time-reversal and the rotation operators are written as
\begin{equation}
    \tilde{{\cal T}} = -\tau_y \sigma_0 {\cal K} ~,~
    \tilde{C}_\theta = \left(
    \begin{array}{cccc}
         e^{i\frac{\theta}{2}} & & & \\
         & e^{-i\frac{\theta}{2}} & & \\
         & & e^{i\frac{\theta}{2}} & \\
         & & & e^{-i\frac{\theta}{2}} \\
    \end{array}
    \right) .
\end{equation}
%\begin{equation}
%    \tilde{I} = \tau_z \sigma_y = \left(
%    \begin{array}{cccc}
%         & & -i & \\
%         & & & i \\
%         i & & & \\
%         & -i & & 
%    \end{array}
%    \right)
%\end{equation}
%Inversion eigensystem,
%\begin{equation}
%    \begin{split}
%        \xi=1 ~&:~ \psi=\frac{1}{\sqrt{2}} \left( \begin{array}{c} 1 \\ 0 \\ i \\ 0 \end{array} \right) , \frac{1}{\sqrt{2}} \left( \begin{array}{c} 0 \\ 1 \\ 0 \\ -i \end{array} \right) \\
%        \xi=-1 ~&:~ \psi=\frac{1}{\sqrt{2}} \left( \begin{array}{c} 1 \\ 0 \\ -i \\ 0 \end{array} \right) , \frac{1}{\sqrt{2}} \left( \begin{array}{c} 0 \\ 1 \\ 0 \\ i \end{array} \right) 
%    \end{split}
%\end{equation}

By assigning $\tilde{H}_{\mathrm{BHZ}}(\bm{k})$ to $h_{\bm{r}}^{\alpha\sigma,\beta'\sigma'}(\bm{k})$ in Eq. (\ref{eq.Model.tFT2}), we obtain a Hamiltonian of the twisted BHZ model. By tuning $M$ for three sampling points, we can design a Moir\'e superlattice system with topological and normal insulator domains.

Note that the basis ordering makes a physical difference in a strict sense due to the momentum twisting in Eqs. (\ref{eq.Model.Uk})(\ref{eq.Model.Lk})(\ref{eq.Model.Tk}), although it makes no physical difference in an untwisted case. However, the difference is negligible in the small twist angle limit when we consider an effective model around the $\Gamma$ point in the atomic BZ. We calculate with $\tilde{H}_{\mathrm{BHZ}}(\bm{k})$, but it is confirmed that almost same results can be obtained by calculating with $H_{\mathrm{BHZ}}(\bm{k})$.

Figure \ref{fig:SM_tBHZ} shows an example of the twisted BHZ model with $\theta=0.80^\circ$, $A=1$, $B=0.1$, and stacking-dependent $M$ that is set to $(M_{AA},M_{AB},M_{AC})=(0.2,-0.3,0.4)$. Because $MB<0$ only in the AB-stacking area, the topological insulator domain should appear around the AB-stacking area, as is in twisted bilayer Bi$_2$(Te$_{1-x}$Se$_x$)$_3$. It can be seen that the expected ring-shape edge-state-originated states VB1-3 and their angular momentum ordering are obtained as shown in Fig.\ref{fig:SM_tBHZ}(b)-(d). There are also Moir\'e topological bands VB6 and VB7 (Fig.\ref{fig:SM_tBHZ}(g)), and a corresponding Moir\'e-scale helical edge state is obtained as shown in Fig.\ref{fig:SM_tBHZ}(h). Because the wave functions of the VB6 (Fig.\ref{fig:SM_tBHZ}(e)) and VB7 (Fig.\ref{fig:SM_tBHZ}(f)) are a edge-state-originated and a bulk-state-originated, respectively, the obtained Moir\'e-scale helical edge state is a ``edge state from edge state".
These results indicate that the twisted BHZ model is a good simple model that reproduces the properties of twisted bilayer Bi$_2$(Te$_{1-x}$Se$_x$)$_3$. Furthermore, it is also shown that the ``edge state from edge state" is not specific for twisted bilayer Bi$_2$(Te$_{1-x}$Se$_x$)$_3$, but is a general phenomenon in similar Moir\'e systems.

\bibliography{reference}% Produces the bibliography via BibTeX.
%\bibliography{reference}

\end{document}